\documentclass[10pt,smallextended,apalike]{svjour3}       

\usepackage{graphicx} %
\usepackage{mathptmx,natbib}      
\usepackage{epsfig,epic,eepic,color,graphicx,wrapfig ,amsmath,amsbsy,amssymb,amsfonts}
\usepackage{enumitem,appendix,wrapfig}
\usepackage{geometry} 
\usepackage{setspace,subfigure,bm,float,lscape}
\RequirePackage{fix-cm}
\smartqed  

\bibpunct{(}{)}{,}{a}{}{;} 

 \journalname{J Math Biol}

\numberwithin{equation}{section}
\numberwithin{figure}{section}

\newcommand{\bsigma}{{\mbox{\bm{$\sigma$}}}}

\newcommand{\bfu}{{\bm{u}}}
\newcommand{\bff}{{\mbox{\bm{$f$}}}}

 \newcommand{\bfG}{\bm{G}}
 \newcommand\bfF{\bm{F}} 
\newcommand{\bfO}{{\bm{\Omega}}}
\newcommand{\bfL}{{\bm{\Lambda}}}

\newcommand{\stress}{\mbox{$\mathbb{T}$}}

\newtheorem{Principle}{Principle}

\numberwithin{equation}{section}
\numberwithin{figure}{section}

\begin{document}

\title{Analysis of a model  microswimmer with  applications to  blebbing
  cells and mini-robots
\thanks{Supported in part by NSF Grants DMS 0817529 and 1311974 and a
 grant from the Simons Fdn to H. G. Othmer, and by NIH grant R01GM107264 and
  NSF grant DMS1562176 to Qing Nie}
}


\author{Qixuan Wang       \and
        Hans G. Othmer   
}

\institute{Qixuan Wang \at
              540R Rowland Hall\\
              University of California, Irvine\\
              Tel.: (949) 824-3217\\
               \email{qixuanw@uci.edu}     
           \and
          Hans G. Othmer  \at
              School of Mathematics, 270A Vincent Hall \\
              University of Minnesota \\
              Tel.: (612) 624-8325\\
              Fax: (612) 626-2017\\
              \email{othmer@math.umn.edu}      
}

\date{\today}

\maketitle

\setcounter{tocdepth}{2}
\tableofcontents

\bigskip
\begin{abstract}
  Recent research has shown that motile cells can adapt their mode of propulsion
  depending on the environment in which they find themselves.  One mode is
  swimming by blebbing or other shape changes, and in this paper we analyze a
  class of models for movement of cells by blebbing and of nano-robots in a
  viscous fluid at low Reynolds number. At the level of individuals, the shape
  changes comprise volume exchanges between connected spheres that can control
  their separation, which are simple enough that significant analytical results
  can be obtained. The goal is to understand how the efficiency of movement
  depends on the amplitude and period of the volume exchanges when the spheres
  approach closely during a cycle. Previous analyses were predicated on wide
  separation, and we show that the speed increases significantly as the
  separation decreases due to the strong hydrodynamic interactions between
  spheres in close proximity.  The scallop theorem asserts that at least two
  degrees of freedom are needed to produce net motion in a cyclic sequence of
  shape changes, and we show that these degrees can reside in different swimmers
  whose collective motion is studied. We also show that different combinations
  of mode sharing can lead to significant differences in the translation and
  performance of pairs of swimmers.
\end{abstract}

\section{Introduction}

\begin{wrapfigure}[]{r}{3in}
\label{blebbing} 
\centering
\psfig{figure=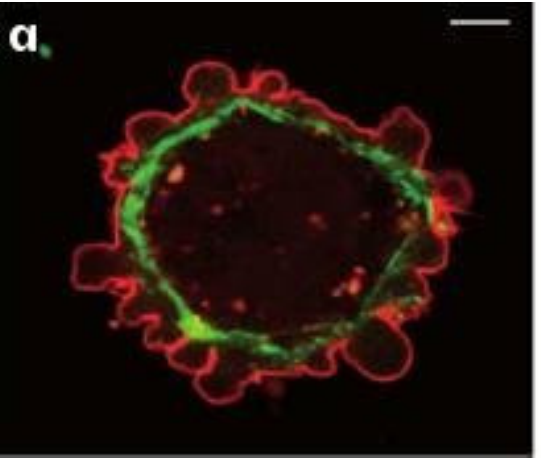,width=1.55in}\psfig{figure=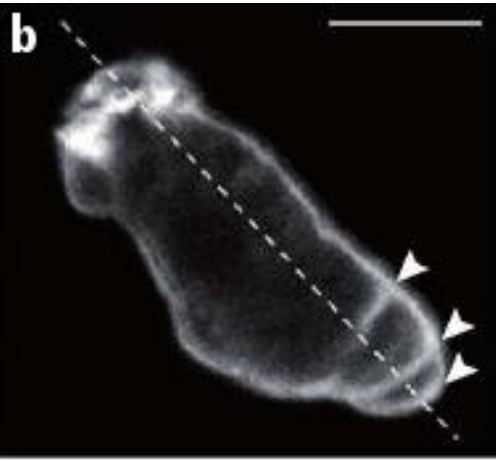,width=1.45in}
\includegraphics[width=3in]{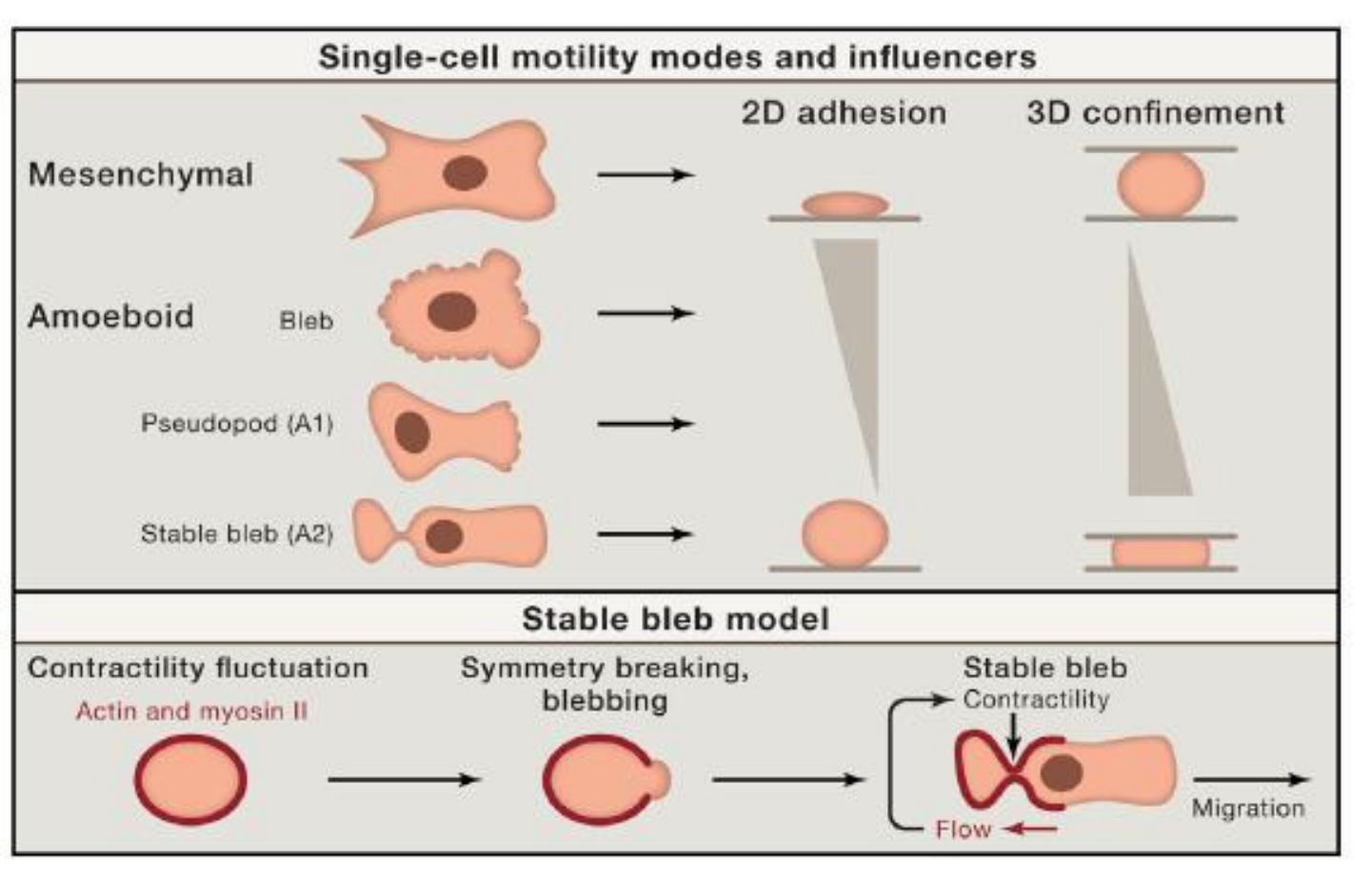} \vspace*{-10pt}
\caption{ (a) Blebbing on a melanoma cell: myosin (green) localizes under the
blebbing membrane (red) (b) The actin cortex of a blebbing Dd cell migrating to
the lower right. Arrowheads indicate the successive blebs and arcs of the actin
cortex (from \protect \cite{Charras:2008:BLW}).  (c) A schematic of the various
modes of movement (from \protect \cite{Welch:2015:CMF}).}
\end{wrapfigure}

Locomotion of cells, both individually and collectively, is an important process
in development, tissue regeneration, the immune response, cancer metastasis, and
wound healing.  The motion of an individual cell is classified as either
mesenchymal or amoeboid, depending on how it interacts mechanically with its
environment \citep{Biname:2010:WMC}.  The mesenchymal mode is used by cells such
as fibroblasts that have a well-organized cytoskeleton, which comprises the
actin filaments, intermediate filaments, and microtubules, and use strong
adhesions to transmit force to their surroundings via integrin-mediated adhesion
complexes. Mesenchymal movement usually involves the extension of broad flat
lamellipodia and/or pseudopodia and is driven by actin polymerization at the
leading edge.  Amoeboid motion involves a less structured cytoskeleton and
weaker surface interactions, and leads to speeds up to forty times faster than
those resulting from mesenchymal motion \citep{Renkawitz:2010:MFG}.  In this
mode cells may use pseudopodia, but can also use protrusions such as blebs (
Figure \ref{blebbing}) which involve blister-like extensions of the membrane.
Leukocytes, which normally use the mesenchymal mode in the extracellular matrix
(ECM), can migrate {\em in vivo} in the absence of integrins, using a 'flowing
and squeezing' mechanism \citep{Lammermann:2008:RLM}. Cells of the slime mold
{\em Dictyostelium discoideum} (Dd) can move either by extending pseudopodia or by
blebbing, and they monitor the stiffness of their surroundings to determine the
mode: pseudopodia in a compliant medium and blebbing in stiffer media
\citep{Zatulovskiy:2014:BDC}.  Furthermore, blebbing cells are efficient in
their chemotactic response to cyclic-AMP, producing nearly all of their blebs
up-gradient.  In certain tumor cells, knockdown of secreted MMPs, which are
enzymes that degrade the ECM, produces only a small reduction in speed because
cells compensate for the decreased proteolysis by undergoing a
`mesenchymal-to-amoeboid transition (MAT)
\citep{Wolf:2003:CMT,Friedl:2003:TCI}. The MAT can also be triggered by changes
in the adhesiveness of the ECM \citep{Friedl:2011:CIM,vanZijl:2011:ISM}.
Moreover, cells such as Dd  and neutrophils can swim in a fluid
environment \citep{Barry:2010:DAN}, and a model of swimming by such cells
appears in \citet{Wang:2015:CAA}. In fact, some cells move only by
blebbing. Certain carcinoma cells in suspension spontaneously polarize and forms
blebs at the leading edge, and while they cannot move on 2D substrates, they can
move in 3D \citep{Bergert:2012:CMC}.

  Thus numerous cell types display enormous plasticity in the choice of
  locomotory mode, in that they sense the mechanical properties of the
  environment and adjust the balance between the modes by adjusting the balance
  between signal transduction pathways that control the structure of the
  cytoskeleton \citep{Renkawitz:2009:AFT,Bergert:2012:CMC,Fackler:2008:CMT}.
  Crawling and swimming are the extremes on a continuum of strategies, but cells
  sense their environment and use the most efficient strategy in a given
  context.  While blebbing is frequently thought of as a 'push-pull' mechanism
  in which a cell expands at the front, followed by contraction at the rear,
  another type of blebbing called 'stable-bleb migration', has recently been
  observed in progenitor cells of the gastrulating zebrafish embryo
  \citep{Maiuri:2015:AFM,Ruprecht:2015:CCT}.  In this mode cells form a
  balloon-like protrusion at their front, and these cells can move more rapidly
  than other cells in the embryo.

  The fact that some cells can use very complicated shape changes for locomotion
  leads to a question posed by experimentalsts, which is `How does deformation
  of the cell body translate into locomotion?'  \citep{Renkawitz:2010:MFG}.  Two
  examples are shown in Fig.~\ref{blebbing}. In (a) is shown  a cell that blebs
  and moves very little, and in (b) is a Dd cell that uses a combination of
  blebs at the front and contraction at the rear to move in a tissue-like
  environment.  Fig. \ref{blebbing} (c) shows the different modes used by cells
  in different environments.  In this paper we analyze a simple model of the
  push-pull type for movement in a fluid by blebbing, motivated by the recent
  experimental findings mentioned above.  Protrusions and other shape changes
  require forces that must be correctly orchestrated in space and time to
  produce net motion -- protrusions on cells in (a) are not, while those in (b)
  are -- and to understand this orchestration one must couple the cellular
  dynamics with the dynamics of the surrounding fluid or ECM.

At the spatial scale of cells and the speeds at which they move, the exterior
fluid problem is a low Reynolds number (LRN) flow.  LRN flows have been
extensively studied in the context of bacterial and sperm movement.  Taylor
\citeyearpar{Taylor:1952:AWC} treated the flagellum as an infinite cylinder
executing small-amplitude oscillations, and Hancock
\citeyearpar{Hancock:1953:SMO} developed a large-amplitude theory using singular
solutions to the Stokes equations situated along the center line of a flagellum,
to describe the motion in three dimensions.  This 'slender-body theory' was
further developed by \citet{Higdon:1979:HFP} to account for hydrodynamic
interactions between the flagellum and the cell head, and \citet{Phan:1987:BEA}
used the boundary-element method to allow for non-spherical heads and
non-slender flagella.  The current state of knowledge is reviewed  in \cite{Elgeti:2015:PMS}.

A separate path in cell locomotion at LRN was stimulated by Purcell's seminal
article \citep{Purcell:1977:LLR} and by interest in mini-robots.  Several models
of LRN swimmers that do not rely on slender-body theory exist. There is
\textit{Purcell's two-hinge swimmer}, the Najafi-Golestanian \textit{ accordion}
model in which three spheres of fixed size are connected by movable links
\citep{Najafi:2004:SSL,Alexander:2009:HLS} , the \textit{push-me-pull-you}
swimmer (PMPY) \citep{Avron:2005:PEM}, in which two spheres that can expand or
contract radially are connected by an extensible arm, and a three-sphere
volume-exchange or \textit{breather} model in which the spheres are linked by
rigid connectors but exchange volume \citep{Wang:2012:MLR}. In
\citet{Wang:2015:PDM} we compared the performance of these models, and showed
that generally the PMPY model is the most efficient. \footnote{The measures of
  performance that are used here and in the literature do not include the work
  needed to move material between spheres in the PMPY and breather models, and
  thus the accordion model suffers by comparison in this respect.}

However, in all analyses of two- or three-sphere models to date, the spheres
were treated as point particles that generate point forces, which in effect
means that the separation between them is large enough that hydrodynamic
interactions between the spheres can be neglected. Thus the conclusions reached
in these analyses are not directly applicable to models of cells that swim by
blebbing, nor to realistic robotic swimmers. Hydrodynamic interactions between
spheres of fixed size have been studied for almost a century, and analytical
results are available for a pair of spheres connected by a rigid rod (a
dumbbell) \citep{Stimson:1926:MTS}. Approximation methods for other
configurations center on either a truncated multipole expansion of the Green's
function or on the method of reflections, both described in
\citet{Kim:1991:MPS}.

In this paper we use the basic PMPY swimmer as a model of blebbing or for
mini-robots to study the swimming behaviors of solitary and group swimmers at
LRN.  Our objective is to understand the effect of higher-order hydrodynamic
interactions between spheres in a PMPY swimmer, and for this we use the reflection method
\citep{Kim:1991:MPS}.  In section \ref{Sec.Wide-Alone} we investigate the
difference between the higher-order solution and the asymptotic solution both
analytically and numerically, and show  that the asymptotic solution may severely
underestimate the effectiveness of a PMPY swimmer. There  we also compare its swimming
behavior to existing experimental data on swimming Dd amoebae. In section
\ref{Sec.Wide-Obstacle} we apply the reflection method to a system consisting of
one active PMPY swimmer and a passive buoyant object. We discuss the
hydrodynamic effect of the velocity field generated by the PMPY swimmer on the
passive object and vice versa, we numerically investigate how effectively the
swimmer can pursue the passive object, and we study the higher-order
hydrodynamic effects on a tracer trajectory. In section \ref{Sec.2PMPY} we apply
the reflection method to a system of two PMPY models, one in which the swimmers
are collinear and the other a planar system, and we discuss the higher-order
hydrodynamic interactions between the two active swimmers. In section
\ref{Sec.ExtScallopThm} we review the \textit{scallop theorem}, which is the
fundamental principle of LRN swimming, and discuss how this extends to the
hydrodynamic interactions amongst several swimmers, each unable to swim on its
own. We also discuss how different combinations of shape change modes affect the
collective swimming behavior of such swimmers.  Throughout we only consider the
regime in which there is sufficient spacing between units so as to justify the
neglect of the lubrication effects that arise when objects in relative motion
are in close proximity. Such effects are discussed by various authors
\citep{Brenner:1961:SMS,Cooley:1969:SMT,Cooley:1969:SMG} in other contexts.


\section{Swimming at low Reynolds number}
\label{Sec.MD}
Hereafter we suppose that the swimmer is immersed in an infinite, incompressible
fluid of density $\rho$ and viscosity $\mu$, that is at rest at infinity.  The
Navier-Stokes equations for the fluid velocity $\bfu$ are
\citep{Childress:1977:MSF}
\begin{align}
\label{NSeqn}
\rho\dfrac{\partial \bfu}{\partial t} +\rho (\bfu \cdot\nabla) \bfu &=  \nabla
 \cdot  \stress  + \bff_{\textrm{ext}}
=- \nabla p  + \mu \Delta \bfu + \bff_{\textrm{ext}} , \\
   \nabla \cdot \bfu &= 0
\end{align}
where 
$$
\stress = -p \bm{\delta} + \mu (\nabla \bfu + (\nabla \bfu)^T)
$$ 
is the Cauchy stress tensor and ${\bf f}_{\textrm{ext}} $ is the external force
field.  We further assume that the swimmer is self-propelled and does not rely
on any exterior force, and therefore we require that $\bff_{\textrm{ext}} =
0$ throughout. Either the swimmer is neutrally buoyant or the gravitational force
is included in the pressure.

When converted to dimensionless form and the symbols re-defined, these
equations read
\begin{eqnarray}
\label{Eq.ReSl_Stokes}
\textrm{ReSl} \dfrac{\partial \bfu}{\partial t} + \textrm{Re} (\bfu \cdot\nabla) = - \nabla p + \mu \Delta \bfu, 
\quad \quad \nabla \cdot \bfu = 0,
\end{eqnarray}
where the Reynolds number based on a characteristic length scale $L$ and a
characterisic speed scale $U$ is Re = $\rho$LU /$\mu$.  In addition,  $\textrm{Sl} =
\omega L/U$ is the Strouhal number and $\omega$ is a characteristic frequency of
the shape changes. When $\textrm{Re} \ll 1$ the convective momentum term in
equation (\ref{Eq.ReSl_Stokes}) can be neglected, but the time variation
requires that $\textrm{ReSl} = \omega \rho L^2/\mu $.  When both terms are
neglected, which we assume throughout, the low Reynolds number (LRN) flow is
governed by the Stokes equations
\vspace*{-5pt}
\begin{equation} 
 \mu \Delta \bfu - \nabla p = {\bf 0}, \qquad \qquad \nabla
\cdot \bfu = 0.
\label{creep}
\end{equation}

Throughout we consider small  cells such as Dd, whose small
size and low speeds lead to LRN flows \citep{Wang:2015:PDM}, and in this regime time does
not appear explicitly, there are no inertial effects, and bodies move by
exploiting the viscous resistance of the fluid. As a result, time-reversible
deformations produce no motion, which is known as the "scallop theorem"
\citep{Purcell:1977:LLR}.  Under the assumptions of an infinite fluid domain
with $\bfu = \bf 0$ at infinity and the absence of external forces, there is no
net force or torque on a self-propelled swimmer in the Stokes regime, and
therefore movement is a purely geometric process: the net displacement of a
swimmer during a stroke is independent of the rate at which the stroke is
executed, as long as the Reynolds number remains small enough.

Let $D (t) \subset\mathbb{R}^3$ be a closed, compact set occupied by the swimmer
at time $t$, and let $\partial D (t)$ denote its prescribed time-dependent
boundary. In reality amoeboid swimming cells may take up or release fluid, but
we assume that the prescribed motion of the boundary is such that the volume of
the swimmer is conserved under all deformations. A \textit{swimming stroke} is
specified by a time-dependent sequence of shapes, and it is \textit{cyclic} if
the initial and final shapes are identical, i.e., $\partial D(0) = \partial D
(T)$, where $T$ is the period.  The swimmer's boundary
velocity $\mathbf{V}$ relative to fixed coordinates can be written as a part
$\mathbf{v}$ that defines the intrinsic shape deformations, and a rigid motion
part $\mathbf{U} + \bfO \times \mathbf{x}$, where $\mathbf{U}, \bfO$ are the
rigid translation and rotation, resp..  If $\mathbf{u}$ denotes the
velocity field in the fluid exterior to $D$, then a standard LRN self-propulsion
problem is: \textit{given a cyclic shape deformation specified by $\mathbf{v}$,
  solve the Stokes equations (\ref{creep}) subject to }
\begin{eqnarray*}
\int_{\partial D} \bsigma \cdot \mathbf{n} = 0, \quad 
\int_{\partial D} \mathbf{x} \times (\bsigma \cdot \mathbf{n} ) = \mathbf{0}, \quad
\mathbf{u}|_{\mathbf{x} \in \partial D} = \mathbf{V} = \mathbf{v} + \mathbf{U} + \bfO \times \mathbf{x}, \quad
\mathbf{u} |_{\mathbf{x} \rightarrow \infty} = \mathbf{0}
\end{eqnarray*}
where $\mathbf{n}$ is the exterior normal, and the integrals are the force- and torque-free conditions.


\setcounter{equation}{0}
\renewcommand{\theequation}{\ref{Sec.Wide-Alone}.\arabic{equation}}

\section{The solitary PMPY swimmer }
\label{Sec.Wide-Alone}

\begin{figure}[htbp] \centering
\includegraphics[width=0.5\textwidth]{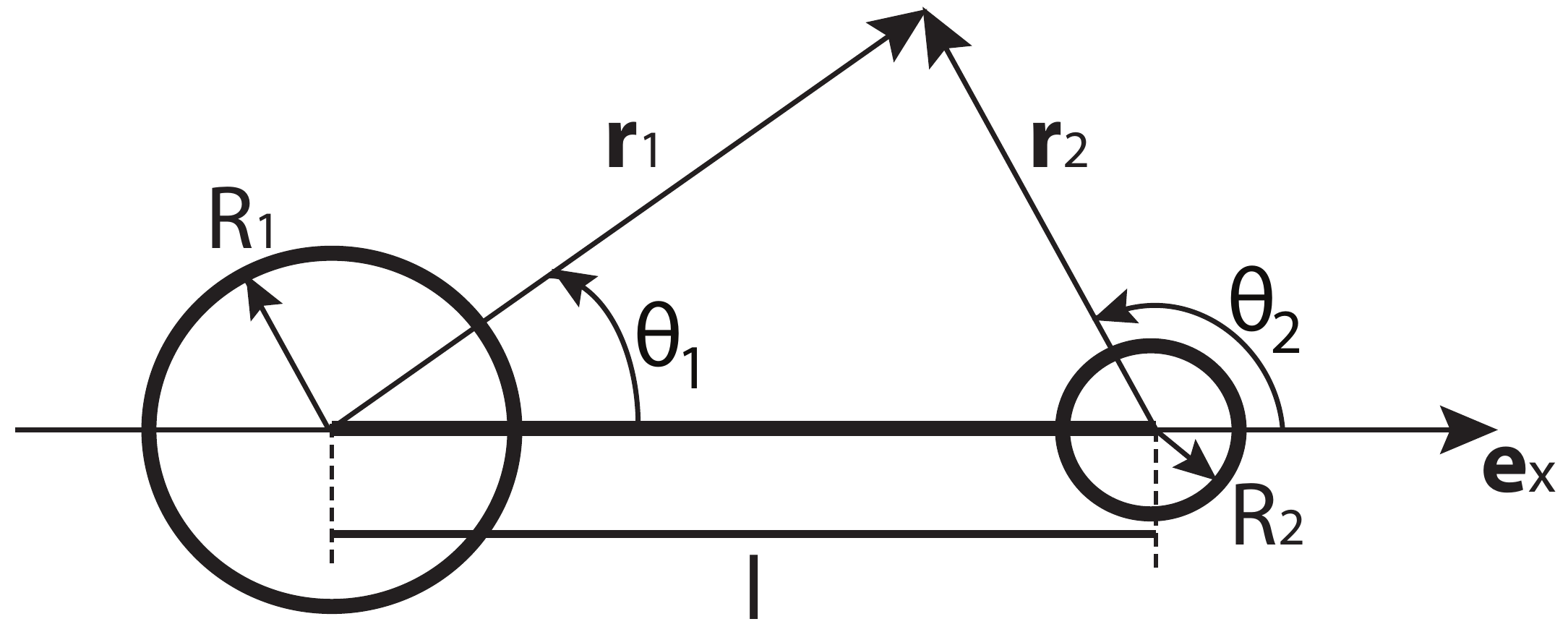}
\caption{Geometry of the pushmpullyou model.}
 \label{fig.PMPY_1}
\end{figure}

The simplest PMPY model consists of two spheres that can expand or contract
radially, and an extensible, massless rod connecting them
(Figure~$\ref{fig.PMPY_1}$).  
The standard assumptions on the domain and the fluid, which is described by the
Stokes' equations (\ref{creep}), apply here.
Let $R_i (t)$ be the radius of the $i$th sphere ($i=1,2$) and $l(t)$ the length
of the rod.  The prescribed motion of each sphere consists of two parts: a rigid
translation along the $x$-axis at velocity  $\mathbf{U}_ i = U_i \mathbf{e}_x$, and a radial
expansion or contraction: $\dot{R}_i$. Thus the no-slip boundary conditions on
the surfaces of the two spheres can be expressed as:
\begin{eqnarray}\label{eq.2sphere_BDCond} \mathbf{u} (\mathbf{r}_i;t) = U_i (t)
\mathbf{e}_x + \dot{R}_i (t) \hat{\mathbf{r}}_i \quad \textrm{at} \quad
|\mathbf{r}_i| = R_i (t), \qquad i=1,2
\end{eqnarray}
where $\mathbf{r}_i$ is the radius vector with origin at the center of the $i$th
sphere (Figure~$\ref{fig.PMPY_1}$) and $\hat{\mathbf{r}}_i$ is the outward unit
vector along the $\mathbf{r}_i$ direction.  The
instantaneous velocity of the swimmer is defined as the average of the
velocities of the two spheres
\begin{eqnarray}
\label{eq.2sphere_TotV} 
\overline{\mathbf{U}} =
  \dfrac{\mathbf{U}_1 + \mathbf{U}_2}{2} = \dfrac{U_1 + U_2}{2} \mathbf{e}_x,
\end{eqnarray}
although other measures such as the velocity of the center of mass could also be
used. The instantaneous velocity 
would change in the
latter case, but the net translation during a cycle would not. 

The rate of change of the length of the connecting rod is $\dot{l} (t)$,
and thus the velocities of the two spheres are related by:
\begin{eqnarray}
\label{eq.2sphere_U12_l} 
U_2 - U_1 = \dot{l}.
\end{eqnarray}
We assume that the total volume $V_1 +V_2 \equiv V_T$ is conserved during the
motion, and thus the radii of the spheres satisfy the constraint
\begin{eqnarray}
\label{eq.2shpere_volume} 
R_1^3 (t) + R_2^3 (t) \equiv \dfrac{3}{4 \pi} V_{\textrm{T}}
\end{eqnarray}
In addition, the PMPY swimmer is force- and torque-free, and while the swimmer's
linear geometry automatically guarantees that it is torque-free in the absence
of asymmetric shear forces, the force-free constraint is non-trivial.  Let
$\mathbf{F}_i (t)$ be the hydrodynamic force due to drag and expansion exerted
on the $i$th sphere at time $t$ -- then the constraint is that
\begin{eqnarray}
\label{eq.2sphere_force_free} 
\sum_{i=1,2} \mathbf{F}_i (t) \equiv \mathbf{0}.
\end{eqnarray}
It is clear that extension and contraction of the rod produces a direct effect
on the translation of either sphere, whereas the expansion only has an indirect
effect. If sphere one expands this has no direct effect on its movement since
the expansion is radially symmetric, but the flow generated affects the second
sphere. Simultaneously, the conservation of volume condition produces a
reduction in size of the second sphere, which induces a flow that affects the
first sphere.  In the LRN regime these effects are felt
instantaneously.\footnote{The reader can easily show that in the absence of
  volume exchange the net translation after a periodic extension and contraction
  of the rod produces no motion. This provides the simplest example of the
  scallop theorem discussed later.}

The shape changes for a PMPY swimmer are described by $\dot{l}, \dot{R}_1$ and
$\dot{R}_2$, but in view of (\ref{eq.2shpere_volume}) two degrees of freedom
define its motion.  A cyclic stroke of a PMPY swimmer is determined by a
periodic profile $(\dot{l} (t), \dot{R}_1 (t))$ for $ \ t \in [0, T]$, and a
solution of the swimmer problem entails finding the relation between the
swimmer's velocity $\overline{\mathbf{U}}$ and the controls $(\dot{l},
\dot{R}_1)$. In the general analysis that follows we assume that the controls 
are chosen so that the motion is not time reversible. An example of how the 
 choice of phase difference between $\dot{l} (t))$  and $\dot{R}_1 (t)$ affects
 the efficiency when both vary sinusoidally  is given later. 


\subsection{Scaling the  PMPY problem}
\label{Sec.SLOA}

The PMPY model contains three length scales: the radii of the spheres $R_1,
R_2$ and the length of the connecting arm $l$. The geometry of the model
requires that the two spheres never overlap, hence $R_1 + R_2 < l$.  As
previously defined, $Re = \rho L U /\mu \ll 1$, and therefore all lengths in the
model must be small enough  to ensure that 
\begin{eqnarray*} R_1, R_2, l \ll \dfrac{\mu}{\rho V}
\end{eqnarray*}
However even if this LRN pre-condition is satisfied, the relations among the
lengths $R_1, R_2, l$ are also crucial in the swimming problem, and different
relations may lead to different regimes of interaction \citep{Kim:1991:MPS}.  We
assume that the radii of the two spheres are comparable, i.e., $O (R_1) \sim O
(R_2)$, which rules out the possibility of interactions between a large sphere
and a very small one (this case is discussed in \cite{Kim:1991:MPS}).  This
leaves three major scenarios: the spheres are in close proximity in part of the
cycle, {\em i.e.}, $R_1(t)+R_2(t) \sim l(t) $ in part of the cycle; the spheres
are widely-separated spheres throughout the cycle, where $R_1(t) + R_2/ \ll l(t)
$, and an intermediate regime. In the first regime the flow in the gap region
dominates when the spheres nearly touch in part of the cycle, and the
lubrication approximation provides the leading terms in an asymptotic expansion
\citep{Kim:1991:MPS}. If the spheres are also well-separated in part of the
cycle this leads to a difficult matching problem that is not attempted here. We
only consider the intermediate regime in which the separation never enters the
lubrication regime, but is also not in the infinitely-separated regime
throughout the cycle, as this has been studied
previously \citep{Avron:2005:PEM}. Our objective is to use the reflection method
to determine corrections to the problem with very large separation.

Since the length scales involved are time-dependent, some care is needed in
setting an appropriate scaling for the lengths. The primary criterion that must
be met is that $(R_1(t)+R_2(t))/l(t) < 1 $ throughout the cycle. 
%
%
%
%
Define
\begin{eqnarray*} R_M = \max_{t} \{ R_i (t) \}_{i=1,2}, \qquad L_m = \min_{t} \{
l (t) \}
\end{eqnarray*}
and 
$$
\delta = R_M/L_m,
$$ 
 We then nondimensionalize the radii and rod lengths
by $R_M$ and $L_m$, resp., 
\begin{eqnarray}\label{eq.ScaleRL_Range}
 \hat{R}_i = \dfrac{R_i}{R_M} \leq 1, \quad \hat{l} =
\dfrac{l}{L_m} \geq 1,
\end{eqnarray}
and we assume that the amplitudes of both rod displacement and radius changes
are of the same order of $R_M$. Thus we let $\xi = \dot{l}$ and $\zeta_i =
\dot{R}_i$,  and apply the following scaling:
\begin{eqnarray*} \hat{\xi} = \dfrac{T}{R_M} \xi, \qquad \hat{\zeta}_i =
\dfrac{T}{R_M} \zeta_i
\end{eqnarray*}
Next, we nondimensionalize other length scales by $R_M$ as well, and time by
the period $T$ to obtain
\begin{eqnarray*} \hat{\mathbf{x}} = \dfrac{\mathbf{x}}{R_M}, \quad \hat{\nabla}
= R_M \nabla, \quad \hat{t} = \dfrac{t}{T}, \quad \hat{\mathbf{u}} =
\dfrac{T}{R_M} \mathbf{u}, \quad \hat{\mathbf{U}}_i = \dfrac{T}{R_M}
\mathbf{U}_i, \quad \hat{\bfO}_i = T \bfO_i.
\end{eqnarray*}
Finally, the drag force $\mathbf{F}$ exerted on a sphere of radius $R$ is related to the
sphere velocity $\mathbf{U}$ via  $\mathbf{F} = 6 \pi \mu R
\mathbf{U}$, which leads to the following scaling for forces:
\begin{eqnarray}\label{eq.ScaleForce} \hat{\mathbf{F}} = \dfrac{T}{6 \pi \mu
R_M^2} \mathbf{F} = \hat{R} \hat{\mathbf{U}}.
\end{eqnarray}


\subsection{The reflection method and the Rotne-Prager-Yamakawa approximation}
\label{Sec.RPY}

In previous analyses of the PMPY swimmer the spheres are infinitely-separated,
and thus treated as the source of point forces. The  free space Green's
function, or Stokeslet, for the Stokes problem  is
\begin{equation}
\label{green}
\bfG(\bm{x},\bm{x}_0)  =  \dfrac{1}{r}\left[\bm{I} +
  \dfrac{\bm{r}\bm{r}}{r^2}\right]=  \dfrac{1}{r}\left[\bm{I} +
  {\bm{\hat r}\bm{\hat r}}\right] .
\end{equation}
Here $\bm{I}$ is the unit second-rank tensor, $\bm{r} = \bm{x}-\bm{x}_0 $, $r =
|\bm{x}-\bm{x}_0|$ and $\bm {\hat r} = \bm{r}/|{\bm{r}}|$. Thus the velocity field
  generated at a point $\bm{x}$ by a point force $\bfF$ at $\bm{x}_0$ is
\begin{equation}
\label{poinf}
\bm{u}(\bm{x}) = \dfrac{\bfG(\bm{x},\bm{x}_0) }{8 \pi \mu}\cdot \bfF(\bm{x}_0 ).
\end{equation}
When combined with the flow field generated by expansion of the spheres ({\em
  cf.} Appendix A), this leads to the following approximation for the velocity
  of the swimmer \citep{Avron:2005:PEM}.
\begin{eqnarray}
\label{eq.AvronSoln} 
\overline{U} = \dfrac{R_1 - R_2}{2 (R_1 +
R_2)} \dot{l} + \big( \dfrac{R_1}{l} \big)^2 \dot{R}_1
\end{eqnarray}
The  first and second terms are the leading order terms that result from the
cyclic change of the rod length and the contraction and expansion of the
spheres, resp.. If we  nondimensionalize this solution we obtain 
\begin{eqnarray}
\label{eq.NonDim-AvronSoln} 
\hat{\overline{U}} =
\dfrac{\hat{R}_1 -\hat{ R}_2}{2 (\hat{R}_1 + \hat{R}_2)} \hat{\xi} + \big(
\dfrac{\hat{R}_1}{\hat{l}} \big)^2 \delta^2 \hat{\zeta}_1.
\end{eqnarray}
Here we see that the first term in equation~($\ref{eq.AvronSoln}$) is $O(1)$
while the second is $O(\delta^2)$. Therefore we expect that at least the $O
(\delta)$ and $O (\delta^2)$ terms that result from the movement of the spheres
should be taken into account.  Said otherwise, equation~($\ref{eq.AvronSoln}$)
is only accurate to $O (1)$.

One approach to obtaining the higher-order effects of the interactions between
the spheres in the PMPY model, both from their relative motion and the volume
changes, is the reflection method \citep{Kim:1991:MPS}.  A general description
of the the algorithm underlying this method is as follows.
 \begin{itemize}
 \item[$\bullet$] \textbf{The 0-th reflection.} The 0-th reflection for either
   sphere is simply the superposition of this sphere alone in whatever
   background flow, call it $\mathbf{u}_{\infty}$, exists.  In another words, in
   this step no interactions between the spheres are taken into
   consideration. We denote the velocity field that results from translation and
   expansion/contraction of the $i$-th sphere by $\mathbf{u}_i^{(0)} (\mathbf{x})$,
   and the sum of these is the new field $\mathbf{u}^{(0)} (\mathbf{x})$. 
   This step leads to the $O(1)$ term in (\ref{eq.AvronSoln}). 
%
 \item[$\bullet$] \textbf{The 1st reflection.} However, in the combined field
   $\mathbf{u}^{(0)} (\mathbf{x})$ the no-slip boundary conditions on each sphere are
   not met and to correct this one computes a new field by solving two new
   Stokes problems: one with the boundary value -$\mathbf{u}_1^{(0)} (\mathbf{x})$ on
   sphere 2, and one with the value -$\mathbf{u}_2^{(0)} (\mathbf{x})$ on sphere
   1. This leads to two new fields $\mathbf{u}_{12}^{(1)} (\mathbf{x})$ and
   $\mathbf{u}_{21}^{(1)} (\mathbf{x})$, and the first reflection field is
   $\mathbf{u}^{(1)} (\mathbf{x}) = \mathbf{u}_{12}^{(1)} (\mathbf{x}) + \mathbf{u}_{21}^{(1)}
   (\mathbf{x})$.
\smallskip
\item[$\bullet$] \textbf{The 2nd reflection.} In the second reflection two new
  corrections that satisfy -$\mathbf{u}_{12}^{(1)} (\mathbf{x})$ and
  -$\mathbf{u}_{21}^{(1)} (\mathbf{x})$, on sphere 1 and 2,  respectively, are computed,
  and are called $\mathbf{u}_{121}^{(2)} (\mathbf{x})$, and $\mathbf{u}_{212}^{(2)}
  (\mathbf{x})$. The resulting second reflection field is $\mathbf{u}^{(2)} (\mathbf{x}) =
 \mathbf{u}_{121}^{(2)} (\mathbf{x}) + \mathbf{u}_{212}^{(2)}  (\mathbf{x})$.
\smallskip
 \item[$\bullet$]  Repeat until the desired accuracy is achieved. 
 \end{itemize}
 This process has been proven to converge very rapidly for finite domains
 \citep{Luke:1989:CMR}.

 In a LRN self-propulsion problem the flows vanish at infinity  and the velocity
 generated by a sphere of radius $R$, centered at $\mathbf{x}_0$, subject to a
 force $\mathbf{F}$, and `expanding' at the rate $\dot{R} = \textrm{d} R /
 \textrm{d} t$ is (Appendix~\ref{Appendix.A}):
\begin{eqnarray}
\label{eq.VelocityField_SingleSphere} 
\mathbf{u} \big(
\mathbf{r}; R, \mathbf{F}, \dot{R} \big) = \dfrac{1}{24 \pi \mu r} \Big[ \big( 3
+ \dfrac{R^2}{r^2} \big) \mathbf{F} + 3 \big( 1 - \dfrac{R^2}{r^2} \big) \big(
\mathbf{F} \cdot \widehat{ \mathbf{r} } \big) \widehat{\mathbf{r}} \Big]
+\dot{R} \Big( \dfrac{R}{r} \Big)^2 \widehat{\mathbf{r}}.
\end{eqnarray}
The first term and second terms in
equation~($\ref{eq.VelocityField_SingleSphere}$) result from the drag force
$\mathbf{F}$ and the radial change $\dot{R}$, resp., and we denote them
by $\mathbf{u} \{ \mathbf{F} \}$ and $\mathbf{u} \{ \dot{R} \}$:
\begin{eqnarray*}
\mathbf{u} \big( \mathbf{r}; R, \mathbf{F}, \dot{R} \big) &=& \mathbf{u} \{ \mathbf{F} \} (\mathbf{r} ;R )
+ \mathbf{u} \{ \dot{R} \} (\mathbf{r} ; R) \\
 \mathbf{u} \{ \mathbf{F} \} (\mathbf{r} ;R ) &=& \dfrac{1}{24 \pi \mu r} \Big[ \big( 3
+ \dfrac{R^2}{r^2} \big) \mathbf{F} + 3 \big( 1 - \dfrac{R^2}{r^2} \big) \big(
\mathbf{F} \cdot \widehat{ \mathbf{r} } \big) \widehat{\mathbf{r}} \Big] \\
\mathbf{u} \{ \dot{R} \} (\mathbf{r} ; R) &=& \dot{R} \Big( \dfrac{R}{r} \Big)^2 \widehat{\mathbf{r}}.
\end{eqnarray*}

The symmetric geometry of the PMPY model  implies that the angular
velocities $\bfO_1$  and $\bfO_2 = \mathbf{0}$  both vanish, while each
reflection contributes to the translational velocity. We denote the translational velocity of sphere $i$
that results from the $n$-th reflection as $\mathbf{U}_i^{(n)}$. In the zeroth
reflection, i.e., when we consider the $i$th sphere ($i=1,2$) of the PMPY model
alone immersed in the fluid, the radial expansion will not result in any rigid
motion, but the drag force $\mathbf{F}_i$ leads to the  translational component
\begin{eqnarray}\label{eq.RPY_U_Lead} \mathbf{U}_{i}^{(0)} = \dfrac{1}{6 \pi \mu
R_i} \mathbf{F}_i
\end{eqnarray}

The first reflection is computed as follows. Given the velocity of sphere $j$,
we have to find the solution of a Stokes problem with velocity $-\mathbf{u}_j^{(0)}$ on
the surface of sphere $i$ and vanishing at infinity, and from that compute the
translational and rotational velocities at the first reflection. We use the
reciprocal theorem as used in \citep{Stone:1996:PMS} to do this. Let
$(U_{i}^{(1)},\stress_{i}^1)$ be the first-reflection translational velocity of,
and stress on, sphere $i$, and similarly for sphere j. Then a version of the
reciprocal theorem used here states that $\nabla \cdot \stress_{j}^0\cdot
\mathbf{U}_i^{1} = \nabla \cdot \stress_{i}^1\cdot \mathbf{u}_j $ and from this
and a similar equivalence for the torques it follows that
\begin{eqnarray}
\label{eq.RecLaw_Force} 
\mathbf{U}_{i}^{(1)} &=& \dfrac{1}{4 \pi
R_i^2} \int_{S_i} \mathbf{u}_j^{(0)} ( \mathbf{x}) \ \textrm{d} S ( \mathbf{x} )
\\ 
\label{eq.RecLaw_Torque} \bfO_{i}^{(1)} &=& \dfrac{3}{8 \pi R_i^3} \int_{S_i}
\mathbf{n} \times \mathbf{u}_j^{(0)} ( \mathbf{x}) \ \textrm{d} S ( \mathbf{x} ).
\end{eqnarray}
Since we are considering spheres,  equivalent expressions  obtained by use of 
Fax\'{e}n's  law \citep{Kim:1991:MPS} are 
\begin{eqnarray}
\label{eq.RecLaw_Force_Faxen} 
\mathbf{U}_{i}^{(1)} &=&\Big( 1 +
\dfrac{R_i^2}{6} \nabla^2 \Big) \mathbf{u}_j^{(0)} \Big|_{\mathbf{x} = \mathbf{x}_i}
\\ \label{eq.RecLaw_Torque_Faxen} \bfO_{i}^{(1)} &=& \dfrac{1}{2} \nabla \times
\mathbf{u}_j^{(0)} \Big|_{\mathbf{x} = \mathbf{x}_i}
\end{eqnarray}
Once again, $\bfO_{i}^{(1)} = \mathbf{0}$ due to the symmetry of the model, and
by using (\ref{eq.VelocityField_SingleSphere}) in (\ref{eq.RecLaw_Force}), the 
translational component is
\begin{eqnarray}
\label{eq.RPY_U_X_1}
 \mathbf{U}_{i}^{(1)} = \dfrac{1}{4 \pi
R_i^2} \int_{S_i} \mathbf{u}_j^{(0)} \ \textrm{d} S = \dfrac{1}{4 \pi R_i^2}
\int_{S_i}  \mathbf{u} \{ \mathbf{F} \} (\mathbf{r} ;R )  \ \textrm{d} S + \dfrac{\dot{R}_j}{4
\pi } \Big( \dfrac{R_j}{R_i} \Big)^2 \int_{S_i} \dfrac{\mathbf{x} -
\mathbf{x}_j}{| \mathbf{x} - \mathbf{x}_j |^3 } \ \textrm{d} S (\mathbf{x})
\end{eqnarray}
We denote the first and second terms in equation~($\ref{eq.RPY_U_X_1}$) as
$\mathbf{U}_{i}^{(1,f)}$ and $\mathbf{U}_{i}^{(1,e)}$, resp., and then 
equation~($\ref{eq.RPY_U_X_1}$) can be written as
\begin{eqnarray*} \mathbf{U}_{i}^{(1)} = \mathbf{U}_{i}^{(1,f)} +
\mathbf{U}_{i}^{(1,e)}
\end{eqnarray*}

$\mathbf{U}_{i}^{(1,f)}$ is the result of  the drag force $\mathbf{F}_j$, and it has
been well studied. The result is given by the Rotne-Prager-Yamakawa (RPY)
approximation \citep{Yamakawa:1970:TPP,Wajnryb:2013:GRP,
Zuk:2014:RPY,Liang:2013:FMM}:
\begin{eqnarray}
\label{eq.RPY_U_X_Force_1} 
\mathbf{U}_{i}^{(1,f)} = \dfrac{1}{8
\pi \mu l} \Big[ \Big( 1 + \dfrac{R_1^2 + R_2^2}{3 l^2} \Big) \mathbf{I} + \Big(
1 - \dfrac{R_1^2 + R_2^2}{ l^2} \Big) \dfrac{ (\mathbf{x}_i - \mathbf{x}_j)
\otimes (\mathbf{x}_i - \mathbf{x}_j) }{|\mathbf{x}_i - \mathbf{x}_j|^2} \Big]
\mathbf{F}_j \quad
\end{eqnarray}
and together with the relations $\mathbf{U}_{i}^{(1,f)} = U_{i}^{(1,f)}
\mathbf{e}_x$ and $\mathbf{F}_j = F_j \mathbf{e}_x$, we find that 
\begin{eqnarray}
\label{eq.RPY_U_X_Force_2} 
\mathbf{U}_{i}^{(1,f)} = \dfrac{1}{4
\pi \mu l} \Big[ 1 - \dfrac{R_1^2 + R_2^2}{3 l^2} \Big] F_j.
\end{eqnarray}
In principle, the RPY approximation is a first order correction to the Oseen
hydrodynamic interaction tensor, and thus allows for closer separation of the
spheres.  However, since the idea of the RPY approximation comes from the reflection
method \citep{Kim:1991:MPS}, the distance between any two spheres should still
be kept within the regime in which the method can be applied.
 
$ \mathbf{U}_{i}^{(1,e)}$ is due to the expansion $\dot{R}_j$, which  can be
calculated directly (Appendix \ref{Appendix.C}) or by Faxen's law, 
\begin{eqnarray}\label{eq.RPY_U_X_Exp} \mathbf{U}_{i}^{(1,e)} = (-1)^i \dot{R}_j
\Big( \dfrac{R_j}{l} \Big)^2 \mathbf{e}_x
\end{eqnarray}
which is precisely the radial expansion term in Avron's solution
(equation~($\ref{eq.AvronSoln}$)).  This completes the first reflection, and 
equations~($\ref{eq.RPY_U_Lead},\ref{eq.RPY_U_X_Force_2},\ref{eq.RPY_U_X_Exp}$)
give the following approximation of the translational velocities of the spheres
after one reflection
\begin{eqnarray}\label{RPY_Ui} U_i \sim U_i^{(0)} + U_i^{(1)} = \dfrac{F_i}{6 \pi
\mu R_i} + \dfrac{F_j}{4 \pi \mu l} \Big( 1 - \dfrac{R_1^2 + R_2^2}{3 l^2} \Big)
+ (-1)^i \dot{R}_j \Big( \dfrac{R_j}{l} \Big)^2.
\end{eqnarray}
Using the nondimensionalization given earlier,  the nondimensional version  of
equation~($\ref{RPY_Ui}$) is:
\begin{eqnarray}\label{eq.Scale_RPY_Ui} \hat{U}_i \sim
\dfrac{\hat{F}_i}{\hat{R}_i} + \dfrac{3}{2} \delta \dfrac{\hat{F}_j}{\hat{l}}
\Big( 1 -\delta^2 \dfrac{\hat{R}_1^2 + \hat{R}_2^2}{3 \hat{l}^2} \Big) + (-1)^i
\delta^2 \Big( \dfrac{\hat{R}_j}{\hat{l}} \Big)^2 \hat{\zeta}_j
\end{eqnarray}
together with the
equations~($\ref{eq.2sphere_force_free},\ref{eq.2sphere_U12_l}$) we obtain  the
following closed system: 
\begin{eqnarray}
\label{eq.RPY_System} 
\left(
\begin{array}{cccc} 
-1 & 0 & \dfrac{1}{ \hat{R}_1} & \Gamma  \\ 0 & -1 & \Gamma  & \dfrac{1}{
\hat{R}_2} \\ 1 & -1 & 0 & 0 \\ 0& 0& 1 & 1
\end{array} 
\right)
\left(
\begin{array}{r} \hat{U}_1 \\ \hat{U}_2 \\ \hat{F}_1 \\ \hat{F}_2
\end{array} \right) = \left(
\begin{array}{c}   \Gamma_2 \\  -\Gamma_1 \\ -\hat{\xi} \\0 
\end{array} \right)  \quad
\end{eqnarray} 
where
$$
\Gamma  \equiv \delta\dfrac{3}{2\hat{l}} \Big( 1 - \delta^2 \dfrac{\hat{R}_1^2 + \hat{R}_2^2}{3
\hat{l}^2} \Big) \qquad \textrm{and} \qquad \Gamma_i  \equiv  \delta^2\Big( \dfrac{\hat{R}_i}{\hat{l}} \Big)^2
\hat{\zeta}_i .
$$

It is easy to see that the matrix in (\ref{eq.RPY_System}) is non-singular and
can be inverted explicitly. Therefore the solution of ($\ref{eq.RPY_System}$)
can be expressed as a power series in $\delta$, and since this solution 
represents an approximation stemming from the first reflection, we must
determine how accurate it is and whether further reflections are justified.  If
we compute a second reflection, the translational velocity that results from
drag forces from the $0^{th}$ up to the $2^{nd}$ reflections for sphere 1 is 
found to be \citep{Kim:1991:MPS}:
\begin{eqnarray} \label{eq.RM_U_Force_O4} \sum_{n=0}^2 U_1^{(n,f)} =
\dfrac{F_1}{6 \pi \mu R_1} \Big( 1 - \dfrac{15 R_1 R_2^3 }{4 l^4} \Big) +
\dfrac{F_2}{4 \pi \mu l } \Big( 1 - \dfrac{R_1^2 + R_2^2}{3 l^2} \Big) + O \big(
\dfrac{1}{l^5} \big),
\end{eqnarray}
and after nondimensionalization this reads 
\begin{eqnarray} \label{eq.Scale_RM_U_Force_O4} \sum_{n=0}^2 \hat{U}_1^{(n,f)} =
\dfrac{\hat{F}_1}{ \hat{R}_1} \Big( 1 - \dfrac{15 }{4 } \delta^4
\dfrac{\hat{R}_1 \hat{R}_2^3}{\hat{l}^4} \Big) + \dfrac{3}{2} \delta
\dfrac{\hat{F}_2}{ \hat{l} } \Big( 1 - \delta^2 \dfrac{ \hat{R}_1^2 +
\hat{R}_2^2}{3 \hat{l}^2} \Big) + O \big( \delta^5 \big).
\end{eqnarray}
One can show that a second reflection for the component due to radial expansion
will lead to the correction $U_i^{(2,e)}$ of $O (\delta^5)$ ({\it cf } Appendix
\ref{Appendix.C}). Hence with one reflection, the solution is accurate up to $O (\delta^4)$.

The perturbation expansions of the solution to equation~($\ref{eq.RPY_System}$)
 to  $O (\delta^4)$ order are:\footnote{In the remainder we omit the $~
  \hat{}~ $ on nondimensionalized quantities for simplicity, but since $\delta$
  appears in the equations this should not lead to any confusion.}
\begin{eqnarray}\nonumber U_1 &\sim& - \dfrac{R_2}{R_1 + R_2} \xi + \delta
\dfrac{3 R_1 R_2 (R_1 - R_2)}{2 (R_1 + R_2)^2 l } \xi + \dfrac{\delta^2}{l^2}
\Big[ \dfrac{9 R_1^2 R_2^2 (R_1 - R_2 )}{2 (R_1 + R_2)^3} \xi - R_2^2 \zeta_2
\Big] \\ \label{eq.PT_U1} & & + \delta^3 \dfrac{R_1 R_2 (R_1 - R_2)}{2 (R_1 +
R_2)^4 l^3} \Big( 25 R_1^2 R_2^2 - R_1^4 - R_2^4 - 2 R_1^3 R_2 - 2 R_1 R_2^3
\Big) \xi + O (\delta^4 ) \qquad \\ \nonumber U_2 &\sim& \dfrac{R_1}{R_1 + R_2}
\xi + \delta \dfrac{3 R_1 R_2 (R_1 - R_2)}{2 (R_1 + R_2)^2 l} \xi +
\dfrac{\delta^2}{l^2} \Big[ \dfrac{9 R_1^2 R_2^2 (R_1 - R_2 )}{2 (R_1 + R_2)^3}
\xi + R_1^2 \zeta_1 \Big] \\ \label{eq.PT_U2} & & + \delta^3 \dfrac{R_1 R_2 (R_1
- R_2)}{2 (R_1 + R_2)^4 l^3} \Big( 25 R_1^2 R_2^2 - R_1^4 - R_2^4 - 2 R_1^3 R_2
- 2 R_1 R_2^3 \Big) \xi + O (\delta^4 ) \qquad \\ \nonumber
%
 %
- F_1 = F_2 &\sim& \dfrac{R_1 R_2}{ R_1 + R_2} \xi +
3 \dfrac{\delta}{l} \Big( \dfrac{R_1 R_2}{ R_1 + R_2} \Big)^2 \xi + 9
\dfrac{\delta^2}{l^2} \Big( \dfrac{R_1 R_2}{ R_1 + R_2} \Big)^3 \xi
\\ \label{eq.PT_Force} & & + \dfrac{\delta^3}{l^3} \dfrac{ ( R_1 R_2 )^2}{ (R_1
+ R_2)^4} \Big( 25 R_1^2 R_2^2 - R_1^4 - R_2^4 - 2 R_1^3 R_2 - 2 R_1 R_2^3 \Big)
\xi+ O (\delta^4 ) \qquad
 \end{eqnarray}
 Finally, the velocity of the PMPY model,  correct to $O (\delta^3)$, and subject to  the pair of controls
$(\xi, \zeta_1)$, is 

\begin{eqnarray} \nonumber \overline{U} &=& \dfrac{R_1 - R_2}{2 (R_1 + R_2)}
\Big[1 + 3 \dfrac{\delta}{l} \dfrac{R_1 R_2}{R_1 + R_2 } + 9
\dfrac{\delta^2}{l^2} \Big( \dfrac{R_1 R_2}{R_1 + R_2 } \Big)^2
\\ \label{eq.PT_AvgU} & & + \dfrac{\delta^3}{l^3} \Big( \dfrac{R_1 R_2}{R_1 +
R_2 } \Big)^3 \Big( 25 - \dfrac{R_1^2 }{R_2^2} - \dfrac{R_2^2 }{R_1^2} - 2
\dfrac{R_1 }{R_2} - 2 \dfrac{R_2 }{R_1} \Big) \Big] \xi+ \delta^2
\Big(\dfrac{R_1}{l} \Big)^2 \zeta_1 + O(\delta^4).
 \end{eqnarray}
 If we compare equation~($\ref{eq.PT_AvgU}$) with
 equation~($\ref{eq.NonDim-AvronSoln}$), we see that the earlier analysis
 ignores all effects due to finite separation of the spheres, and thus
 ($\ref{eq.NonDim-AvronSoln}$) is only valid to zeroth-order in $\delta$.


 \subsection{Power expenditure and the performance of a PMPY}
 \label{Sec.Perf_PMPY}
 
 Next we consider the power $P (t)$ required to propel the swimmer. For a PMPY
model, the power $P$ comprises two parts: $P_{\,D} $ that results from
the drag force on the spheres, and $P_{\,V} $ that results from the
radial expansion of the spheres.  $P_{\,D}  $ is given by
\begin{eqnarray*} P_{\,D}  = F_1 U_1 + F_2 U_2
 \end{eqnarray*}
 which can be simplified by the force-free condition
(equation~$\ref{eq.2sphere_force_free}$) and the geometric relation between the
two spheres (equation~$\ref{eq.2sphere_U12_l}$) to the form 
\begin{eqnarray*} P_{\,D} = F_2 \big( U_2 - U_1 \big) = F_2 \dot{l}.
 \end{eqnarray*}

The stress on the surface of a sphere expanding at the rate $\dot{R} (t)$ in a
 Newtonian fluid is obtained as follows \citep{Brennen:2013:CBD}. The continuity equation in the exterior
 fluid implies that the radial velocity $v_r = dR/dt $ has the form 
$$
v_r = \dfrac{F(t)}{r^2}. 
$$
The forces per unit area on the 
 sphere acting at the sphere-fluid interface are the interior pressure $p_i$,
 the  fluid stress force 
$$
\stress_{rr}  = \Big( -p + 2 \mu \dfrac{\partial v_r}{\partial r} \Big) \Big|_{r = R} 
$$
due to the exterior fluid motion, and a  tension T due to interfacial forces equal to 
$$
\dfrac{2T}{R}.
$$
We neglect both the pressure difference across the interface 
and the interfacial tension, and therefore the force is 
$$
\stress_{rr}  =  2 \mu \dfrac{\partial v_r}{\partial r} = -4\dfrac{\mu}{R}\dfrac{dR}{dt}.
$$
Therefore  the power required to expand a sphere is
\begin{eqnarray*} -\int_S \mathbf{\stress_{rr}} v_r \ d s = 16 \pi \mu R \dot{R}^2
 \end{eqnarray*}
 Therefore for   the PMPY model we have
\begin{eqnarray*} P_{\,V}  = 16 \pi \mu \big( R_1 \dot{R}_1^2 + R_2
\dot{R}_2^2 \big)
 \end{eqnarray*}
 and thus the  the power expended to propel a PMPY at time $t$ is
\begin{eqnarray}\label{eq.PMPY_Power} P = P_{\,D}  + P_{\,V} =  F_2 \dot{l} + 16 \pi \mu \big( R_1
\dot{R}_1^2 + R_2 \dot{R}_2^2 \big).
 \end{eqnarray}

 We nondimensionalize $P$ as
\begin{eqnarray*} \hat{P} = \dfrac{1}{6 \pi \mu} \dfrac{T^2}{R_M^3} P
 \end{eqnarray*}
 so that while $P = FU$ in dimensional form, after nondimensionalization we also have $\hat{P} =
\hat{F} \hat{U}$. Thus the nondimensional version of equation~(\ref{eq.PMPY_Power}) is
\begin{eqnarray}\label{eq.PMPY_Power_Scale}
P = F \xi^2 + \dfrac{8}{3} \big( R_1 \zeta_1^2 + R_2 \zeta_2^2 \big)
\end{eqnarray}
with hat notation omitted. While $ P_{\,V}  $ is determined,
$P_{\,D}  $ depends on the perturbation, which in turn depends on
$\delta$. A first order approximation to $ P $ is given in
\cite{Avron:2005:PEM}, which after nondimentionalization reads
\begin{eqnarray}\label{eq.PMPY_Power_Avron} 
P = \dfrac{R_1 R_2}{R_1 + R_2} \xi^2 + \dfrac{8}{3} \big( R_1 \zeta_1^2 + R_2 \zeta_2^2 \big) + O(\delta)
 \end{eqnarray}
 while higher-order approximations can be obtained by substituting the results
 of $F$ obtained from equations~(\ref{eq.RPY_System}) or (\ref{eq.PT_Force})
 into equation~(\ref{eq.PMPY_Power_Scale}).
   
 Finally we define the performance $\mathcal{P}$ of a stroke as the ratio of the
translation per cycle to the energy expended in a cycle, {\em viz.}
\begin{eqnarray} \mathcal{P} = \dfrac{\big| \int_0^T \overline{U} (t) \ dt, 
\big|}{ \int_0^T P (t) \ dt}.
 \end{eqnarray}
which has the units of ${force}^{-1}$. $\mathcal{P}$ measures the
energy required for the PMPY to swim a certain distance in a cycle, and a  large
value of $\mathcal{P}$ indicates an energy-saving stroke. The nondimensional form of
$\mathcal{P}$ is
\begin{eqnarray*} \hat{\mathcal{P}} = \dfrac{6 \pi \mu R_M^2}{T} \mathcal{P} =
\dfrac{| \int_0^1 \hat{\overline{U}} d \hat{t} |}{\int_0^1 \hat{P} d \hat{t}}
\end{eqnarray*}
%

 
\subsection{A comparison of the solutions}
\label{Sec.CompSoln}

\begin{figure}[htbp] \centering
\includegraphics[width=1\textwidth]{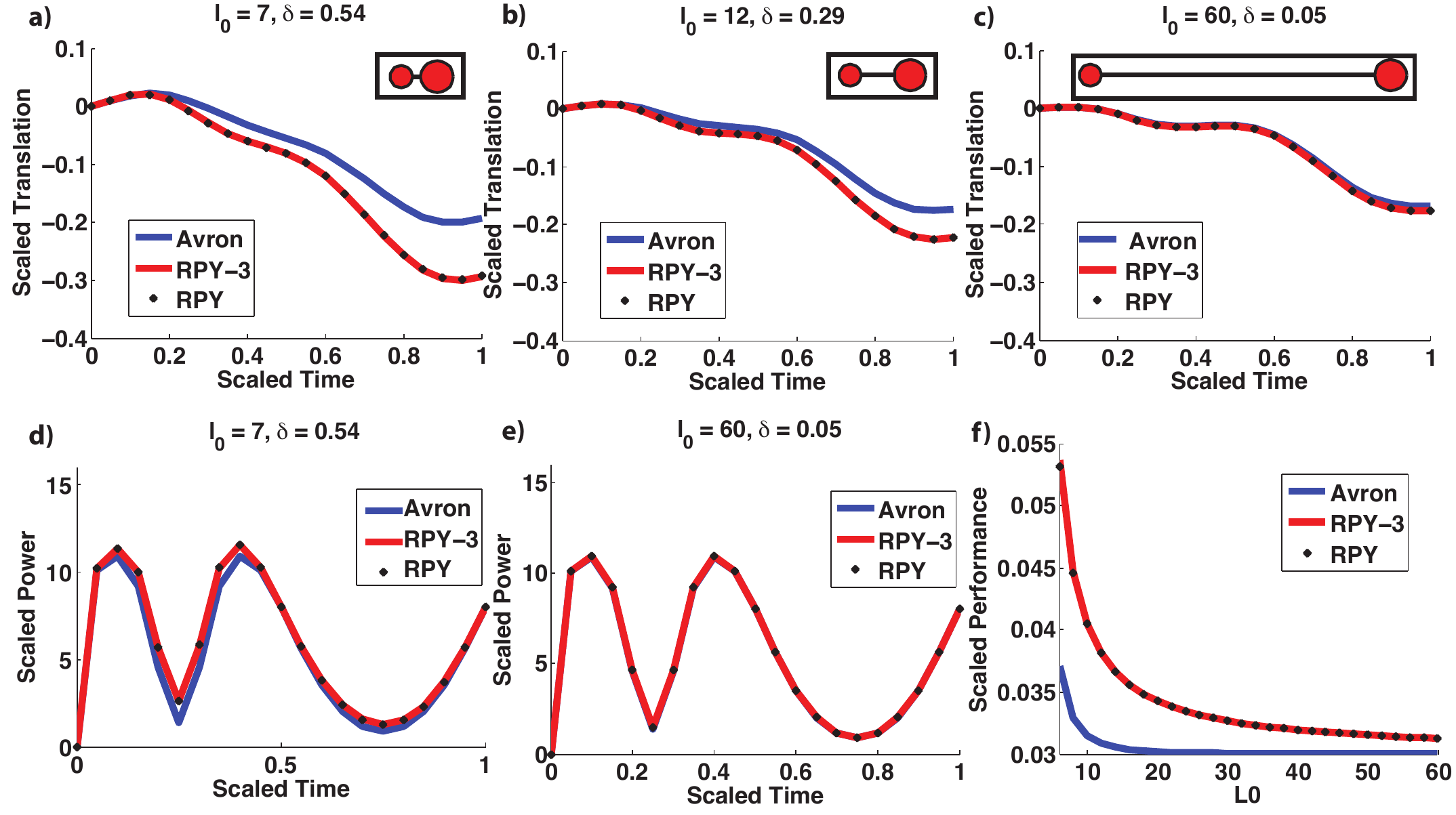}
\caption{A comparison of the asymptotic Avron solution (blue lines), the
  solution obtained by the reflection method (black dots) and its perturbation
  analysis up to the $O (\delta^3)$ order (red lines).  (a-c) The scaled
  translation $\hat{X} (\hat{t}) = \int_0^{\hat{t}} \hat{\overline{U}} \ d
  \hat{t}$ within a period for $l_0 = 7,12,60$.  The initial profile of the
  swimmer is shown in the box.  (d,e) The scaled power $\hat{P} (\hat{t})$
  within a period, with $l_0=7, 60$.  (f) The scaled performance
  $\hat{\mathcal{P}}$ of PMPY with respect to $l_0$.}
 \label{fig.CompPMPY}
\end{figure}

Next we compare the asymptotic  solution given by Eq. ($\ref{eq.AvronSoln}$),
the solution obtained by the reflection method (equation~. $\ref{eq.RPY_System}$)) and the
solution that results from  the $O (\delta^3)$ order approximation to the
latter, given by (equation~($\ref{eq.PT_AvgU}$)), for a prescribed loop in the
control space described by  $(\dot{l}, \dot{R}_1)$. We use the sinusoidal
circuits 
\begin{eqnarray}\label{eq.SinLoop} R_1 (t) = 2 + \sin 2 \pi t, \qquad R_2 (0) =
3, \qquad l = l_0 + \cos 2 \pi t, \qquad \textrm{for} \quad t \in [0,1]
\end{eqnarray}
with typical length unit $\mu$m and time unit $\min$ -- these set the length and
time scales for biological LRN swimmer, such as Dd amoebae
\citep{van2011amoeboid}, as will be seen later in this section.  Thus the
nondimensionalization is given by:
\begin{eqnarray*}
R_M \sim 3.24 \mu \textrm{m}, \quad L_m = ( l_0 - 1) \mu\textrm{m}, \quad \delta \sim \dfrac{3.24}{l_0 - 1}, \quad T= 1 \min
\end{eqnarray*}
The results for scaled translation $\hat{X} (\hat{t}) = \int_0^{\hat{t}}
\hat{\overline{U}} \ d \hat{t}$ and scaled power $\hat{P} (\hat{t})$ in a period
for different values of $l_0$ are shown in Figure~$\ref{fig.CompPMPY}a-e $, and
the relation between the scaled performance $\hat{\mathcal{P}}$ and $l_0$ is
given in Figure~$\ref{fig.CompPMPY}f$.

To maintain the correct geometry, the relation $l > R_1 + R_2$ must hold at all
times. In Figure~\ref{fig.CompPMPY} we see that both the scaled translation
$\hat{X}$ and the power $\hat{P}$ computed via the first reflection and its $O
(\delta^3)$ approximation (black dots and red solid lines, resp.,) agree very
well. On the other hand, the asymptotic approximation for the translation given
by Eq. ($\ref{eq.AvronSoln}$) and shown in Figure~\ref{fig.CompPMPY}a-c, (blue
lines) deviates from them significantly when $l_0$ is small.  
The scaled power of the Avron approximation coincides closely with that of
the higher-order approximations (Figure~\ref{fig.CompPMPY}d-e).
As a result, the scaled performance deviates from the reflection results
significantly when the spheres are relatively close (Figure~\ref{fig.CompPMPY}f).

As pointed out earlier \citep{Wang:2015:PDM}, PMPY adopts a mixed control
strategy in $(\dot{l},\dot{R})$, which makes it superior to other linked-sphere
models that adopt combinations of a single type of control. In fact, PMPY is the
only model studied there for which the net translation $\hat{X} = \int_0^1
\hat{\overline{U}} \ dt \sim O (1)$ over a period.  This can be seen in
Figures~$\ref{fig.CompPMPY}$a-c, in that the net translation $\hat{X}$ does not
vanish as the length of the connecting rod increases.  For $l \rightarrow \infty$, we
have $\delta \rightarrow 0$, and equation~($\ref{eq.PT_AvgU}$) gives the
estimate
\begin{eqnarray*} \lim_{ \delta \rightarrow 0}   \hat{\overline{U} }= \dfrac{\hat{R}_1 - \hat{R}_2}{2
(\hat{R}_1 + \hat{R}_2)} \hat{\xi} \sim O (1), \qquad \lim_{\delta \rightarrow 0} \hat{X} = \lim_{\delta \rightarrow 0} \int_0^T \hat{\overline{U}} (t) \ dt \sim O (1) 
\end{eqnarray*}
For the stroke prescribed by equation~($\ref{eq.SinLoop}$), we have the estimate
$\hat{X} \sim - 0.17$ as $ l_0 \rightarrow \infty, \ \delta \rightarrow 0$.  On
the other hand, the power $\hat{P}$ does not change much as $l_0$ changes, and
taken together, we have the estimate for the performance of PMPY in the limit $
l_0 \rightarrow \infty, \ \delta \rightarrow 0$ of about $\hat{\mathcal{P}} \sim
0.03$.

Finally, we check if the flow regime for the PMPY model computations satisfies
$\textrm{Re} \ll 1$ and $\textrm{ReSl} \ll 1$, which is required for a LRN
swimmer. We assume that  the medium is water ($\rho \sim 10^3 \textrm{kg} \cdot
\textrm{m}^{-3}$, $\mu \sim 10^{-3} \textrm{Pa} \cdot \textrm{s}$), and test two
sets of $L $ and $U$ from our simulations ($L = 6 \mu \textrm{m}, \ U \sim - 1.2
\mu \textrm{m} / \min$ from Figure~\ref{fig.CompPMPY}a, or $L = 60 \mu
\textrm{m}, \ U \sim - 0.6 \mu \textrm{m} / \min$ from
Figure~\ref{fig.CompPMPY}c). In either case we have $\textrm{Re} \ll 1 $ and
$\textrm{ReSl} \ll 1$ ($\textrm{Re} \sim O (10^{-6}) $, $\textrm{ReSl} \sim O
(10^{-6}) $ in Figure~\ref{fig.CompPMPY}a, and $\textrm{Re} \sim O (10^{-6}) $,
$\textrm{ReSl} \sim O (10^{-5}) $ in Figure~\ref{fig.CompPMPY}c).

\begin{figure}[ht!] 
\centering
\includegraphics[width=.85\textwidth]{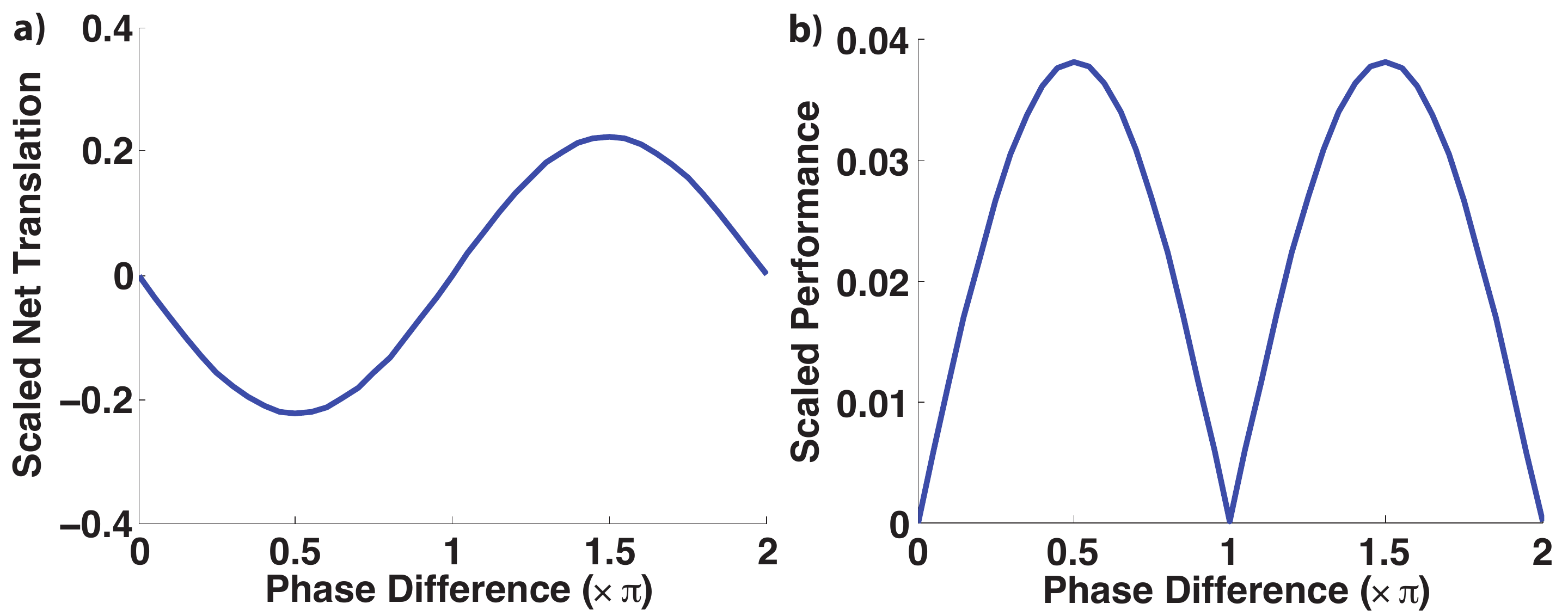}
\caption{The scaled net translation (a) and performance (b) with respect to phase difference between tthe two controls $\dot{l}$ and $\dot{R}_1$, 
with $l_0 = 12$, simulated by the RPY approximation.}
 \label{fig.PMPY_PhaseDiff}
\end{figure}

The foregoing results are for a fixed phase difference, and next we investigate
the effect of changing the phase difference between the two controls $\dot{l}$
and $\dot{R}_1$. We consider the following type of sinusoidal cycles
\begin{eqnarray*}
R_1 (t) = 2 + \sin 2 \pi t, \qquad R_2 (0) = 3, \qquad l = l_0 + \sin (2 \pi t + \phi), \qquad \textrm{for} \ t \in [0,1]
\end{eqnarray*} 
where $\phi \in [0, 2 \pi]$ is the phase difference. The scaled net translation
and performance with respect to $\phi$ is shown in
Figure~\ref{fig.PMPY_PhaseDiff}, from which we see that the maxima of both
scaled net translation and performance are reached at a phase difference of
$\phi = k \pi + \pi/2$, $k \in \mathbb{Z}$. In contrast, when $\phi = k \pi, \ k
\in \mathbb{Z}$, the net translation after one cycle equals zero, which
naturally leads to zero performance as well. This stems from the fact that in
these cases the shape deformation become time reversible, and according to the
scallop theorem, no net translation results.

To compare our analysis with experimental observations, we use the data on
swimming amoebae from \cite{van2011amoeboid}, where it is reported that
Dd amoebae move in a fluid environment by side protrusions. Typically
the cell body is elongated, and single or multiple protrusions cyclically
propagate along the cell \cite{Wang:2015:CAA}. Although the shape deformation
mode of these amoebae cell is not exactly the same as a PMPY model, however
similar to a PMPY, such a traveling protrusion mode does exploit mass
transfer  along an elongated body. It is reported that amoebae using this
swimming mode have maximum cell body length $\sim 25 \mu \textrm{m}$, average
cell body width $\sim 6 \mu \textrm{m}$, a typical stroke has period $\sim 1
\min$, and a typical swimming velocity $\sim 3 \mu \textrm{m} / \min$.  Using
this data, we approximate
\begin{eqnarray*}
L \sim 25 \mu \textrm{m}, \quad T \sim 1 \min, \quad R \sim 6 \mu \textrm{m}, \quad U \sim 3 \mu \textrm{m} /  \min
\end{eqnarray*}
thus $\delta \sim R/L \sim 0.24$, $\hat{U} \sim T U/R \sim 0.5$ and the scaled
net translation within a period $\hat{X} = X/ R \sim 0.5$, which is about the
same as for a PMPY. In fact, as can be seen from Figure~\ref{fig.CompPMPY}a, if
the spheres are not too separated in the model, the scaled net translation of a
PMPY can reach $\hat{X} \sim 0.3$. On the other hand, other linked-sphere models
that also have elongated shape and adopt large scale of shape deformations can
only result in $\hat{X} \sim O (\delta^2)$ \citep{Wang:2015:PDM}, which is far
less than a PMPY or a swimming amoebae as observed.  Hence we believe the PMPY
model is suitable in the study of swimming cells at LRN.


\section{A PMPY swimmer in the presence of a passive buoyant obstacle}
\label{Sec.Wide-Obstacle}

\setcounter{equation}{0}
\renewcommand{\theequation}{\ref{Sec.Wide-Obstacle}.\arabic{equation}}

\begin{figure}[htbp] \centering
\includegraphics[width=0.3\textwidth]{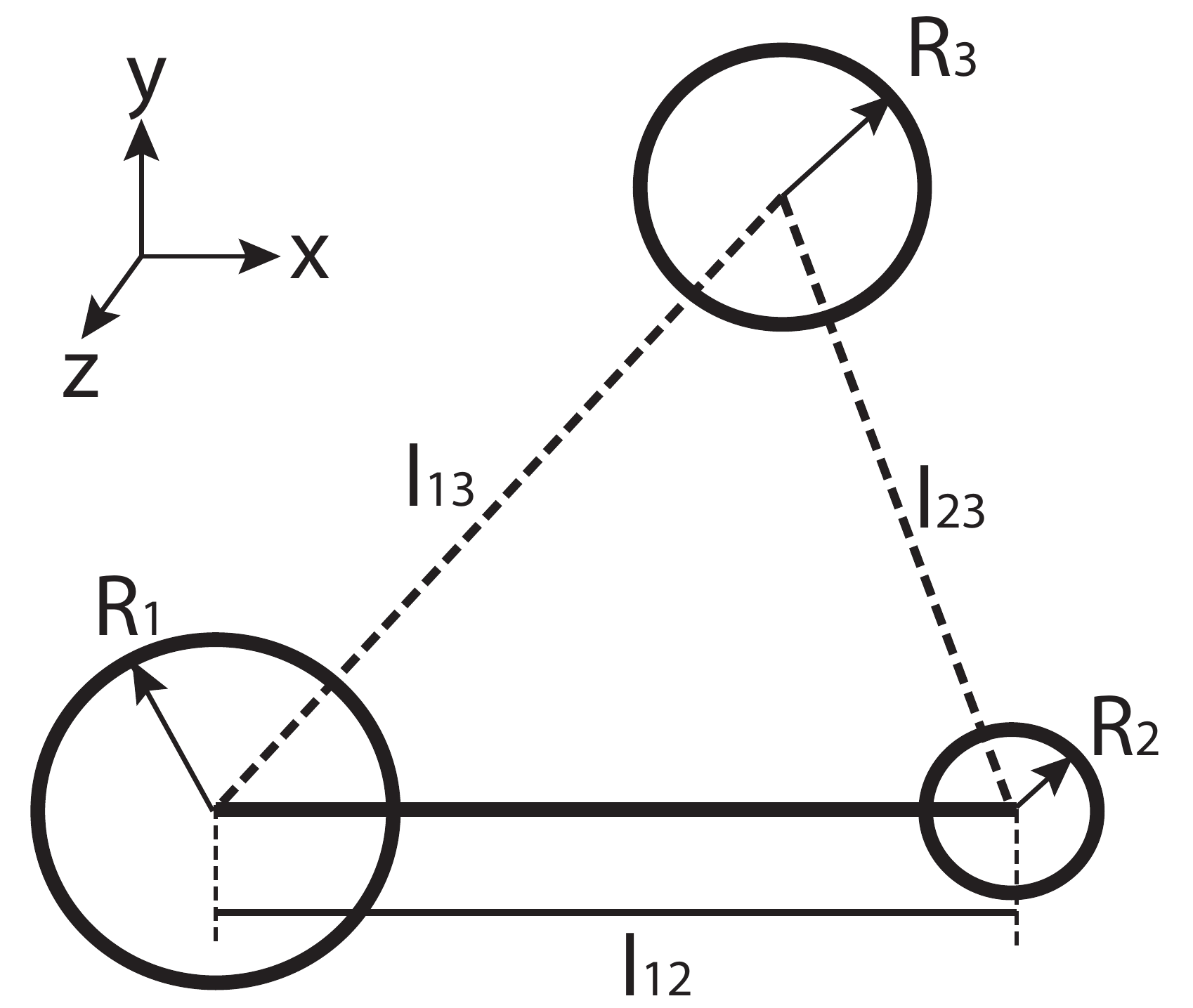}
\caption{The geometry of the system of a PMPY swimmer and a passive, neutrally-buoyant spherical
object.}
 \label{fig.3sphere_geometry}
\end{figure}

In an extension of the previous results that leads to several interesting
applications, we next analyze a PMPY swimmer that interacts with an untethered,
passive, neutrally-bouyant object nearby.  To simplify the computation of the
interaction, we suppose that the object is a rigid sphere, and that there is no
external force or torque imposed on it. The geometry of the system is shown in
Figure~$\ref{fig.3sphere_geometry}$, where one sees that there are six
characteristic lengths: the radius of each sphere ( $R_i,\, i=1,2,3)$ and the lengths between any pair of spheres: ($l_{ij}, \,i, j =
1,2,3)$. For scaling purposes we define
\begin{eqnarray*}
R_M = \max_t \{R_i (t) \}_{i=1,2,3}, \quad L_m = \min_t \{ l_{12} (t) \}, \quad \delta = \dfrac{R_M}{L_m}
\end{eqnarray*}
To apply the reflection method to this system, we require that $\delta <1$ as
before, and in addition, we only consider the regime $l_{13}, l_{23} >
L_m$. Thus all non-dimensionalization relations in section \ref{Sec.SLOA} apply.
The following discussion is similar to that in section~$\ref{Sec.RPY}$, except
that one more sphere is now involved in the system.  We assume that the centers
of all three spheres lies in the $xy$-plane, and that the active swimmer moves
along the $x$-axis (Figure~$\ref{fig.3sphere_geometry}$) initially.  In this
configuration the torque and angular velocities must be taken into account
unless the three spheres are co-linear. However, the motion of the spheres will 
remain in the plane defined by their initial positions since the axis of
rotation of a sphere is orthogonal to that plane. Thus the problem remains
effectively two-dimensional, but this plays no role in the analysis. 

%


\subsection{The linear and angular velocities after the first reflection}
\label{Sec.PMPY_1Ob_Analysis}

At the zeroth-order reflection, in which no hydrodynamic interactions between  the
spheres are considered, the results for the swimmer are the same as in the
absence of a passive object, and the velocity field for each sphere is given by
equations~($\ref{eq.VelocityField_SingleSphere},\ref{eq.RPY_U_Lead}$).  At this
order the linear and angular velocities  of sphere 3 are
\begin{eqnarray}
\label{eq.FreeOb_U3_0} 
\mathbf{U}_{3}^{(0)} = \bfO_{3}^{(0)} =
\mathbf{0}, \quad \mathbf{u}_3^{(0)} \equiv \mathbf{0}
\end{eqnarray}
At the first reflection, each sphere is  subject to  the flows generated by the other 
spheres. Since $\mathbf{u}_3^{(0)} \equiv \mathbf{0}$, the calculation of rigid
motions for sphere 1 and 2 in the first reflection are identical to
section~$\ref{Sec.RPY}$ -- thus we have: 
\begin{eqnarray*} \mathbf{U}_1^{(1)} &=& \Big( 1 + \dfrac{R_1^2}{6} \nabla^2
\Big) \big( \mathbf{u}_2^{(0)} + \mathbf{u}_3^{(0)} \big) \Big|_{\mathbf{x} = \mathbf{x}_1}
=   \dfrac{\mathbf{F}_2}{4 \pi \mu l_{12}}
\Big( 1 - \dfrac{R_1^2 + R_2^2}{3 l_{12}^2} \Big) - \dot{R}_2 \Big(
\dfrac{R_2}{l_{12}} \Big)^2 \\ \bfO_1^{(1)} &=& \dfrac{1}{2} \nabla \times \big(
\mathbf{u}_2^{(0)} + \mathbf{u}_3^{(0)} \big) \Big|_{\mathbf{x} = \mathbf{x}_1} = \mathbf{0}
\\ \mathbf{U}_2^{(1)} &=& \Big( 1 + \dfrac{R_2^2}{6} \nabla^2 \Big) \big(
\mathbf{u}_1^{(0)} + \mathbf{u}_3^{(0)} \big) \Big|_{\mathbf{x} = \mathbf{x}_2} =
  \dfrac{\mathbf{F}_1}{4 \pi \mu l_{12}}
\Big( 1 - \dfrac{R_1^2 + R_2^2}{3 l_{12}^2} \Big) + \dot{R}_1 \Big(
\dfrac{R_1}{l_{12}} \Big)^2 \\ \bfO_2^{(1)} &=&\dfrac{1}{2} \nabla \times \big(
\mathbf{u}_1^{(0)} + \mathbf{u}_3^{(0)} \big) \Big|_{\mathbf{x} = \mathbf{x}_2} = \mathbf{0}.
\end{eqnarray*}
From this one sees that at the first reflection, the presence of the passive
sphere at a sufficient distance does not affect the swimmer, but the converse is
not true --  the effect of the swimmer on the passive sphere is non-zero after the
first reflection. Its translational velocity $\mathbf{U}_3^{(1)}$ is given by
the sum of the contributions from translation and expansion of the swimmer's
spheres, {\em viz.},
\begin{eqnarray}\label{eq.U3_1} \mathbf{U}_3^{(1)} = \sum_{i=1,2}
\mathbf{U}_{3,i}^{(1,f)} + \mathbf{U}_{3,i}^{(1,e)}
\end{eqnarray}
where 
%
\begin{eqnarray}\label{eq.U3_1f} \mathbf{U}_{3,i}^{(1,f)} &=& \Big( 1 +
\dfrac{R_3^2}{6} \nabla^2 \Big) \mathbf{u} \{ \mathbf{F}_i \} \Big|_{\mathbf{x}
= \mathbf{x}_3} = \dfrac{1}{8 \pi \mu l_{i3}} \Big[ \Big( 1 + \dfrac{R_i^2 +
R_3^2}{3 l_{i3}^2} \Big) \mathbf{F}_i + \Big( 1 - \dfrac{R_i^2 + R_3^2}{
l_{i3}^2} \Big) \big( \mathbf{F}_i \cdot \mathbf{d}_{i3} \big) \mathbf{d}_{i3}
\Big] \quad \\ [15pt] 
\label{eq.U3_1e} \mathbf{U}_{3,i}^{(1,e)} &=& \dfrac{1}{2} \nabla
\times \Big( R_i^2 \dot{R}_i \dfrac{\mathbf{x} - \mathbf{x}_i}{|\mathbf{x} -
\mathbf{x}_i|^3} \Big) \Big|_{\mathbf{x} = \mathbf{x}_3} = \dot{R}_i \Big(
\dfrac{ R_i}{l_{i3}} \Big)^2 \mathbf{d}_{i3}
\end{eqnarray} 
and \vspace*{-5pt}
\begin{eqnarray*} \mathbf{d}_{i3} = \dfrac{\mathbf{x}_3 -
\mathbf{x}_i}{|\mathbf{x}_3 - \mathbf{x}_i|}, \qquad l_{i3} = |\mathbf{x}_3 -
\mathbf{x}_i|.
\end{eqnarray*}
Although sphere 1 and 2 are
rotation-free  after the first reflection, 
one finds that the translation
of sphere 1 and 2 contributes to the rotation of $\bfO_3^{(1)}$,  while  their expansion has
no effect on $\bfO_3^{(1)}$. The detailed calculation of $\bfO_3^{(1)}$ is
straightforward  and is given in Appendix~$\ref{Appendix.D}$. The result is that
\begin{eqnarray}\label{eq.Omega3_1} \bfO_3^{(1)} = \sum_{i=1,2} \dfrac{F_i}{8
\pi \mu l_{i3}^3} \Big[ \big( \mathbf{x}_3 - \mathbf{x}_i \big) \cdot \mathbf{e}_y \Big] \mathbf{e}_z.
\end{eqnarray}
To summarize the analysis for the first reflection, we first solve for the 
 motion of the swimmer  and the forces it exerts  using  the system
($\ref{eq.RPY_System}$), and use the results in  equations~($\ref{eq.U3_1},
\ref{eq.Omega3_1}$) to obtain the  motion of sphere 3.


\subsection{Accuracy of the system}
\label{Sec.PMPY1OB_Accuracy}

For the PMPY, as we mentioned earlier, up to the first reflection, the
passive buoyant sphere 3 has no effect on the PMPY model.  It follows from
section~$\ref{Sec.RPY}$, that after the first reflection the translational
velocities of the two spheres of the PMPY ($U_1, U_2$), together with the
translational velocity of the PMPY ($\overline{\mathbf{U}}$), is accurate up to
the $\delta^3$ order.  The leading order of $\hat{\overline{\mathbf{U}}}$ is of
$O(1)$, as shown by equation~(\ref{eq.PT_AvgU}).  Because $\bfO_{1,2}^{(0)} =
\bfO_{1,2}^{(1)} = \mathbf{0}$, we neglect the rotation effect of the PMPY.

Similar to the analysis of the PMPY, neglecting the
second and higher reflections for the passive sphere 3 results in the translational velocity
$\mathbf{U}_3 \sim \mathbf{U}_3^{(0)} + \mathbf{U}_3^{(1)}$ accurate up to order
$\delta^3$ . Moreover, the effects from the self-deformable PMPY model on sphere
3 starts to show up from the first reflection ($\mathbf{U}_3^{(0)} =
\mathbf{0}$), thus an estimate of the leading term of $ \hat{ \mathbf{U}}_3 $ is
\begin{eqnarray*} \sum_{i=1,2} \dfrac{3}{4}
\dfrac{ \hat{\mathbf{F}}_i + \big( \hat{\mathbf{F}}_i \cdot \mathbf{d}_{i3} \big) \mathbf{d}_{i3}}{\hat{l}_{i3}} \delta
\sim O (\delta)
\end{eqnarray*}
For the angular velocity we have $\bfO_3 \sim \bfO_3^{(0)} + \bfO_3^{(1)}
=\bfO_3^{(1)} $, and thus the leading order of $\hat{\bfO}_3$  can be estimated
as follows:
\begin{eqnarray*}
\sum_{i=1,2}   \dfrac{3 \hat{  \mathbf{F}}_i}{4 \hat{l}_{i3}^3}  \Big[ \big( \hat{\mathbf{x}}_3 - \hat{\mathbf{x}}_i \big) \cdot \mathbf{e}_y \Big] \mathbf{e}_z \delta^3
\sim O (\delta^3)
\end{eqnarray*}
In the wide-separation regime, i.e., when $\delta \ll 1$, it is clear that the
angular velocity of the passive sphere 3 ($\hat{\bfO}_3 \sim O
(\delta^3)$) can be neglected compared to its translational velocity
$\hat{\mathbf{U}}_3 \sim O (\delta)$ or the translational velocity of PMPY
($\hat{\overline{\mathbf{U}}} \sim O (1)$).

To investigate the second reflection, we must first compute all the velocities 
--- $\mathbf{u}_{ij}^{(1)}$ for $i,j = 1,2,3, \ i \neq j$ --- that result from
putting sphere $j$ into the flow $\mathbf{u}_i^{(0)}$. Since $\mathbf{u}_3^{(0)} \equiv
\mathbf{0}$ (equation~($\ref{eq.FreeOb_U3_0}$)), we have $\mathbf{u}_{3j}^{(1)} \equiv
\mathbf{0}$ for $j=1,2$.  On the other hand, when we put sphere 2 or 3 into
$\mathbf{u}_1^{(0)}$, the resulting flow $\mathbf{u}_{1j}^{(1)}$ is a superposition of two
parts:
\begin{eqnarray*} \mathbf{u}_{1j}^{(1)} = \mathbf{u}_{1j}^{(1)} \{ \mathbf{F}_1 \} +
\mathbf{u}_{1j}^{(1)} \{ \dot{R}_1 \}
\end{eqnarray*}
where $\mathbf{u}_{1j}^{(1)} \{ \mathbf{F}_1 \} $ results from the drag force
$\mathbf{F}_1$ exerted on sphere 1, and its leading term is of the order $
\delta^4$ \citep{Kim:1991:MPS}; $\mathbf{u}_{1j}^{(1)} \{ \dot{R}_1 \}$ results from
the radial deformation of sphere 1, and its leading term is of the order
$\delta^5$.  Hence $\mathbf{u}_{1j}^{(1)} \sim O (\delta^4)$ and similarly,
$\mathbf{u}_{2j}^{(1)} \sim O (\delta^4)$ as well.  Thus for sphere 1:
\begin{eqnarray*} \mathbf{U}_1^{(2)} &=& \Big( 1 + \dfrac{R_1^2}{6} \nabla^2
\Big) \big( \mathbf{u}_{12}^{(1)} + \mathbf{u}_{13}^{(1)} + \mathbf{u}_{23}^{(1)} +
\mathbf{u}_{32}^{(1)} \big) \Big|_{\mathbf{x} = \mathbf{x}_1} + \sim O(\delta^4) \\
\bfO_1^{(2)} &=& \dfrac{1}{2} \nabla \times \big( \mathbf{u}_{12}^{(1)} +
\mathbf{u}_{13}^{(1)} + \mathbf{u}_{23}^{(1)} + \mathbf{u}_{32}^{(1)} \big) \Big|_{\mathbf{x} =
\mathbf{x}_1} + \sim O(\delta^5)
\end{eqnarray*}
with similar results for sphere 2 and 3 as well. In conclusion, the results
we obtained from section~$\ref{Sec.PMPY_1Ob_Analysis}$ are accurate up to the $\delta^3$ term.

Finally, it is easily seen that the results obtained in
section~$\ref{Sec.PMPY_1Ob_Analysis}$ can be applied to a system consisting of a
PMPY model and $N$ passive neutrally-buoyant spheres, as long as the spheres are separated
sufficiently.  We number the two spheres in the PMPY model as sphere 1 and 2, as
usual, and the others from sphere 3 to sphere $N+2$. For each of the $N$
passive  spheres, $\mathbf{u}_i^{(0)} \equiv 0 \ (i=3,4, \cdots, N+2)$ in
the zeroth reflection, from which we conclude that: up to the first reflection,
\begin{enumerate}
\item The PMPY model does not "see" the other spheres:
\begin{eqnarray*} \bfL_i^{(1)} = \bfL_{i,j}^{(1,f)} + \bfL_{i,j}^{(1,e)} +
\sum_{n=3}^{N+2} \Big( \bfL_{i,n}^{(1,f)} + \bfL_{i,n}^{(1,e)} \Big) =
\bfL_{i,j}^{(1,f)} + \bfL_{i,j}^{(1,e)}
\end{eqnarray*}
where $i,j =1,2,  \ i \neq j$ and $\bfL $ stands for $\mathbf{U}$ or $\bfO$.
\item Each passive sphere only "sees" the PMPY model and "sees" no other
spheres:
\begin{eqnarray*} \bfL_n^{(1)} = \sum_{i=1,2} \Big( \bfL_{n,i}^{(1,f)} +
\bfL_{n,i}^{(1,e)} \Big) + \sum_{3 \leq m \leq N+2}^{m \neq n} \Big(
\bfL_{n,m}^{(1,f)} + \bfL_{n,m}^{(1,e)} \Big) = \sum_{i=1,2}
\Big(\bfL_{n,i}^{(1,f)} + \bfL_{n,i}^{(1,e)} \Big)
\end{eqnarray*}
where $n \in {3,4,\cdots, N+2}$ and $\bfL $ stands for $\mathbf{U}$ or $\bfO$.
\end{enumerate}


\subsection{Chasing an object}

The scenario of a micro-swimmer swimming with a passive object has many
applications.  For example, can a microorganism that locates a target object
(nutrient, a bacteriuum, etc.), capture the object within a reasonable time
period? To be specific, the micro-swimmer should be able to swim fast enough to
reach the target object, and the object, as it is passive, should not be
pushed away faster than the micro-swimmer swims, especially when they are close.
 
We consider a scenario in which a PMPY swims toward a passive sphere
directly in front of it, in simulating a microorganism chasing an object. As we
discussed in section \ref{Sec.CompSoln}, a PMPY is an effective swimmer --- with
a translational velocity $\hat{\overline{\mathbf{U}}}$ scales as $O (1)$ --- and
approximates the swimming behavior of Dd amoebae
\citep{van2011amoeboid}, which feed on bacteria. This indicates that a PMPY can
swim a distance within a reasonable time period by consuming a reasonable amount
of energy.  For the movement of the passive sphere 3 that results from
the hydrodynamic interaction with the PMPY, as we discussed in section
\ref{Sec.PMPY1OB_Accuracy}, the leading term of $\hat{\mathbf{U}}_3$ is
\begin{eqnarray}\label{PMPY_Ob_Asymp}
\hat{\mathbf{U}}_3 \sim \hat{\mathbf{U}}_3^{(1)} \sim \sum_{i=1,2} \dfrac{3}{4} \dfrac{ \hat{\mathbf{F}}_i + \big( \hat{\mathbf{F}}_i \cdot \mathbf{d}_{i3} \big) \mathbf{d}_{i3}}{\hat{l}_{i3}} \delta
\end{eqnarray}
Since the spheres are collinear we have $\mathbf{d}_{13} = \mathbf{d}_{23}$, and the
force-free constraint of the PMPY gives $\mathbf{\hat{\mathbf{F}}_1} = -
\hat{\mathbf{F}}_2$, both of which lie along $\mathbf{d}_{13}$. Therefore we
have the following estimate of the translational velocity $\hat{\mathbf{U}}_3$.
\begin{eqnarray*}
\hat{\mathbf{U}}_3 \sim - \dfrac{3}{2} \hat{\mathbf{F}}_1 \big( \dfrac{1}{l_{13}} - \dfrac{1}{l_{23}} \big) \delta
= - \dfrac{3}{2} \hat{\mathbf{F}}_1 \dfrac{l_{12}}{l_{13} l_{23}} \delta,
\end{eqnarray*}
and by using equation (\ref{eq.PT_Force}) we have 
\begin{eqnarray}\label{eq.U3Lead}
\hat{\mathbf{U}}_3 \sim  \dfrac{3}{2} \dfrac{l_{12}}{l_{13} l_{23}} \dfrac{R_1 R_2}{R_1 + R_2} \delta \xi.  
\end{eqnarray}
From equation (\ref{eq.U3Lead}) we see that when the passive sphere 3 is far
from the PMPY, i.e., $l_{13} \gg 1, \ l_{23} \gg 1$, $\hat{\mathbf{U}}_3$ scales
much less than $O (\delta)$, and only when the PMPY is close to the target can
$\hat{\mathbf{U}}_3$ increase to $O (\delta)$. Therefore, with the estimates of
the translational velocities of the PMPY and the passive sphere giving
$\hat{\overline{\mathbf{U}}} \sim O(1)$ and $\hat{\mathbf{U}}_3 < O (\delta)$,
we confirm that a PMPY can easily capture its passive target. Moreover, we note
that although in problems like nutrient supply, the target objects are likely to
be much smaller than the microswimmer, this size relation is not required
here. As we can see from the equation (\ref{eq.U3Lead}), the size of the passive
sphere ($R_3$) does not enter the leading order term of $\hat{\mathbf{U}}_3$. In
fact, from equations (\ref{eq.U3_1f}, \ref{eq.U3_1e}), $R_3$ only shows up in
the $O (\delta^3)$ term, and as long as $R_3/l_{i3} \sim O(1), \ i=1,2$, the
above analysis and conclusions still hold.

\begin{figure}[htbp] \centering
\includegraphics[width=1\textwidth]{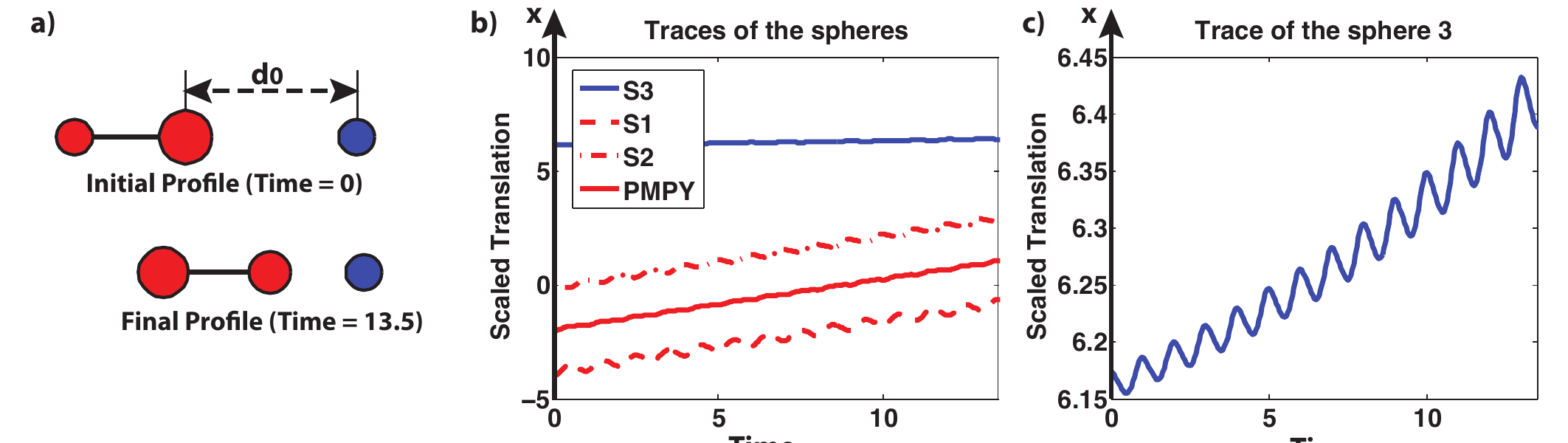}
\caption{Simulation of a PMPY swimmer and a passive sphere, arranged collinearly. (a) Initial and final profiles
of the system. (b) Translation of all components in the 13.5 cycles.  (d) Translation of S3 within 13.5 cycles.}
 \label{fig.Linear_PMPY_1Ob}
\end{figure}

To numerically investigate the effects of the PMPY on the passive sphere 3, we
consider the following system (Figure~\ref{fig.Linear_PMPY_1Ob}a). At $t = 0$
the passive sphere 3 is at a distance $d_0$ from the leading sphere (i.e.,
sphere 2), and for $t >0$ the PMPY swimmer executes the following cyclic
deformations.
\begin{eqnarray}\label{eq.PMPY1Ob_Stroke}
R_1 (t) = 2 - \sin 2 \pi t, \quad R_2 (0) = 3, \quad R_3 \equiv 2,  \quad l_{12} = l_0 + \cos 2 \pi t, \quad \textrm{for } t \in [0,1]
\end{eqnarray}
With this stroke the PMPY swimmer moves in the positive $x$ direction and pushes
the passive sphere in that direction. We take $l_0 =12, \ d_0 = 20$, which
allows the PMPY swimmer to execute a few cycles before it gets too close to
sphere 3, that is, when $l_{23} \leq L_m$.  The system profile gives $R_M \sim
3.24, \ L_m = 11, \ \delta \sim 0.29$.  The translation of the PMPY is computed
by equation (\ref{eq.RPY_System}), while that of sphere 3 is computed from
equations (\ref{eq.U3_1f}, \ref{eq.U3_1e}).  After 13.5 cycles, the PMPY model
swims a scaled distance of $\hat{X}_{\textrm{PMPY}} \sim 3.13$, while the
passive sphere only moves a scaled distance $\hat{X}_3 = 0.29$
(Figure~$\ref{fig.Linear_PMPY_1Ob}$b\&c).  Thus $\hat{X}_3 /
\hat{X}_{\textrm{PMPY}} \sim 0.087$, which is even less than the estimated ratio
$\hat{U}_3 / \hat{\overline{U}} \sim O (\delta)$.  One sees in (b) that the
trajectory of the passive sphere (blue solid line in
Figure~$\ref{fig.Linear_PMPY_1Ob}$b) is only slightly tilted as compared to that
of the PMPY swimmer (red solid line in Figure~$\ref{fig.Linear_PMPY_1Ob}$b).
This reflects the fact that sphere 3 oscillates back and forth, and the
oscillations grow as the swimmer approaches the sphere, as seen in
Figure~$\ref{fig.Linear_PMPY_1Ob}$c. Thus the swimmer can easily catch up to the
passive object, but lubrication effects arise when they are in close proximity.

%


\subsection{Tracer trajectories}

Another important application of this swimmer-object interaction problem is
swimmer-tracer scattering \citep{dunkel2010swimmer}, in which  the swimming of
micro-organisms stirs the surrounding fluid. This  is 
important in controlling and enhancing nutrient uptake, and  leads to   enhanced tracer
diffusion observed in swimmer suspensions \citep{wu2000particle,leptos2009dynamics,
  sokolov2009enhanced,kurtuldu2011enhancement,mino2011enhanced}.  Experimental
observations show that trajectories of tracers in a suspension of swimmers are often
nearly-closed loops \citep{leptos2009dynamics}, and theoretical arguments and
simulation predictions have emerged since to elucidate  this phenomenon
\citep{underhill2008diffusion,rushkin2010fluid,
  ishikawa2010fluid,lin2011stirring,zaid2011levy}. In particular, the PMPY model
has been used in the study of swimmer-tracer scattering, where an asymptotic
analysis  based on the stroke-averaged behavior of the PMPY, together
with some simulations, illustrate the near closed-loop of a triangular shape
of the tracer \citep{dunkel2010swimmer}.  In this section we  apply the  reflection
analysis elaborated in  sections \ref{Sec.PMPY_1Ob_Analysis} and
\ref{Sec.PMPY1OB_Accuracy} to further investigate the tracer trajectories
induced by a PMPY swimmer.

\subsubsection{The instantaneous  velocity of the tracer sphere}

\begin{figure}[htbp] \centering
\includegraphics[width=0.4\textwidth]{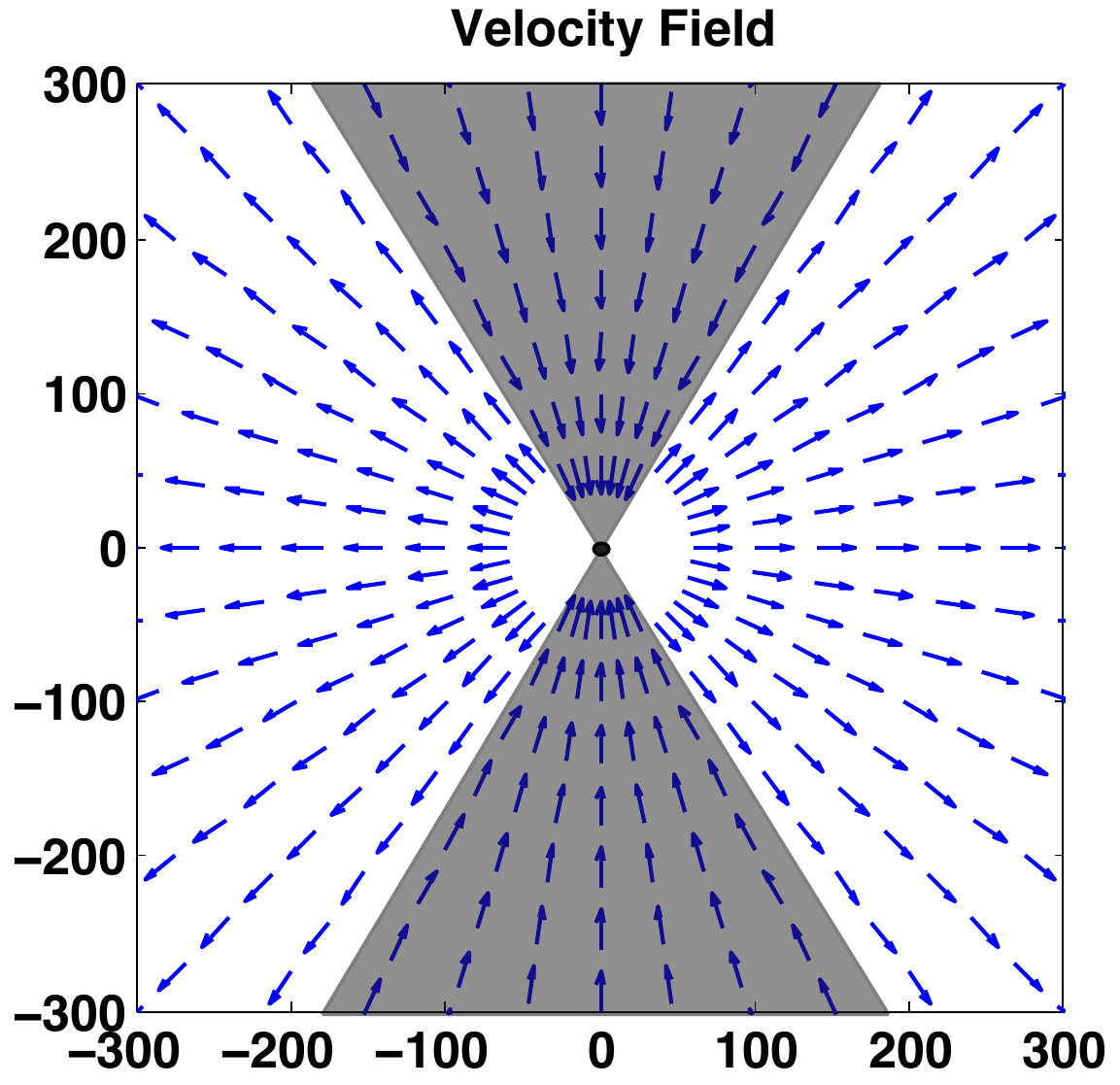}
\caption{A snapshot of the instantaneous velocity field generated by a PMPY
  swimmer and felt by a tracer sphere when the latter is far away from the
  swimmer.  In this simulation the PMPY is located at the origin, with an
  instantaneous profile $R_1 = R_2 = 2, \ l=6$, the connecting rod is expanding
  (i.e., $\dot{l} >0$), and the expansion/contraction of the spheres in the PMPY
  is neglected as it only generates $O(\delta^2)$ order terms.  The tracer
  sphere has radius $R_3 = 2.5$.  Arrows only show the direction, not the
  magnitude, of the velocity.}
 \label{fig.PMPY_Far_VelocityField}
\end{figure}

The swimmer-tracer interaction is easily found by asymptotic analysis when the
swimmer and the tracer are far apart
\citep{dunkel2010swimmer,pushkin2013fluid,yeomans2014introduction}. As is the
case for a PMPY swimmer, when a spherical object is far away from it, i.e.,
$l_{13}, l_{23} \gg 1$, the asymptotic estimate is given by equation
(\ref{PMPY_Ob_Asymp}), and the instantantaneous velocity of the tracer sphere as
a function of its location, is shown in Figure \ref{fig.PMPY_Far_VelocityField}.
In this case, the velocity field is the same as the asymptotic behavior of the
velocity field generated by a single PMPY \citep{dunkel2010swimmer}.
In this snapshot the connecting rod is expanding ($\dot{l} > 0$), and one sees
that the velocity field is divided into four domains.  A tracer sphere located
in the two shaded domains will be attracted to the PMPY and repulsed if it is
located in the two white domains.

\begin{figure}[htbp] \centering
\includegraphics[width=0.95\textwidth]{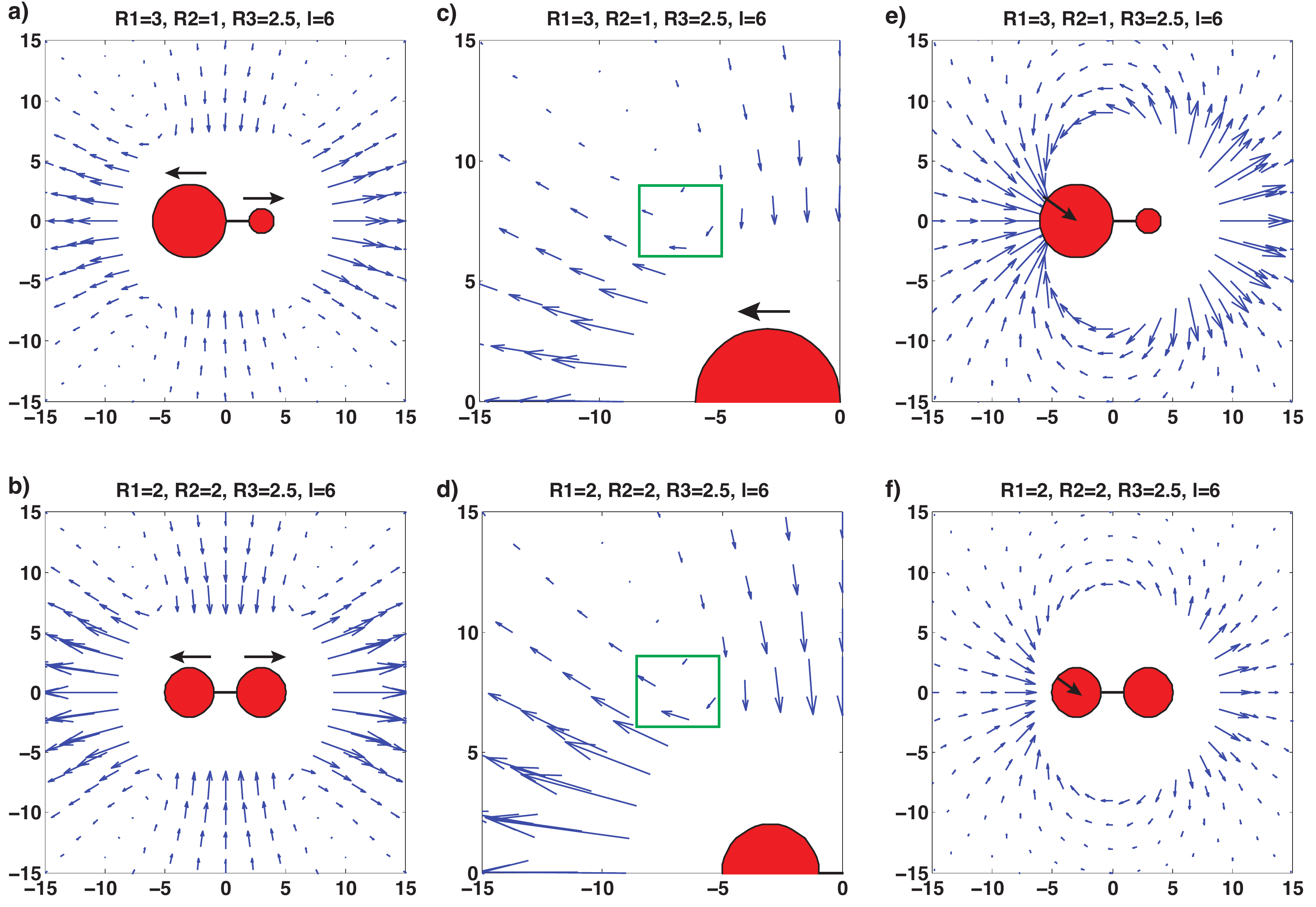}
\caption{The instantaneous velocity of a tracer sphere that results from a
  swimming PMPY, when the tracer sphere is close to the PMPY.  The arrows in all
  panels are scaled uniformly, thus arrows show both direction and magnitudes.
  a,b) 
  The
connecting
  rod is instantaneously expanding at $\dot{l} = 2 \pi$, and there is no 
  expansion/contraction in the PMPY. 
  c/d) A blown-up view of panel a/b) near sphere 1 in the PMPY.  
 e,f) Sphere 1 is instantaneously shrinking at
  $\dot{R}_1 = - 2 \pi$ while sphere 2 is expanding. No connecting
  rod expansion/contraction in the PMPY. }
 \label{fig.PMPY_VF_Uf-Ue}
\end{figure}
The reflection analysis allows us to observe the situation when the PMPY and the
tracer sphere get close to each other, as long as $\delta < 1$ still holds.
Moreover, as we showed in section \ref{Sec.PMPY_1Ob_Analysis}, the velocity of
the tracer sphere $\mathbf{U}_3$ consists of two parts: $\mathbf{U}_3^{(f)}$
that results from the drag forces on the two spheres in the swimmer (equation
(\ref{eq.U3_1f})), and $\mathbf{U}_3^{(e)}$, which results from the radial
changes (equation (\ref{eq.U3_1e})). To further investigate the separate effects
of drag and expansion when the tracer sphere is close to the PMPY, we show the
instantaneous velocity of the tracer sphere in different stages of movement of
the swimmer in Figure \ref{fig.PMPY_VF_Uf-Ue}.  Figure
\ref{fig.PMPY_VF_Uf-Ue} a\&b show the velocity field when the PMPY connecting rod
is expanding without radial changes, and the spheres of the PMPY are either of
unequal sizes (a) or equal sizes (b). Comparing Figure
\ref{fig.PMPY_VF_Uf-Ue} a\&b with Figure \ref{fig.PMPY_Far_VelocityField} we see
that even at small separations, the local velocity varies little from the
far-field behaviors. The reason for this similarity is that other than the
leading $O (1)$ term, the next term  in $\mathbf{U}_3^{(f)}$ (equation
(\ref{eq.U3_1f})) is an $O (\delta^3)$ term, which decreases very rapidly as
$l_{13} ~\&  ~l_{23}$ increase. However, a blown-up view of the velocity fields shows
that the $O (\delta^3)$ term does give rise to a  tangential velocity when the tracer
sphere is close to the swimmer, as it must.  This is  clear at the border of the 
shaded \& white regions  shown in  Figure \ref{fig.PMPY_Far_VelocityField}
(shown by arrows in the green boxes in Figure \ref{fig.PMPY_VF_Uf-Ue} c\&d). 
This rotation of the velocity filed is larger when the PMPY have unequal-sized
spheres and the tracer sphere is near the larger one (Figure
\ref{fig.PMPY_VF_Uf-Ue}c). Figure \ref{fig.PMPY_VF_Uf-Ue} e\&f show the velocity
field when the PMPY undergoes  radial changes without changes in the length of
the connecting rod. Again we consider unequal-sized spheres (Figure
\ref{fig.PMPY_VF_Uf-Ue}e) and equal-sized spheres (Figure
\ref{fig.PMPY_VF_Uf-Ue}f) at one instant.  In the asymptotic analysis (Figure \ref{fig.PMPY_Far_VelocityField}) the
effects on the tracer sphere that result from radial changes in the swimmer are  not
considered, as they only give rise to $O (\delta^2)$ terms (equation
(\ref{eq.U3_1e})). However, as we can see from Figure \ref{fig.PMPY_VF_Uf-Ue}ef,
 expansion and contraction of the swimmer  have significant  effects on
the tracer sphere, and therefore  should be taken into account. 

\begin{figure}[htbp] \centering
\includegraphics[width=1\textwidth]{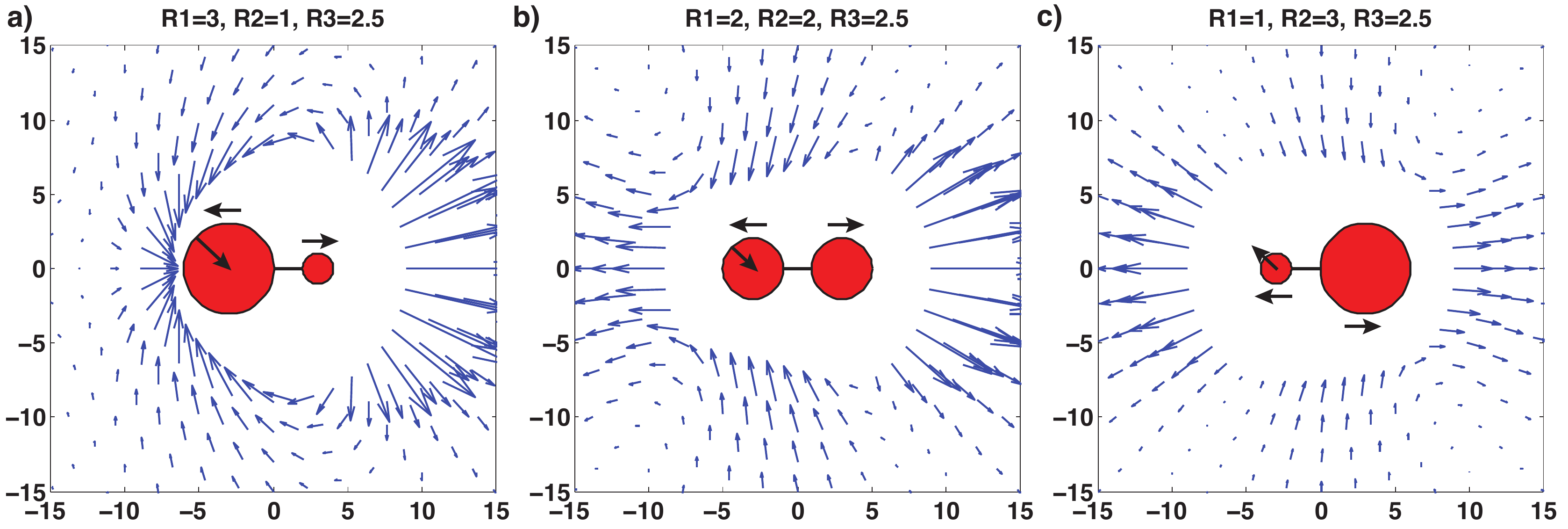}
\caption{The instantaneous velocity of a tracer sphere that results from a PMPY
  swimmer when the tracer sphere is close to the PMPY and higher-order effects
  are included. The arrows in all panels
  are scaled uniformly, thus arrows show both direction and magnitudes.
   a) 
  The connecting
  rod is instantaneously expanding at $\dot{l} = 2 \pi$ and sphere 1 is
  shrinking at $\dot{R}_1 = -2 \pi$. 
  figure.  
  b) $\dot{l} = 2   \pi, \ \dot{R}_1 = -2 \pi$.
  c) $\dot{l} = 2 \pi, \ \dot{R}_1 = 2 \pi$.}
 \label{fig.PMPY_VF}
\end{figure}

Finally, when the two shape changes governed by $\dot{l}$ and $\dot{R}_1$ are
combined, as occurs in most of the simulations, the $O (\delta^2)$ term in the
tracer's velocity clearly gives rise to a large change from the asymptotic
solution when the tracer is close to the PMPY (compare Figure
\ref{fig.PMPY_Far_VelocityField} and \ref{fig.PMPY_VF}). In particular,
depending on the instantaneous system profile, the effects resulting from the
sphere expansion/contraction might overcome that from the rod length changes
(Figure \ref{fig.PMPY_VF}a), and therefore an asymptotic analysis is not
sufficient for the study of swimmer-tracer interactions when they are close
together.

\subsubsection{The long-term behavior of the tracer}

\begin{figure}[htbp] \centering
\includegraphics[width=1\textwidth]{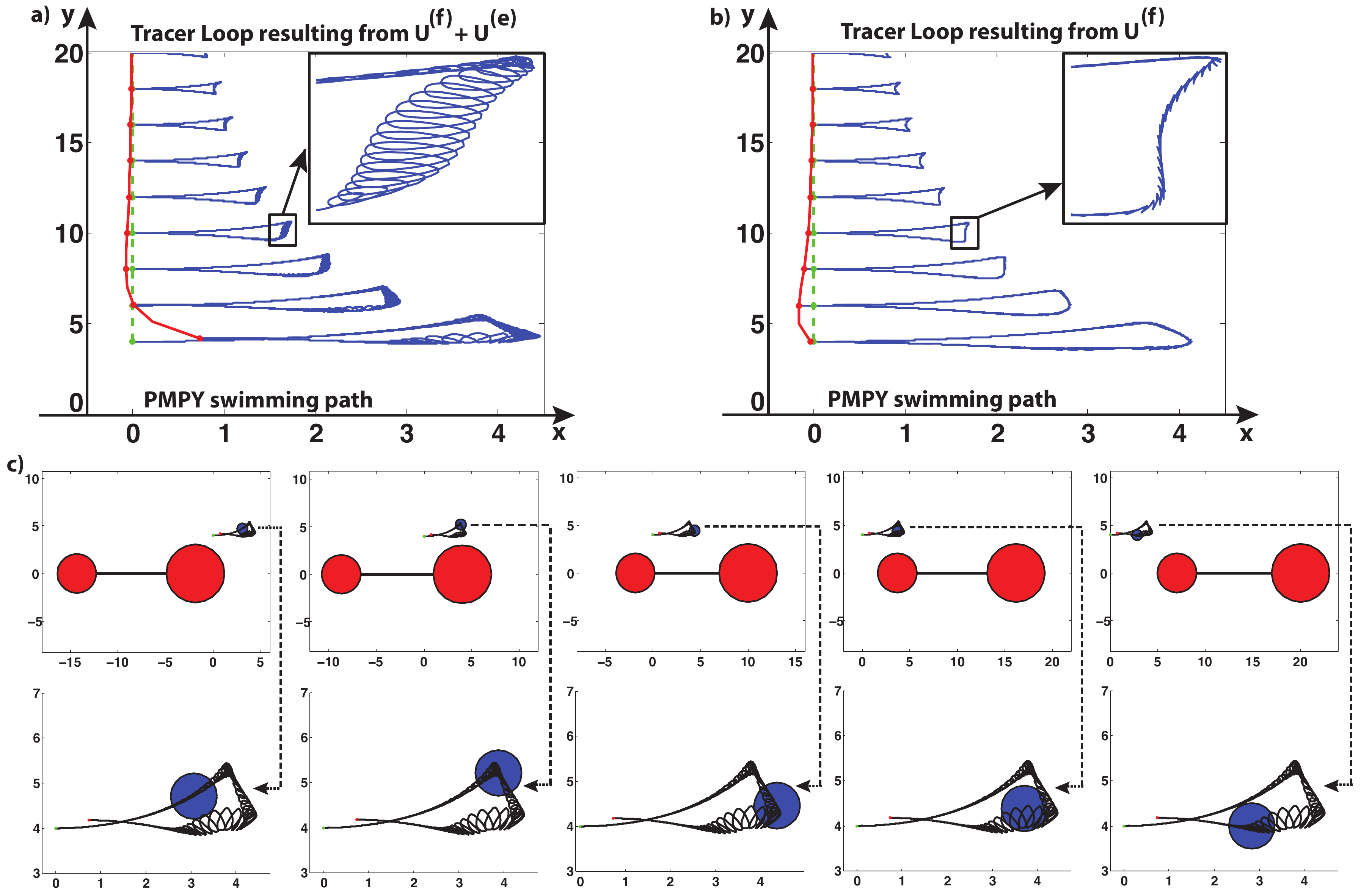}
\caption{The tracer loop when a PMPY swimmer moves in a straight line. The
system profile is: $l (t) =12 + \cos (2 \pi t), \ R_1 = 2 - \sin (2 \pi t), \
R_2 (0) = 3, \ R_3 \equiv 0.5$. The swimmer moves along the $x$-axis, from $x
= -1000$ to $x =1000$. The tracer sphere 3 is originally located at point $(0, Y_0)$, where
$Y_0 \in [4,20]$ is the vertical distance of the tracer to the swimmer's path. The
green dashed lines give the starting points of the tracer while the red solid
line gives the end points. These curves are computed at increments of $\Delta
Y_0 = 1$.  We also show several tracer trajectories (blue solid lines), drawn
for $ \Delta Y_0 = 2$.  a) The tracer loop of sphere 3, with $\mathbf{U}_3 =
\mathbf{U}_3^{(f)} + \mathbf{U}_3^{(e)} $.
b) The tracer loop of sphere 3, computed as $\mathbf{U}_3
\sim \mathbf{U}_3^{(f)} $ only.
c) Snap-shots of the system in  (a) with $Y_0 = 4$ and the enlarged view of sphere
3 showing its location in the loop.
}
 \label{fig.TracerLoop}
\end{figure}

As we mentioned
earlier, experimental observations show that trajectories of tracers in a
suspension of swimmers often look loop-like \citep{leptos2009dynamics}, but 
a loop-like trajectory will not enhance nutrient supply. In fact, simulations of
Rhodobacter sphaeroids \citep{shum2010modelling} have shown that when the tracer
is far away from the straight path of the swimmer, the loop-like trajectory is
approximately true with the net displacement between the initial and final
locations considerably shorter than the characteristic trajectory size. On the
other hand, when the tracer is close to the swimmer path, it is clearly pulled
forward by the swimmer \citep{pushkin2013fluid,yeomans2014introduction}.
Here we demonstrate a similar effect for  a PMPY swimmer. The
results, again with the translation of the PMPY computed by equation
(\ref{eq.RPY_System}) and that of sphere 3 by equations (\ref{eq.U3_1f},
\ref{eq.U3_1e}) are  given in Figure \ref{fig.TracerLoop}a.
The result is qualitatively similar to what is obtained from the simulations of
Rhodobacter sphaeroids \citep{shum2010modelling}, that is, the tracer is pushed
backwards slightly when more distant from the PMPY swimming path, and clearly
pulled forward when close to the path. Its trajectory has three distinct
branches, and a close-up view of a part of the trajectory shows that the tracer
undergoes a spiral oscillatory motion, which is qualitatively similar to
existing results \citep{dunkel2010swimmer,
  pushkin2013fluid,yeomans2014introduction}.  According to an asymptotic
analysis \citep{pushkin2013fluid}, the tracer trajectory should approximate a
closed loop, but when it is close to the PMPY Figures \ref{fig.PMPY_VF_Uf-Ue}
and \ref{fig.PMPY_VF} show that the $O (\delta^2)$ term resulting from sphere
expansion and contraction induces a significant distortion to the instantaneous
velocity field of the tracer. The blow-up in Figure \ref{fig.TracerLoop}c shows
the position of the tracer sphere relative to the swimmer's position in the
upper panels and the corresponding position along the quasi-loop in the lower
panels. These are instantaneous snapshots, but the swimmer undergoes many
cycles in the entire sequence. One sees there that the oscillations are smaller when the swimmer is
approaching the tracer, during which the tracer is pushed away from the swimmer;
on the other hand. Conversely, the oscillations are larger when the tracer is in the wake of
the swimmer, when the tracer is being pulled toward the swimmer. While the net 
translation direction  of the tracer per swimmer cycle is determined by its relative position to the swimmer
(tracer movement:forward/downward/backward $\equiv$ location: in front of/above/behind) , the amplitude
of the tracer oscillations is determined by its distance to the swimmer, as can be expected.

To investigate  the effects of rod or/and sphere  expansion/contraction we ran
the long-term swimmer-tracer simulation again with the same strokes but with the
translational velocity of the tracer computed from the flow generated by rod
length changes only, i.e., $\mathbf{U}_3 \sim \mathbf{U}_3^{(f)}$ (equation
(\ref{eq.U3_1f})). The result is shown in Figure \ref{fig.TracerLoop}b, from
which we see that the tracer is still pushed backward slightly, even when close
to the PMPY swimming path. Thus Figure \ref{fig.TracerLoop}b shows that at least
under the current system profile, the effect of the drag force, which
approximates the asymptotic behavior, indeed produces an essentially closed
loop-like trajectory of the tracer, as was concluded in
\citep{pushkin2013fluid}.  On the other hand, the forward motion of the tracer
(Figure \ref{fig.TracerLoop}a) is primarily due to the sphere
expansion/contraction $O (\delta^2)$ terms. Moreover, a close-up view of the
trajectory shows that the tracer still undergoes an oscillatory motion when
subject only to the velocity field due to rod shortening and lengthening (Figure
\ref{fig.TracerLoop}b, upper-right corner), but not in a spiral
manner. Therefore we conclude that the spiral mode of the tracer motion is
primarily due to the expansion/contraction of the PMPY spheres.

\begin{figure}[htbp] \centering
\includegraphics[width=1\textwidth]{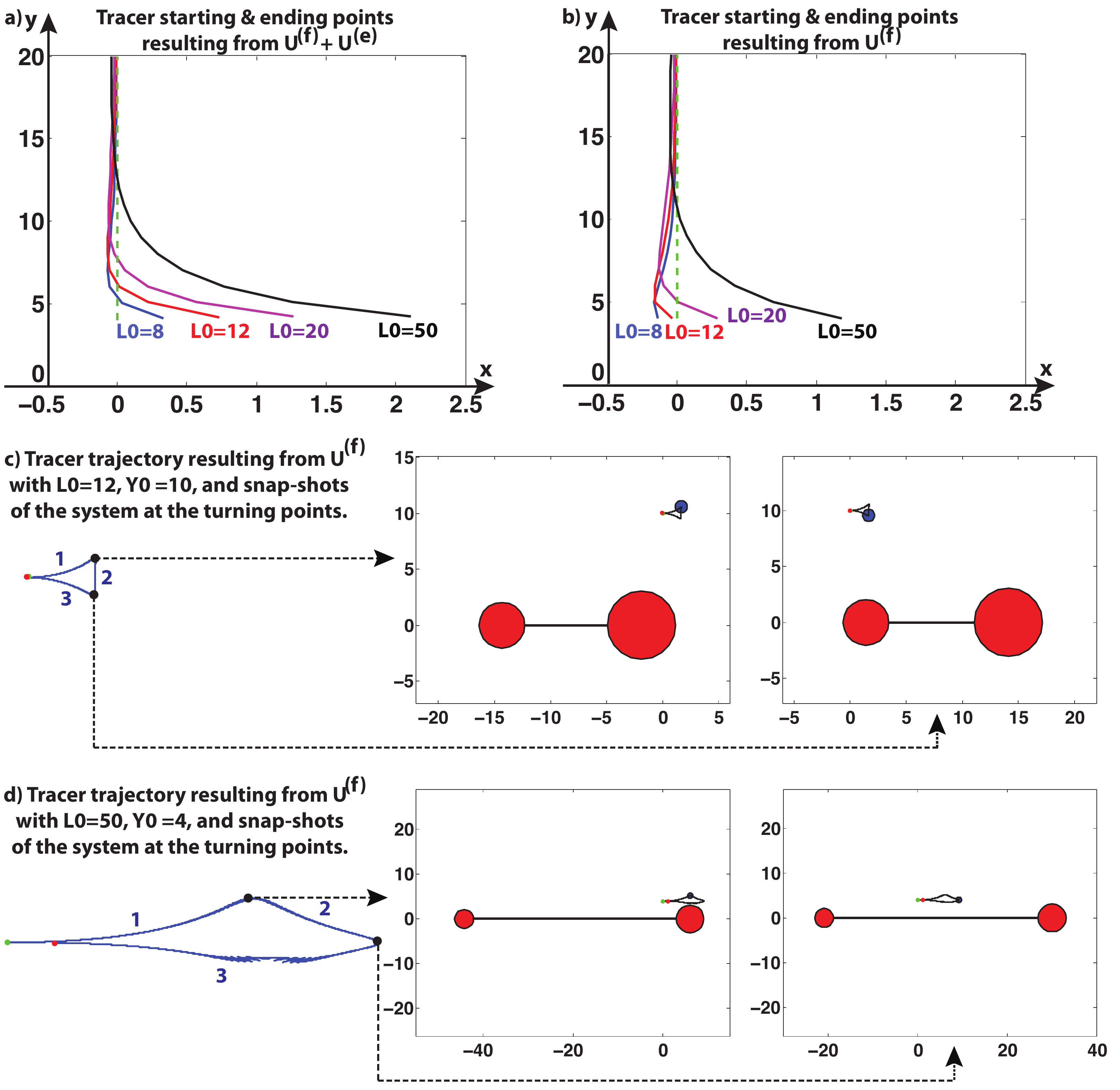}
\caption{a) The tracer sphere's starting (green dashed line) and ending points
  (solid lines) when a PMPY swims along the $x$-axis, from $x = -1000$ to $x =
  1000$. The system profile is: $l (t) =L_0+ \cos (2 \pi t), \ R_1 = 2 - \sin (2
  \pi t), \ R_2 (0) = 3, \ R_3 \equiv 0.5$, with $L_0 = 8, 12, 20, 50$ (blue,
  red, magenda, and black solid lines). The tracer sphere 3 is originally
  located at $(0, Y_0)$, where $Y_0 \in [4,20]$ is the vertical distance of the tracer to
  the PMPY swimming path.  The lines are computed with $\Delta Y_0 = 1$. The
  translational velocity of the tracer sphere is calculated as $\mathbf{U}_3 =
  \mathbf{U}_3^{(f)} + \mathbf{U}_3^{(e)} $. 
  b) $\mathbf{U}_3 \sim \mathbf{U}_3^{(f)} $.
  c)  The tracer trajectory with $L_0 = 12, \
  Y_0 = 10$ and $\mathbf{U}_3 \sim \mathbf{U}_3^{(f)} $ only. Snap-shots of the
  system at the turning points of the tracer trajectory.  d) As in (c) but with 
 $L_0 = 50, \ Y_0 =  4$. }
 \label{fig.TracerLoop_DifL0}
\end{figure}

Next we investigate whether  the shape of the PMPY swimmer also has an effect on the
tracer scattering behavior. We consider swimmers with different rod
lengths. In the following simulations we consider the system profiles
\begin{eqnarray*}
l (t) =L_0+ \cos (2 \pi t), \ R_1 = 2 - \sin (2 \pi t), \ R_2 (0) = 3, \ R_3
\equiv 0.5, 
\end{eqnarray*}
with $L_0 = 8, 12, 20, 50$. The simulation results, with $\mathbf{U}_3 =
\mathbf{U}_3^{(f)} + \mathbf{U}_3^{(e)} $ calculated from equations
(\ref{eq.U3_1f}, \ref{eq.U3_1e})  are given in Figure
\ref{fig.TracerLoop_DifL0}a. There one  sees that the longer the rod is, or
equivalently, the further the two spheres in the swimmer  are apart, the more the
tracer is pulled  forward,  especially when it is close to the  swimmer's path.
Because we found earlier that the PMPY sphere expansion/contraction contributes
significantly to the movement of the tracer, we repeat the simulations in Figure
\ref{fig.TracerLoop_DifL0}a but with $\mathbf{U}_3 \sim \mathbf{U}_3^{(f)} $
computed by equation (\ref{eq.U3_1f} ) only, which approximates the asymptotic
behavior.  The results are shown in Figure \ref{fig.TracerLoop_DifL0}b, from
which we see that even without the sphere expansion/contraction, the tracer is
still clearly dragged forward when it is close to the swimming path of a long
PMPY ($L_0 = 50$).  To understand this behavior, we compare two tracer
trajectories, both of which result from $\mathbf{U}_3 \sim \mathbf{U}_3^{(f)} $
only.  One starts with $L_0 = 12$ and $Y_0 = 10$ (Figure
\ref{fig.TracerLoop_DifL0}c), i.e, the PMPY spheres are not far apart and the
tracer is not very close to the swimmer's path, and the other starts with
$L_0 = 50$ and $Y_0 = 4$ (Figure \ref{fig.TracerLoop_DifL0}d), i.e, the PMPY
spheres are far apart and the tracer is close to the swimmer's path.  In
Figure \ref{fig.TracerLoop_DifL0}c) the trajectory of the  tracer
approximates an isosceles triangle and the base (side 2 in the figure)
corresponds to the part of the trajectory in which the tracer is essentially
directly over the swimmer.  On the other hand, when $L_0$ is large, as in (d),
the trajectory of the tracer is very different, and in particular, we see that
side 2 of the triangle-like trajectory is stretched horizontally, which reflects
enhanced pulling effect on the tracer.  In either case, side 2 of the trajectory
loop essentially reflects the tracer motion when it is above the swimmer, and
therefore the comparison between Figure \ref{fig.TracerLoop_DifL0}c and d
indicates that the separation of spheres in the PMPY results in an increased
pulling effect on the tracer if it is close enough to the swimmer's path.

In conclusion, we find that when the tracer is close to the infinite PMPY
swimming path,
\begin{enumerate}
\item the $O (\delta^2)$ term that results from the PMPY sphere
  expansion/contraction  contributes significantly  to the tracer scattering;
\item increasing the sphere separation in a PMPY swimmer will enhance the tracer
  scattering.

\end{enumerate}


 \section{Swimming with a friend}
\label{Sec.2PMPY}

\setcounter{equation}{0}
\renewcommand{\theequation}{\ref{Sec.2PMPY}.\arabic{equation}}

\begin{figure}[htbp] \centering
\includegraphics[width=0.35\textwidth]{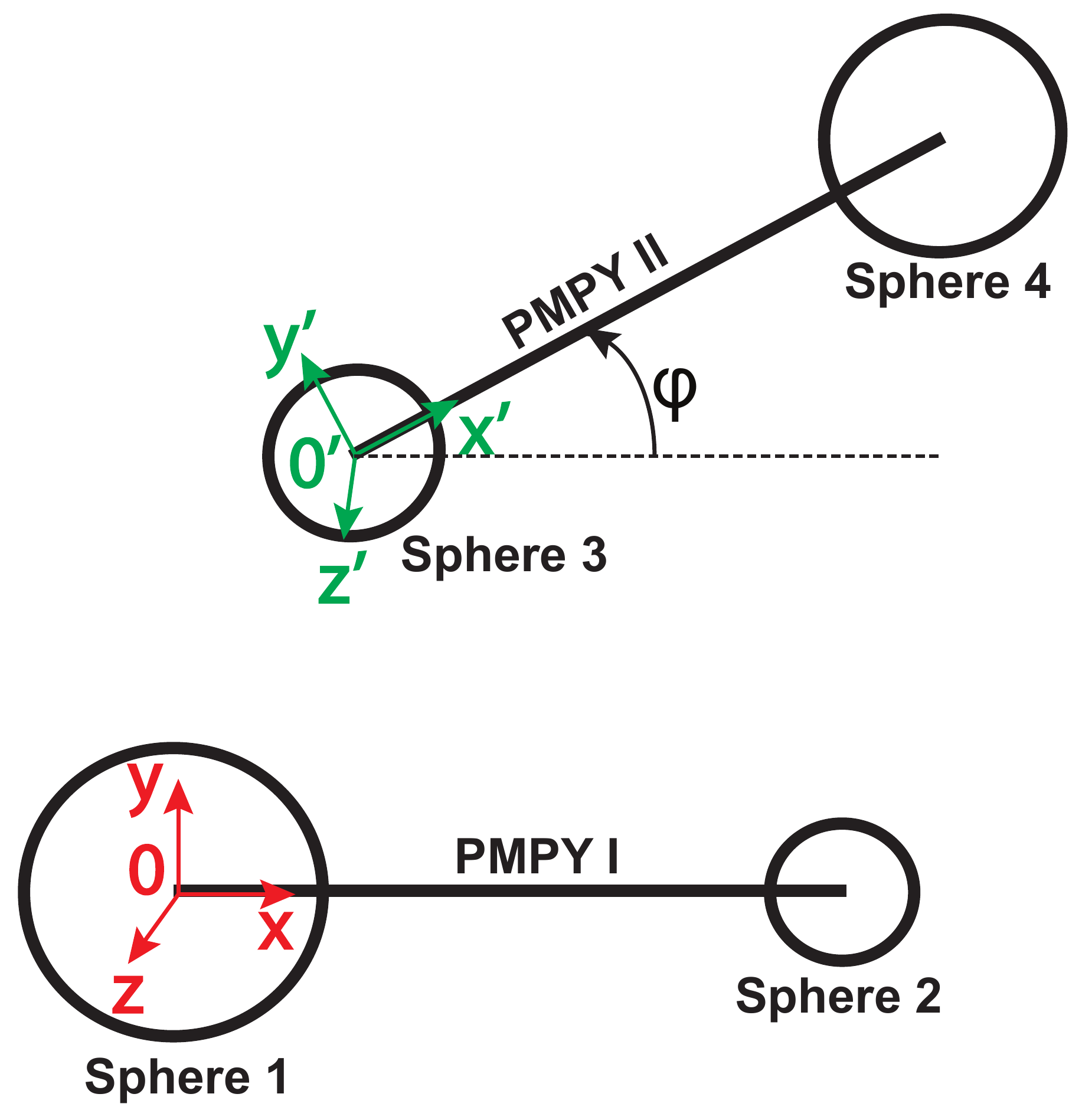}
\caption{The geometry of two  PMPY swimmers in $\mathbb{R}^3$.}
 \label{fig.2PMPY_Geometry}
\end{figure}

In this final section of analysis we consider the hydrodynamic interactions
between two PMPY swimmers in an infinite fluid domain, as shown in
Figure~$\ref{fig.2PMPY_Geometry}$.  To fix the geometry, we introduce a fixed
Cartesian frame and we assign a body frame to each swimmer. $\{\mathbf{0};
\mathbf{e}_x,\mathbf{e}_y,\mathbf{e}_z\}$ is attached to PMPY I and has its
origin $\mathbf{0}$ at the center $\mathbf{x}_1$ of sphere 1, and has
$\mathbf{e}_x$ along the direction of the connecting rod; while $\{\mathbf{0'};
\mathbf{e}_{x'},\mathbf{e}_{y'},\mathbf{e}_{z'} \}$ is attached to PMPY II and
has its origin $\mathbf{0}$ at the center $\mathbf{x}_3$ of sphere 3, and
$\mathbf{e}_{x'}$ along the direction of its connecting rod. The relationship
between the two frames is given by
\begin{eqnarray*} \left\{
 \begin{array}{l} 
\mathbf{e}_{x'} = \cos \phi \mathbf{e}_x + \sin \phi \mathbf{e}_y \\
\mathbf{e}_{y'} = - \sin \phi \mathbf{e}_x + \cos \phi \mathbf{e}_y \\
\mathbf{e}_{z'} = \mathbf{e}_z
 \end{array}
 \right.
 \end{eqnarray*} 
where $\phi $ is the angle between the two connecting rods: $\phi = \arccos  ( \mathbf{d}_{12} \cdot \mathbf{d}_{34}
 )$ and $\mathbf{d}_{ij} = (\mathbf{x}_j -
\mathbf{x}_i)/ | \mathbf{x}_j - \mathbf{x}_i |$.

%
 

 \subsection{Preliminary analysis of the two-swimmer system}
 
 The basic steps in the analysis of the two PMPY swimmer system are the same as
 in previous sections, and therefore we only sketch the analysis up to the first
 reflection. In the zeroth reflection, no hydrodynamic interactions  are
 considered, and therefore 
\begin{eqnarray}\label{eq.2PMPY_0Ref} \mathbf{U}_i^{(0)} = \dfrac{1}{6 \pi \mu
R_i } \mathbf{F}_i, \qquad \bfO_i^{(0)} = \mathbf{0}.
 \end{eqnarray}

The translational velocities after the first reflection are given by
\begin{eqnarray}\label{eq.2PMPY_1Ref_U} \mathbf{U}_i^{(1)} &=& \sum_{j\neq i}
\Big[ \mathbf{U}_{i,j}^{(1,f)} + \mathbf{U}_{i,j}^{(1,e)} \Big] \\\label{eq.2PMPY_1Ref_Uf}
\mathbf{U}_{i,j}^{(1,f)} &=& \dfrac{1}{8 \pi \mu l_{ij}} \Big[ \Big( 1 +
\dfrac{R_i^2 + R_j^2}{3 l_{ij}^2} \Big) \mathbf{F}_j + \Big( 1 - \dfrac{R_i^2 +
R_j^2}{ l_{ij}^2} \Big) \big( \mathbf{F}_j \cdot \mathbf{d}_{ji} \big)
\mathbf{d}_{ji} \Big] \\ \label{eq.2PMPY_1Ref_Ue} \mathbf{U}_{i,j}^{(1,e)} &=&
\dot{R}_j \Big( \dfrac{R_j}{l_{ij}} \Big)^2 \mathbf{d}_{ji}
 \end{eqnarray}
where $l_{ij} = |\mathbf{x}_i - \mathbf{x}_j|$ and $\mathbf{d}_{ji} =
(\mathbf{x}_i - \mathbf{x}_j)/l_{ij}$.  The corresponding  angular velocities
are 
\begin{eqnarray}\label{eq.2PMPY_1Ref_Omega34} \bfO_{\alpha}^{(1)} &=&
\sum_{i=1,2} \dfrac{F_i}{8 \pi \mu l_{i \alpha}^3} \Big[ \big( \mathbf{x}_\alpha -
\mathbf{x}_i \big) \cdot \mathbf{e}_y \Big] \mathbf{e}_z \qquad \textrm{for} \quad \alpha = 3,4
\\ \label{eq.2PMPY_1Ref_Omega12} \bfO_{i}^{(1)} &=& \sum_{\alpha=3,4}
\dfrac{F_\alpha}{8 \pi \mu l_{ \alpha i}^3} \Big[  \big( \mathbf{x}_i -
\mathbf{x}_\alpha \big) \cdot  \mathbf{e}_{y'}  \Big] \mathbf{e}_{z'} \qquad \textrm{for} \quad i=1,2.
\end{eqnarray}

In addition, the system should satisfy the  volume conservation condition 
\begin{eqnarray}\label{eq.2PMPY_Volume} \sum_{i=1,2} R_i^3 (t) \equiv
\dfrac{3}{4 \pi} V_I, \qquad \sum_{\alpha = 3,4} R_\alpha^3 (t) \equiv
\dfrac{3}{4 \pi} V_{II}
\end{eqnarray}
and the force- and torque-free conditions
\begin{eqnarray*}
\textrm{Force-free:} &\quad& \mathbf{F}_1 + \mathbf{F}_2 =  \mathbf{0}, \quad   \mathbf{F}_3 + \mathbf{F}_4 =  \mathbf{0} \\
\textrm{Torque-free:} &\quad&  \mathbf{x}_1 \times \mathbf{F}_1 +  \mathbf{x}_2 \times \mathbf{F}_2  = \mathbf{0}, \quad
\mathbf{x}_3 \times \mathbf{F}_3 +  \mathbf{x}_4 \times \mathbf{F}_4  = \mathbf{0}.
\end{eqnarray*}
The latter  constraints require that 
\begin{eqnarray*}
\mathbf{F}_i = F_i \mathbf{e}_x \ (i=1,2), \qquad \mathbf{F}_\alpha = F_\alpha \mathbf{e}_{x'} \ (\alpha=3,4)
\end{eqnarray*}
and
\begin{eqnarray}\label{eq.2PMPY_ForceFree} \sum_{i=1,2} \mathbf{F}_i =
\sum_{i=1,2} F_i \mathbf{e}_x \equiv \mathbf{0} , \qquad \sum_{\alpha=3,4}
\mathbf{F}_\alpha = \sum_{\alpha =3,4} F_\alpha \mathbf{e}_{x'} \equiv
\mathbf{0}.
\end{eqnarray}
Finally the rigid motions of the spheres in each PMPY should satisfy:
\begin{eqnarray}\label{eq.2PMPY_RigidMotion_U} & &\big( \mathbf{U}_2 -
\mathbf{U}_1 \big) \cdot \mathbf{e}_x = \dot{l}_{12}, \qquad \big( \mathbf{U}_4
- \mathbf{U}_3 \big) \cdot \mathbf{e}_{x'} = \dot{l}_{34}
\end{eqnarray}
Equations~($\ref{eq.2PMPY_0Ref}$ - $\ref{eq.2PMPY_RigidMotion_U}$) define the
system that determines the swimming of the two PMPY models. 

We set
\begin{eqnarray*}
R_M = \max_t \{ R_i (t) \}_{i=1,2,3,4}, \quad L_m = \min_t \{ l_{12} (t), l_{34} (t) \}, \quad \delta = \dfrac{R_M}{L_m},
\end{eqnarray*}
we restrict the analysis to the regime that $l_{ij} > L_m, \ i=1,2, \ j =3,4$,
and we use  the nondimensionalization given  in section~\ref{Sec.SLOA}. A simple
analysis similar to that in section~\ref{Sec.PMPY1OB_Accuracy} gives the following
leading order estimate
\begin{eqnarray*} \mathbf{U}_i^{(0)} \sim   O (1), \quad
\mathbf{U}_i^{(1)} \sim  O (\delta) , \quad \bfO_i^{(1)} \sim
 O (\delta^3)
 \end{eqnarray*}
From this  we see that when  $\delta \ll 1$, the rotation effect of the PMPYs
is much smaller as compared to the translation effect.

 
 \subsection{Hydrodynamic interactions between two PMPY swimmers on a line}
\label{Sec.2PMPY_Line}

 To proceed further we must specify how the pair moves, and we first consider
 the configuration in which they lie in the same line, specifically,
 suppose that the  centers of all spheres lie along the $x$-axis.  In addition,
 we place  PMPY II in front of PMPY I, hence from negative to positive direction along the
$x$-axis, the spheres are ordered from sphere 1 to sphere 4 (Figure \ref{fig.Linear2PMPY_PhaseDiff_s}il).

Equations
(\ref{eq.2PMPY_1Ref_U}, \ref{eq.2PMPY_1Ref_Uf})  show that the hydrodynamic
interactions between the two PMPYs  arise  $O (\delta)$.  A simple perturbation analysis gives the
non-dimensionalized forces exerting on the spheres (again with hat notation
omitted):
\begin{eqnarray*}
- F_1 = F_2 &\sim& \dfrac{R_1 R_2}{R_1 + R_2} \xi_I  + 3 \dfrac{\delta}{l_{12}}
\Big(  \dfrac{R_1 R_2}{R_1 + R_2} \Big)^2 \xi_I \\ 
& & + \dfrac{3}{2} \delta \dfrac{R_1 R_2 R_3  R_4}{ (R_1 + R_2 ) (R_3 + R_4)}
\Big( \dfrac{1}{l_{23}} - \dfrac{1}{l_{24}} - \dfrac{1}{l_{13}} +
\dfrac{1}{l_{14}} \Big) \xi_{II} + O (\delta^2) \\   
- F_3 = F_4 &\sim& \dfrac{R_3 R_4}{R_3 + R_4} \xi_{II}  + 3
\dfrac{\delta}{l_{34}} \Big(  \dfrac{R_3 R_4}{R_3 + R_4} \Big)^2 \xi_{II} \\ 
& & + \dfrac{3}{2} \delta \dfrac{R_1 R_2 R_3  R_4}{ (R_1 + R_2 ) (R_3 + R_4)}
\Big( \dfrac{1}{l_{23}} - \dfrac{1}{l_{24}} - \dfrac{1}{l_{13}} +
\dfrac{1}{l_{14}} \Big) \xi_{I} 
+ O (\delta^2) 
\end{eqnarray*}
where $l_{ij} = |\mathbf{x}_i - \mathbf{x}_j|$, and the translational velocities of the spheres:
\begin{eqnarray*}
U_1 &\sim& - \dfrac{R_2}{R_1 + R_2} \xi_I + \delta \dfrac{R_1 R_2 (R_1 - R_2)}{2
  (R_1 + R_2)^2 l_{12}} \xi_I  \\ 
& & - \dfrac{3}{2} \delta \dfrac{R_3 R_4}{(R_1 + R_2 ) (R_3 + R_4)} \xi_{II}
\Big[ R_1 \Big( \dfrac{1}{l_{13}} - \dfrac{1}{l_{14}} \Big) + R_2 \Big(
\dfrac{1}{l_{23}} - \dfrac{1}{l_{24}} \Big) \Big] + O (\delta^2) \\ 
U_3 &\sim& - \dfrac{R_4}{R_3 + R_4} \xi_{II} + \delta \dfrac{R_3 R_4 (R_3 -
  R_4)}{2 (R_3 + R_4)^2 l_{34}} \xi_{II}  \\ 
& & - \dfrac{3}{2} \delta \dfrac{R_1 R_2}{(R_1 + R_2 ) (R_3 + R_4)} \xi_{I}
\Big[ R_3 \Big( \dfrac{1}{l_{13}} - \dfrac{1}{l_{23}} \Big) + R_4 \Big(
\dfrac{1}{l_{14}} - \dfrac{1}{l_{24}} \Big) \Big] + O (\delta^2)  
\end{eqnarray*}
Reference to  equations (\ref{eq.PT_U1}, \ref{eq.PT_U2}), and taking  into
consideration  the geometric condition at equation (\ref{eq.2PMPY_RigidMotion_U}),
leads to the following estimate of the hydrodynamic interaction effects on the
velocities of the PMPYs. 
\begin{eqnarray}\label{eq.Linear2PMPY_Hydro_21}
U^{\textrm{hydro}}_{II \rightarrow I} &=& - \dfrac{3}{2} \delta \dfrac{R_3
  R_4}{(R_1 + R_2 ) (R_3 + R_4)} \xi_{II} \Big[ R_1 \Big( \dfrac{1}{l_{13}} -
\dfrac{1}{l_{14}} \Big) + R_2 \Big( \dfrac{1}{l_{23}} - \dfrac{1}{l_{24}} \Big)
\Big] + O (\delta^2)  \\ \label{eq.Linear2PMPY_Hydro_12} 
U^{\textrm{hydro}}_{I \rightarrow II} &=& \dfrac{3}{2} \delta \dfrac{R_1
  R_2}{(R_1 + R_2 ) (R_3 + R_4)} \xi_{I} \Big[ R_3 \Big( \dfrac{1}{l_{23}} -
\dfrac{1}{l_{13}} \Big) + R_4 \Big( \dfrac{1}{l_{24}} - \dfrac{1}{l_{14}} \Big)
\Big] + O (\delta^2)  
\end{eqnarray}
The geometric relations give 
\begin{eqnarray*}
 \dfrac{1}{l_{13}} - \dfrac{1}{l_{14}}, \  \dfrac{1}{l_{23}} - \dfrac{1}{l_{24}}, \   \dfrac{1}{l_{23}} - \dfrac{1}{l_{13}} , \  \dfrac{1}{l_{24}} - \dfrac{1}{l_{14}}  > 0 
\end{eqnarray*}
and thus the leading order of $U^{\textrm{hydro}}_{II \rightarrow I} $ has the
opposite sign of $\xi_{II}$ while that of $U^{\textrm{hydro}}_{I \rightarrow
  II} $ has  the same sign as  $\xi_{I}$:
\begin{eqnarray*}
U^{\textrm{hydro}}_{II \rightarrow I}  \xi_{II} < 0 , \quad  U^{\textrm{hydro}}_{I \rightarrow II}  \xi_{I} > 0
\end{eqnarray*}
This can be understood as follows.  Since PMPY II is leading, when it lengthens 
it impedes the swimming of PMPY I, which follows; on the other hand, when PMPY I
lengthens, it enhances the swimming of PMPY II in front of it.

The full expressions of the $O (\delta^2)$ term of $U^{\textrm{hydro}}_{II
  \rightarrow I} , \ U^{\textrm{hydro}}_{I \rightarrow II} $ are lengthy,
however, by taking $\xi_I = \xi_{II} = 0$, we can easily obtain the hydrodynamic
interactions on the sphere velocities that result from the sphere
expansion/contraction of the other PMPY only  (Appendix \ref{Appendix.ControlRR}):
\begin{eqnarray*}
U^{\textrm{hydro}}_{II \rightarrow I} \{ \dot{R} \} &=& -R_3^2 \zeta_3 \delta^2
\Big[ \dfrac{R_1}{R_1 + R_2} \Big( \dfrac{1}{l_{13}^2} - \dfrac{1}{l_{14}^2}
\Big)  +  \dfrac{R_2}{R_1 + R_2}  \Big(   \dfrac{1}{l_{23}^2} - \dfrac{1}{l_{24}^2}
\Big) \Big] + O (\delta^3) \\ 
U^{\textrm{hydro}}_{I \rightarrow II} \{ \dot{R} \}  &=& - R_1^2 \zeta_1
\delta^2 \Big[ \dfrac{R_3}{R_3 + R_4} \Big( \dfrac{1}{l_{23}^2} -
\dfrac{1}{l_{13}^2} \Big)  +  \dfrac{R_4}{R_3 + R_4}  \Big(
\dfrac{1}{l_{24}^2} - \dfrac{1}{l_{14}^2} \Big) \Big] + O (\delta^3)  
\end{eqnarray*}
The geometry of the system also dictates that the leading order term of
$U^{\textrm{hydro}}_{II \rightarrow I} \{ \dot{R} \} $ has the opposite sign of
$\zeta_3$, while that of $U^{\textrm{hydro}}_{I \rightarrow II} \{ \dot{R} \} $
has the opposite sign of $\zeta_1$:
\begin{eqnarray*}
\big( U^{\textrm{hydro}}_{II \rightarrow I}  \{ \dot{R} \} \big)  \zeta_3  < 0 , \quad  \big( U^{\textrm{hydro}}_{I \rightarrow II} \{ \dot{R} \}  \big)  \zeta_1 < 0
\end{eqnarray*}
The first relation is easy to understand: the expansion of sphere 3, which is
the closer of 3 and 4 to PMPY I, will impede the swimming of PMPY I, which
follows PMPY II. For the second relation, recall that the volume conservation
constraint gives $R_1^2 \zeta_1 = - R_2^2 \zeta_2$, thus we have $\big(
U^{\textrm{hydro}}_{I \rightarrow II} \{ \dot{R} \} \big) \zeta_2 > 0$, that is,
the leading order of $U^{\textrm{hydro}}_{I \rightarrow II} \{ \dot{R} \} $ has
the same sign as $\zeta_2$.  In another words, the expansion of sphere 2 which
is the sphere in PMPY I that is near PMPY II, will enhance the swimming of PMPY
II.

When $\delta \ll 1$ the hydrodynamic effect that results from sphere
expansion/contraction ($U^{\textrm{hydro}} \{ \dot{R} \}$ $ \sim O (\delta^2)$)
can certainly be neglected compared to the total hydrodynamic effect due to
expansion/contraction of the rod, which is $U^{\textrm{hydro}} \sim O (\delta)$.
Here we consider the hydrodynamic effects when the spheres are allowed to come
closer, i.e., $\delta <1$ but not too small. We first consider the instantaneous
behavior of the system, and  then the period-average behavior. For example, we
consider an instantaneous system configuration in which  $R_1 = R_2 = R_3 = R_4 = R$ and
$l_{12} = l_{23} = l_{34} = L$, and for this  we have for the leading order terms
\begin{eqnarray*}
\big| U^{\textrm{hydro}}_{II \rightarrow I} \big| = \dfrac{R}{4L} \delta |\xi_{II}|, \quad 
 \big| U^{\textrm{hydro}}_{II \rightarrow I} \{ \dot{R} \}  \big| = \dfrac{4R^2}{9L^2} \delta^2 |\zeta_3 |.
\end{eqnarray*}
Recalling  that the non-dimensionalization of our system guarantees that $R \leq
1, \ L \geq 1$ (equation (\ref{eq.ScaleRL_Range})), in the case of $R = L = 1$, we have the relation
\begin{eqnarray*}
\dfrac{ \big| U^{\textrm{hydro}}_{II \rightarrow I} \{ \dot{R} \}  \big| }{\big| U^{\textrm{hydro}}_{II \rightarrow I} \big| }
\sim \dfrac{16}{9} \delta \Big| \dfrac{\zeta_3}{\xi_{II}} \Big|.
\end{eqnarray*}
For a system with $\delta \sim 0.5$ where the spheres are close to each other,
the coefficient of the above equation is $8/9 $, and in this case the
hydrodynamic effects that result from sphere expansion/contraction must be
included, together with higher order terms in $U^{\textrm{hydro}}$.

\begin{figure}[htbp] \centering
\includegraphics[width=1\textwidth]{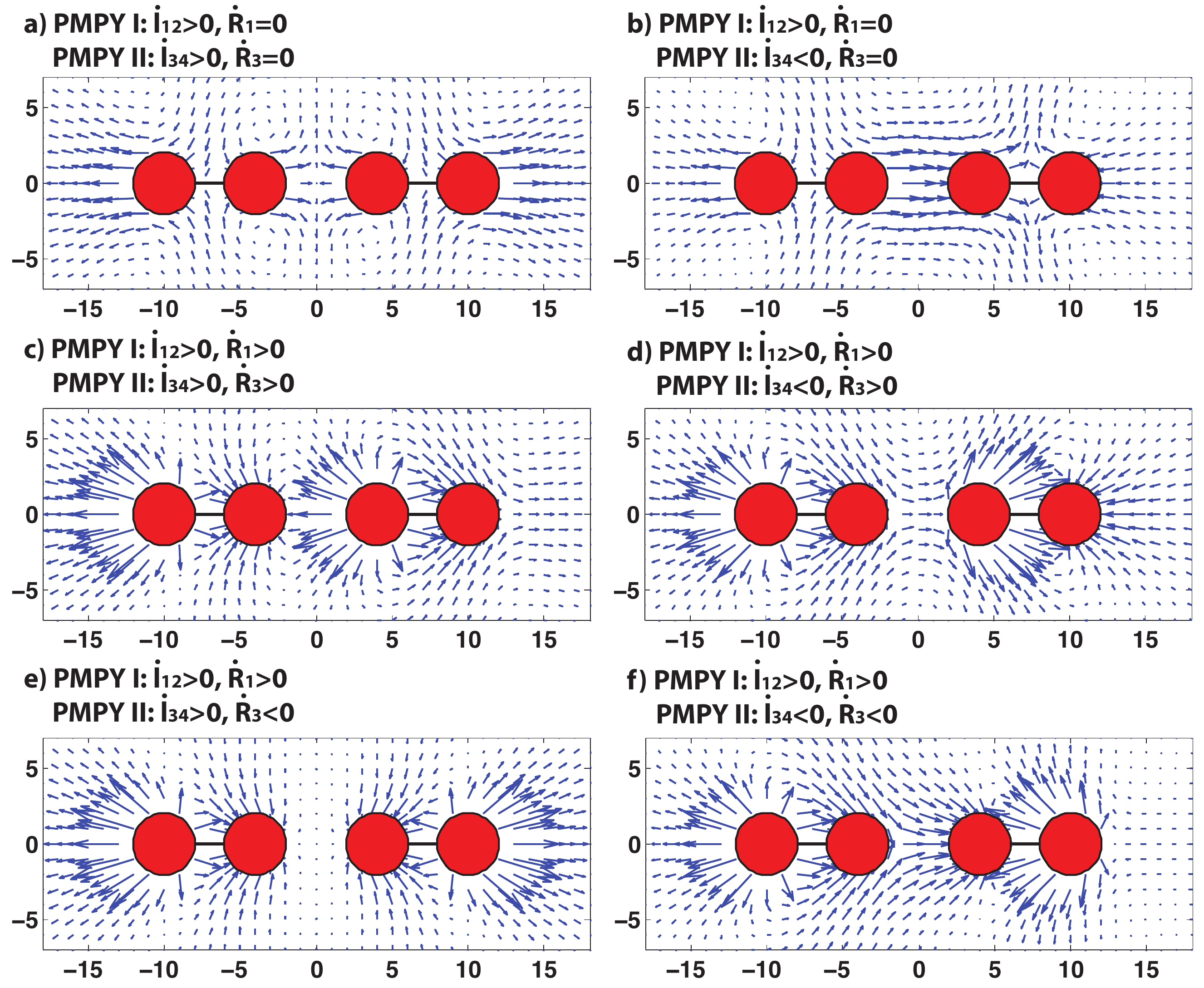}
\caption{The instantaneous fluid velocity field around two linear PMPYs, with
  $R_1 = R_2 = R_3 =R_4 =2$, $l_{12} = l_{34} = 6$, $l_{23} = 8$. ( a) Both
  swimmers undergo expansion and contraction of their connecting rods, but no
  volume changes. For I,  $\dot{l}_{12} =2 \pi, \ \dot{R}_1 = 0$, and for II \,
  $\dot{l}_{34} =2 \pi, \ \dot{R}_3 = 0$; (b) As in (a), but with $\dot{l}_{34} =
  - 2 \pi$); (c)  For   I,  $\dot{l}_{12} =2 \pi, $ and $ \dot{R}_1 = 2
  \pi$,  and for   II $\dot{l}_{34} =2 \pi , \ \dot{R}_3 = 2 \pi$; (d) As in (c), but with
  $\dot{l}_{34} =-2 \pi$; (e) For I,  $\dot{l}_{12} =2 \pi, \ \dot{R}_1 = 2
  \pi$, and for II $\dot{l}_{34} =2 \pi , \  \dot{R}_3 = -2 \pi$; (f) As in (e)
  but  with $\dot{l}_{34} =-2 \pi , \   \dot{R}_3 = -2 \pi$.  }
 \label{fig.Linear2PMPY_VF}
\end{figure}

\begin{figure}[htbp] \centering
\includegraphics[width=1\textwidth]{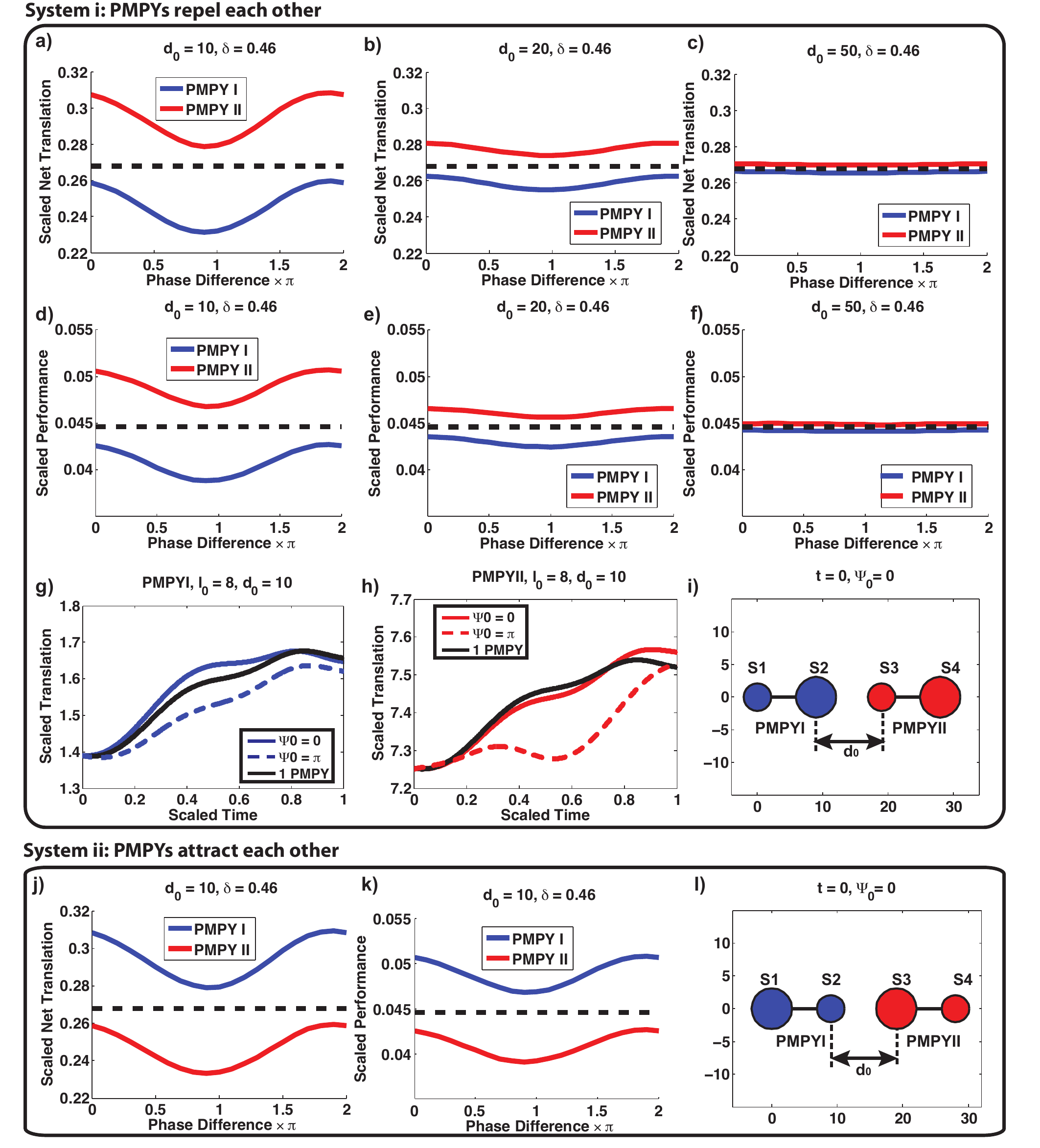}
\caption{Two PMPYs swim in a line. In panels (a-f) and (j,k) the dashed black
  line gives either the scaled net translation $\hat{X}$ or the scaled
  performance $\hat{\mathcal{P}}$ of a single PMPY undergoing the same sequence
  of shape changes. {\bf (a-i)} Simulation results for system i -- {\bf(j-k)} simulation
  results for system ii. (a-c,j) The relation between the scaled net
  translations of the two PMPYs ($\hat{X}_{I}$, PMPY I, blue line;
  $\hat{X}_{II}$, PMPY II, red line) and the phase difference $\Psi_0$ for $d_0
  = 10,20,50$ and $l_0 = 8$. (d-f,k) The relation between the scaled performance
  of the two PMPYs ($\hat{\mathcal{P}}_{I}$, PMPY I, blue line;
  $\hat{\mathcal{P}}_{II}$, PMPY II, red line) and the phase difference $\Psi_0$
  for $d_0 = 10,20,50$ and $l_0 = 8$.  (g) The scaled trajectory $\hat{X}_{I} (0
  \leq t \leq 1)$ of PMPY I with $l_0=8, d_0=10$ and phase difference $\Psi_0 =
  0$ (blue solid line) or $\pi$ (blue dashed line), comparing to the scaled
  trajectory of a single PMPY undergoing  the same sequence of shape deformations
  (black solid line).  (h) As in (g), but for PMPYII.  (i,l) The initial profile of the system with $l_0=8, d_0=10, \Psi_0
  =0$.  }
 \label{fig.Linear2PMPY_PhaseDiff_s}
\end{figure}

Figure \ref{fig.Linear2PMPY_VF} gives the instantaneous fluid velocity field
around a system of two linear PMPYs, where the instantaneous system profile is
$R_1 = R_2 = R_3 =R_4 =2$, $l_{12} = l_{34} = 6$, $l_{23} = 8$. Figure
\ref{fig.Linear2PMPY_VF}a \&b give the velocity fields when only the rod lengths
vary, while in Figure \ref{fig.Linear2PMPY_VF}c-f we include both rod and sphere
changes.  A comparison of the two cases shows that for a linear system in which
$\delta$ is not too small, the $O (\delta^2)$ terms that result from the sphere
changes give rise to a large perturbation of the surrounding fluid near the swimmers.

To obtain some insight into  the swimming behavior of a linear system of two PMPYs over a period, 
we consider  the following two systems.
\begin{itemize}
\item System i:
\begin{eqnarray*}
 & &\textrm{PMPY I:} \quad R_1 (t) = 2 - \sin 2 \pi t, \quad
R_2(0) = 3, \quad l_{12} = l_0 + \cos 2 \pi t \\ & &\textrm{PMPY II:} \quad R_3
(t) = 2 - \sin ( 2 \pi t + \psi_0 ), \quad l_{34} = l_0 + \cos ( 2 \pi t +
\psi_0 )  
\end{eqnarray*}
\item System ii:
\begin{eqnarray*}
 & &\textrm{PMPY I:} \quad R_1(0) = 3, \quad  R_2 (t) = 2 + \sin 2 \pi t, \quad
  l_{12} = l_0 + \cos 2 \pi t \\ & &\textrm{PMPY II:} \quad R_4
(t) = 2 + \sin ( 2 \pi t + \psi_0 ), \quad l_{34} = l_0 + \cos ( 2 \pi t +
\psi_0 ) , 
\end{eqnarray*}
 \end{itemize}
both with the following constraint and initial condition:
\begin{eqnarray*} & &\textrm{Equal volume conservation:} \qquad \dfrac{4 \pi}{3}
\big( R_1^3 + R_2^3 \big) = \dfrac{4 \pi}{3} \big( R_3^3 + R_4^3 \big) \equiv
\textrm{Const.} \\ & &\textrm{Initial distance between the two PMPY models:}
\qquad l_{23} (0) = d_0   
 \end{eqnarray*}
In either  the system, the two PMPYs undergo the same loop in the control
space $(\dot{l}, \dot{R}_1)$ except for a phase difference
$\psi_0$.

We solve the linear system equations~(\ref{eq.2PMPY_0Ref} -
\ref{eq.2PMPY_RigidMotion_U}), with equations (\ref{eq.2PMPY_1Ref_Omega34}, \ref{eq.2PMPY_1Ref_Omega12}) for
the angular motions removed due to the linear
geometry (Appendix~\ref{Appendix.E}). We take $l_0=8$, thus $ R_M \sim 3.24, \
l_m = 7, \ \delta \sim 0.46$.  Simulation results of the two systems for the
translation and performance are shown in
Figure~\ref{fig.Linear2PMPY_PhaseDiff_s}, and for comparison we show the results for a single swimmer undergoing
the same cyclic deformations (black dotted line in Figure~\ref{fig.Linear2PMPY_PhaseDiff_s}a-f\&jk, and black solid line
in Figure~\ref{fig.Linear2PMPY_PhaseDiff_s}gh). From these results  we can draw the following
conclusions. 

\begin{enumerate}

\item Figure~$\ref{fig.Linear2PMPY_PhaseDiff_s}$a-c show that in system i the
  one in front (PMPY II) gets pushed forward (red line), while the one that
  follows (PMPY I, blue line) gets pushed backward, and
  Figure~$\ref{fig.Linear2PMPY_PhaseDiff_s}$d-f show that the performance of the
  PMPY in front increased while that of the one that follows is decreased. On
  the other hand, in system ii we observe the reverse effect: the one in front
  gets pulled back and its performance decreased, while the one that follows
  gets pushed forward and its performance enhanced
  (Figure~\ref{fig.Linear2PMPY_PhaseDiff_s}j,k). Therefore we see that in system
  i the two PMPYs are repelling one another, while in system ii they are
  attracting each other.  In short, \textit{when two PMPYs that are identical
    except for a phase difference in their shape deformations swim in a line,
    they may repel or attract each other with a small amplitude, depending on
    their initial configuration and shape deformations.}

\item Regarding the phase difference, in the repelling system i, when $\psi_0 = 0$, PMPY II gets the
maximum increase while PMPY I gets the minimum decrease in both net translation
and performance; on the other hand when $\psi = \pi$, PMPY II gets the minimum
increase while PMPY I gets the maximum decrease in both net translation and
performance (Figure~\ref{fig.Linear2PMPY_PhaseDiff_s}a,d). In the attracting system ii
we observe the similar effect (Figure~\ref{fig.Linear2PMPY_PhaseDiff_s}j,k).
\end{enumerate}

\begin{figure}[htbp] \centering
\includegraphics[width=1\textwidth]{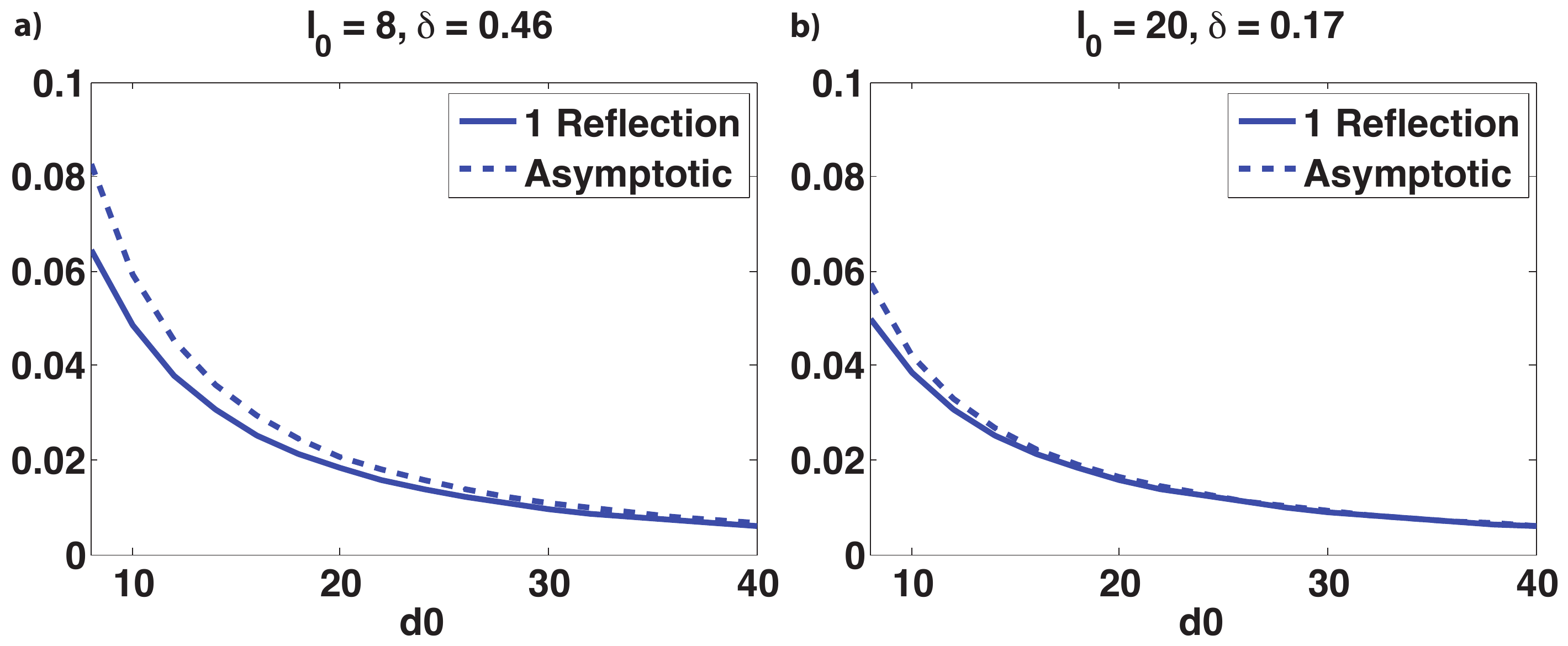}
\caption{$\hat{l}_{23} (1) - \hat{l}_{23} (0)$ for system i, with different values of $l_0 $ and $d_0$,
where the translation is either calculated up to the first reflection (solid lines),
or by asymptotic approximation to $O (\delta) $ (dashed lines).}
 \label{fig.Linear2PMPY_HydroEffect}
\end{figure}

Finally, we compute the scaled distance change $\hat{l}_{23} (1) - \hat{l}_{23}
(0)$ between the two PMPYs after one period for systems with different values of
$l_0 $ and $d_0$.  If there is no hydrodynamic interaction between them, due
either to length changes in the rod or volume changes in the spheres, then this
difference is identically zero. On the other hand, with hydrodynamic
interactions the difference will generally be non-zero, and will reflect the
strength of the interaction. This raises the question as to how accurately the
asymptotic approximation can capture this interaction.  In simulations to
examine this, we allow $l_{23} < L_m$ as long as all spheres are always kept
separated. Figure \ref{fig.Linear2PMPY_HydroEffect} shows the results for $l_0 =
8$ or $20$, and $d_0 \in [8,40]$, with the translation either calculated up to
the first reflection (equations (\ref{eq.2PMPY_0Ref} - \ref{eq.2PMPY_1Ref_Ue}),
Figure \ref{fig.Linear2PMPY_HydroEffect} solid lines), or by asymptotic
approximation to $O (\delta) $ (Figure \ref{fig.Linear2PMPY_HydroEffect} dashed
lines). From this we see that when $l_0$ and $d_0$ are both small, i.e., the
spheres are close, there is a clear difference in magnitude between the
asymptotic solution and the solution that results from one reflection, although
they are qualitatively similar. This and the foregoing results show that the
higher-order terms have a significant effect on both the instantaneous and
long-term translation and performance of the swimmers.

\subsection{Hydrodynamic interaction between two PMPY models swimming in a plane}

To conclude the analysis of interacting swimmers, we consider the scenario in
which the swimmers are swimming parallel to one another (Figure~\ref{fig.2PMPY_Geometry}).  To simplify
the computation, we approximate the equations $\ref{eq.2PMPY_0Ref}$ -
$\ref{eq.2PMPY_RigidMotion_U}$  to order $\delta^2$, hence for the rigid
motions we have the following approximations.
\begin{eqnarray}\label{eq.PlanePMPY_U_delta3} \hat{\mathbf{U}}_i &\sim& \dfrac{\hat{\mathbf{F}}_i }{ \hat{R}_i }
 + \sum_{j \neq i} \Big[ \dfrac{3}{4 \hat{l}_{ij}} \delta \Big(
\hat{\mathbf{F}}_j + \big( \hat{\mathbf{F}}_j \cdot \hat{\mathbf{d}}_{ji} \big) \hat{\mathbf{d}}_{ji}
\Big) + \hat{\zeta}_j \Big( \dfrac{\hat{R}_j}{\hat{l}_{ij}} \Big)^2 \delta^2 \mathbf{d}_{ji} \Big] +
O(\delta^3) \\
\label{eq.PlanePMPY_Omega_delta3} 
\bfO_i &\sim& O (\delta^3)
\end{eqnarray}
This allows us to neglect the angular motion of the PMPYs. Again, we see that
the hydrodynamic interactions contribute from the $O (\delta)$ term.  The
algorithm for the computational scheme is given in Appendix~$\ref{Appendix.E}$.
The complexity of the geometry makes it difficult to gain insight from an
asymptotic analysis, but the velocity field shown in Figure
\ref{fig.PMPY_VF_Uf-Ue} may help in understanding the behavior of the
hydrodynamic interactions between the two PMPYs. For example, when one of the
PMPYs is expanding its connecting rod and the other PMPY is right above or below
it, they will be attracted to each other. Also, when the PMPYs are short (i.e.,
small $l_I, \ l_{II}$) and close to each other, the $O (\delta^2)$ terms that
result from sphere expansion/contraction may have significant effects and have to be
incorporated (Figure \ref{fig.PMPY_VF_Uf-Ue} e \& f).

\begin{figure}[htbp] \centering
\includegraphics[width=1\textwidth]{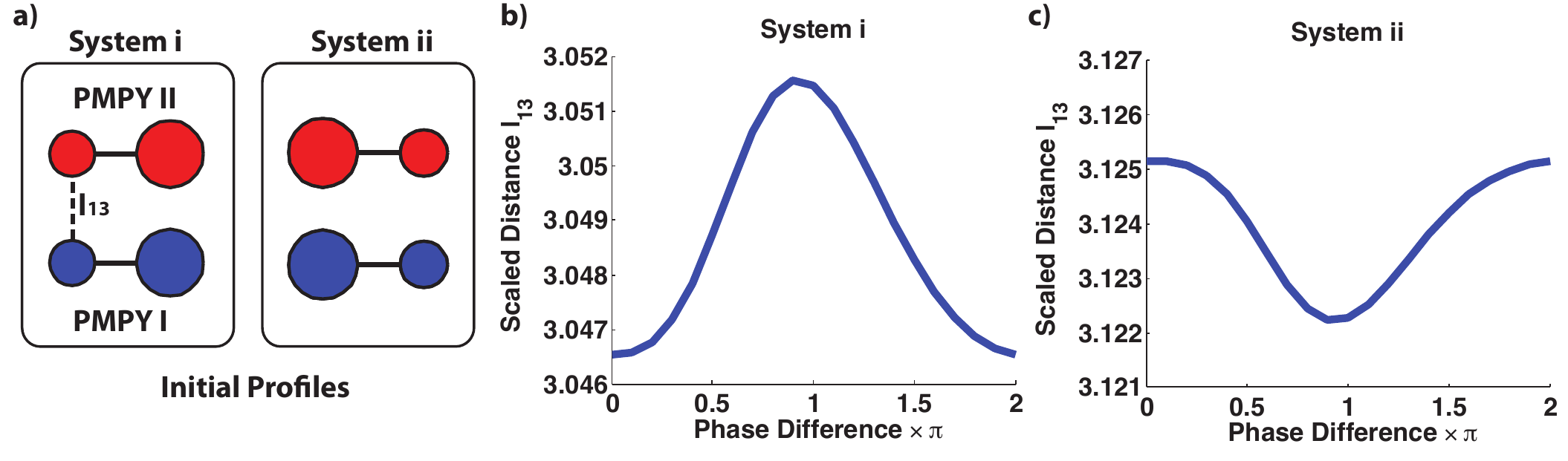}
\caption{Two PMPYs swim from a parallel initial position. (a) With $l_0 = 8, \  l_{13} (0) = 10, \  \psi_0 =
0$, the initial profiles of system i and ii.  (b) The relation  between the scaled distance
in the vertical direction between sphere 1 and 3 in system i after one cycle ($\hat{l}_{13} (1)$) and the
phase difference $\psi_0$. (c) As in (b) but for system ii.}
 \label{fig.Plane2PMPY_Parallel}
\end{figure}

We suppose that the two swimmers start from parallel initial positions -- both
lie along the $x$-direction, with sphere 3 located directly above sphere 1
(Figure~$\ref{fig.Plane2PMPY_Parallel}$a). Again we consider the two protocols
specified in section \ref{Sec.2PMPY_Line}, with the initial vertical distance
between sphere 1 and 3 $l_{13} (0) = 10$ and $l_0 = 8$. Thus the systems have
scales $R_M \sim 3.24, \ l_m =7$ and $\delta \sim 0.46$.  In system i, the
scaled distance between the sphere 1 and 3 decreases from $\hat{l}_{13} (0) \sim
3.086$ to $ \hat{l}_{13} (1) \sim 3.0491 \pm 0.003$
(Figure~\ref{fig.Plane2PMPY_Parallel}b). That is, the hydrodynamic interaction
between the swimmers leads to an attraction effect of scaled distance $\sim
0.037 $, which is of the same order as what was obtained for the linear system
in section \ref{Sec.2PMPY_Line}. The attraction effect is also slightly affected
by the phase difference $\psi_0$, as shown by
Figure~$\ref{fig.Plane2PMPY_Parallel}$b: when $\psi_0 = 0$ it reaches its
maximum while at $\psi_0=\pi $ it is at the minimum. On the other hand, in
system ii, within one cycle, the scaled distance between spheres 1 and 3
increases from $\hat{l}_{13} (0) = 3.086$ to $ \hat{l}_{13} \sim 3.124 \pm 0.002
$ (Figure~\ref{fig.Plane2PMPY_Parallel}c), indicating a small repulsion between
the two PMPYs resulting from the hydrodynamic interaction.  What is interesting
is that prior analysis shows that when laid collinear, the repelling system i
becomes attracting, while the attracting system ii then becomes repelling, when
subjected to the cyclic protocol in section \ref{Sec.2PMPY_Line}.


 \section{Extended scallop theorem and mixed controls}\label{Sec.ExtScallopThm}

 A widely-quoted principle in LRN swimming is that \textit{any reciprocal stroke
   gives no net motion}, which is known as the "scallop theorem"
 \citep{Purcell:1977:LLR}.  An immediate  corollary of this theorem is:
 if a self-propulsion swimmer has only one degree of freedom in its shape
 deformations, any cyclic stroke must be reciprocal and hence it cannot swim at
 LRN.  However, a group of reciprocal swimmers, none of which can swim in isolation,
 may coordinate their shape deformations such that the aggregate shape
 deformations of the group are not reciprocal. Thus, by  taking advantage of the
 hydrodynamic interactions, they may swim. This phenomenon is referred to as "no
 many scallop theorem" in the  existing literature \citep{koiller1996problems,
   lauga2008no, alexander2008dumb,lauga2011life}. We quote from
 \citep{lauga2011life}:
\smallskip
"Although a body undergoing reciprocal motion cannnot swim, two bodies undergoing reciprocal motion 
with nontrivial phase differences are able to take advantage of the unsteady hydrodynamic flows they
create to undergo nonzero collective and relative dynamics; there is thus no many-scallop theorem."
\smallskip

In previous sections each PMPY could swim by itself, but here we modify the PMPY
model into a scallop-type swimmer and study the collective behavior. In
particular, we focus on how well the collection can swim by taking advantage of
hydrodynamic interactions.  As we discussed earlier, a PMPY has two degrees of
freedom in its shape deformations, namely $\dot{l}$ and $\dot{R}_1$, and to make
it a scallop-type swimmer, we can disable either of them. Thus for two hobbled
PMPYs swimming together, there are three possibilities for their controls:
$(\dot{l}_I, \dot{l}_{II})$, $(\dot{R}_{I,1}, \dot{R}_{II,1})$, or a mixed
control $(\dot{l}_I, \dot{R}_{II,1})$. The first case, i.e., $(\dot{l}_I,
\dot{l}_{II})$, in which each PMPY is a simple dumb-bell with an extensible
connecting rod, has been studied previously \citep{lauga2008no,
  alexander2008dumb}, while the other combinations of controls have not.  Recall
that for a single active LRN swimmer, three linked-sphere models have been
designed according to these three different combinations of controls:
Najafi-Golestanian three-sphere model in $(\dot{l}_1, \dot{l}_2)$
\citep{Alexander:2009:HLS,Najafi:2004:SSL}, PMPY in $(\dot{l}, \dot{R}_1)$
\citep{Avron:2005:PEM} and the three-sphere volume-exchange model $(\dot{R}_1,
\dot{R}_3)$ \citep{Wang:2012:MLR}.  It was shown that a PMPY adopting the mixed
control is superior than the other two by order $O (L^2)$ in both net
translation and performance, where $L$ is the typical length of the swimmers
\citep{Wang:2015:PDM}. Here we ask the question: will a mixed control strategy
of deformations lead to better LRN swimming for two hobbled PMPYs, as it does  for
a single active swimmer?

 Again we consider two PMPYs lying along the $x$-axis, with PMPY II in front. We design three systems according to different
 types of controls.
 \begin{itemize}
 \item System A in $(\dot{l}_I, \dot{l}_{II})$ (two dumb-bells):
 $$ R_1 = R_2 = R_3 = R_4 = 3.24, \quad  l_{12} = l_0 + \cos (2 \pi t), \quad l_{34} = l_0 + \sin(2 \pi t)  $$
 \item System B in $(\dot{R}_{I,1}, \dot{R}_{II,1})$:
 $$ R_1(t) = 2 + \cos (2 \pi t), \quad R_2(0) = 2, \quad  R_3(t) = 2 + \sin (2 \pi t), \quad R_4(0) = 3\, \quad  l_{12} = l_{34} = l_0 -1  $$
 \item System C in $(\dot{l}_I, \dot{R}_{II,1})$:
 $$ R_1 = R_2 = 3, \quad R_3 (t) = 2 + \sin (2 \pi t) , \quad R_4 (0) = 3, \quad l_{12} (t) = l_0 + \cos (2 \pi t), \quad l_{34} = l_0 $$
 \item Initial condition: $l_{23} (0) = d_0$
 \end{itemize} 
 Therefore the three systems have the same scales: $R_M \sim 3.24, \ L_m = l_0 -
 1$.  We perform simulation with $l_0 = d_0 = 12$, which gives $\delta \sim
 0.29$, and the scaled net translations $\hat{X}_{I}, \hat{X}_{II}$ and scaled
 performances $\hat{\mathcal{P}}_{I}, \hat{\mathcal{P}}_{II}$ are given in
 Table~\ref{Tab.SimResults_ReviseScallop}.
   \begin{table}[htbp]
  \caption{\textbf{Simulation results of the three systems of two degraded PMPYs.}}
  \label{Tab.SimResults_ReviseScallop}
\begin{center}
\begin{tabular}{|c|c|c|c|c|}
\hline
  & $\hat{X}_{I}$ & $\hat{X}_{II}$ & $\hat{\mathcal{P}}_{I}$ & $ \hat{\mathcal{P}}_{II}$ \\
  \hline
  System A &$- 0.8 \times 10^{-3}$ &  $- 1.1 \times 10^{-3}$ & $5.25 \times 10^{-4}$ &  $6.94 \times 10^{-4}$ \\
  \hline
  System B &$ 3.3 \times 10^{-3}$ &  $ 4.0 \times 10^{-3}$ & $ 6.66 \times 10^{-4}$ &  $ 7.84  \times 10^{-4}$ \\
  \hline  
    System C &$ 0.5 \times 10^{-3}$ &  $ -17.5  \times 10^{-3}$ & $ 3.39 \times 10^{-4}$ &  $ 34 \times 10^{-4}$ \\
  \hline  
\end{tabular}
\end{center}
\end{table}
  
First, from Table~\ref{Tab.SimResults_ReviseScallop} we see that all three
systems can swim, although none is as effective as one single active PMPY, which
typically swims a scaled net translation of $\hat{X} \sim O(10^{-1})$ and with a
scaled performance of $ \hat{\mathcal{P}} \sim O (10^{-2})$ under this protocol
(see section \ref{Sec.CompSoln}). Next, we find that the swimming behaviors of
systems A and B, i.e., the two with the same type of shape deformations, are
quite similar: both PMPYs in both systems have scaled net translations and
scaled performances of the orders $\hat{X} \sim O (10^{-3})$ and $
\hat{\mathcal{P}} \sim O (10^{-4})$. On the other hand, system C, which adopts
the mixed controls, behaves very differently. The PMPY with shape deformations
in $\dot{l}$ (i.e., PMPY I) swims much less effectively than the one using an
$\dot{R}$ control (PMPY II) , with scaled net translations and scaled
performance ratios $\hat{X}_{II}/ \hat{X}_I \sim 35$ and
$\hat{\mathcal{P}}_{II}/ \hat{\mathcal{P}}_I \sim 10$.  Also, PMPY II in system
C clearly swims better than either of  the two PMPYs in system A or B,
while the poorer one (PMPY I) swims slightly worse than the two PMPYs in system A or B .
A detailed asymptotic analysis of the three systems is given in Appendix \ref{Appendix.F}. 

The reasons for this difference in outcome for different  control choices are: 1) $\dot{l}$
gives rise to  lower-order terms in  the translational velocity than does $\dot{R}$ (equation
\ref{eq.Scale_RPY_Ui}); 2) to guarantee that the lower-order terms resulting from
$\dot{l}$ do not vanish in the net translation, it is necessary that  the coefficients
 that depend on the radii be time-dependent, otherwise the temporal integral will become
an exact integral, or at best only result in higher order terms. To be more specific:
\begin{enumerate}
\item Controls in $(\dot{R}_{I,1}, \dot{R}_{II,1})$
will result in terms no lower than $O (\delta^2)$ in velocity and net translation.
\item Controls in $(\dot{l}_I, \dot{l}_{II})$ will result in leading order term
  of $O (1)$ or $O (\delta)$ in velocity, depending on the geometry of the
  model. However, their coefficients are either of the form $\Phi (R_{1,2,3,4})$
  or $\Phi (R_{1,2,3,4}) / l (t)$, where the radii are all constants. In the
  former case, it gives rise to an exact integral when computing the net
  translation; in the latter case, when integrated, it only gives higher order
  terms.  For details,  see the discussion in Appendix \ref{Appendix.F}.1.
\item Only with mixed controls in $(\dot{l}_I, \dot{R}_{II,1})$ and for the PMPY
  with shape deformation $\dot{R}$, the leading order in velocity is of $O
  (\delta)$ order and of the from $\Phi (R_{1,2,3,4} (t)) / l (t)$, which
  neither vanishes nor degrades when integrated.
\end{enumerate} 
 
Finally we summarize previous studies on the \textit{scallop theorem}
\citep{Purcell:1977:LLR,lauga2011life} together with our discussions from
section~$\ref{Sec.Wide-Obstacle},\ref{Sec.2PMPY}$ into the following
\textit{extended scallop principle}:
 \begin{Principle}[Extended Scallop Principle]
\label{ReviseScallopThm} In an LRN Newtonian flow,
\begin{enumerate}
\item a scallop cannot swim;
\item a living scallop surrounded by a  a few dead scallops cannot swim;
\item a group of living scallops can swim, but in an energy-inefficient manner.
\end{enumerate}
 \end{Principle} The precise mathematical interpretation of
Principle~$\ref{ReviseScallopThm}$ is as follows.  At low Reynolds number, 
\begin{enumerate}
\item a self-deformable swimmer with only one degree of freedom cannot swim --- which is the
statement of the original scallop theorem.
\item A self-deformable swimmer with only one degree of freedom cannot swim
  efficiently in  the presence of passive rigid objects.  This can be seen from
  section~$\ref{Sec.Wide-Obstacle}$, as the presence of a rigid object that
  cannot deform itself has no effect on the swimmer up to the  first reflection.
  When further  reflections
  are considered, the presence of the rigid objects nearby will indeed affect
  the swimmer, but the effect would be too small to have a significant effect on
  the swimmer.
\item A group of self-deformable swimmers, all with only one degree of freedom,
  can swim by taking advantage of hydrodynamic interactions.  However, both the
  translation and performance are much worse than one swimmer with multiple
  degrees of freedom.
\end{enumerate}
%


\section{Discussion}
 
Herein we used a basic PMPY model to study the LRN swimming characteristics of
both single and multiple swimmers, so as to understand the effect of
hydrodynamic interactions on the translation and efficiency of such swimmers.
One significant result is that the PMPY model is an efficient LRN swimmer whose
swimming behavior approximates that of swimming Dd amoebae. This suggests that
the PMPY model may provide a good first-order model for the study of
microorganisms swimming at LRN. As was shown, to better approximate the
characteristics of LRN swimming microorganisms one must allow the spheres to
approach more closely than in previous analyses, in which case the asymptotic
solution for $\delta =0$ is inadequate and higher-terms in the interactions must
be included. When a PMPY is swimming with a passive
object, the swimming PMPY has a clear effect on the passive object, while the
existence of the latter has little effect on the PMPY, as long as the size of
this object is comparable to or less than the spheres in the PMPY. If the
passive object is directly ahead of the PMPY, the PMPY can catch up with it
within a few cycles, using a reasonable amount of energy. If the freely buoyant
object is not directly in the PMPY's path, its long-term trajectory approximates
a closed-triangle when it is far away from the PMPY's swimming path, or it will
be pushed forward if it is close to the PMPY's swimming path. In this case the
higher-order terms in the solution of the translational velocity of the PMPY
contribute significantly to this entrainment effect, and again, an asymptotic
solution does not capture this effect. Moreover, a longer PMPY will enhance the
entrainment effect.  

When there are two PMPYs swimming together, either collinearly or not, the
hydrodynamic interactions among them may cause some attraction or repulsion
between them, depending on the stroke.  However, this effect is small compared
with the net translation of the PMPYs, but again, higher-order terms should be
taken into consideration, particularly if one is interested in either the
instantaneous or long-term behavior of the system. A scallop-type swimmer cannot swim at LRN on its own or in the
presence of non-deformable bouyant objects, but by cleverly coordinating their
strokes, two or more of them can swim by taking advantage of hydrodynamic
interactions. Although swimming in this manner is generally not efficient, a
pair of scallop-type swimmers that use a mixed $(\dot{l},\dot{R})$ control may
enhance the swimming of one of the pair as compared to pairs in which both use
either a pure $(\dot{l})$ or $(\dot{R})$ control.

In the biological context the surrounding material may be non-Newtonian, and in
particular, is often viscoelastic. Some results for this case are known,
\citep{Qiu:2014:SRM,Curtis:2013:TSS}, but much remains to be done.

 
 \vspace{0.5in}
 
\appendix 
\addcontentsline{toc}{section}{Appendices}
 \noindent{\bf \normalsize Appendices} \setcounter{equation}{0}
\renewcommand{\theequation}{\ref{Appendix.A}.\arabic{equation}}

 \section{Newtonian flow produced by the translation and radial expansion 
of a sphere} 
\label{Appendix.A}

Here we derive the velocity field $\mathbf{u} (\mathbf{x})$ of a
LRN  flow produced by a sphere of radius $R $, pulled by a force
$\mathbf{F}$ and  expanding radially at a rate $\dot{R}  = \textrm{d} R /
\textrm{d} t$.  Due to the linearity of LRN flows, the velocity field is the sum of two
terms: that  resulting from the drag force   --- $\mathbf{u} \{ \mathbf{F} \}$, 
and that  resulting from the  radial
expansion  --- $\mathbf{u} \{ \dot{R} \}$ :
\begin{eqnarray}\label{VelocityEqn_1} \mathbf{u} =
\mathbf{u} \{ \mathbf{F} \} + \mathbf{u} \{ \dot{R} \}
\end{eqnarray}

The flow produced by the translation of a solid sphere is a classical result 
 \citep{Pozrikidis:1992:BIS}.  It can be represented in terms of a
Stokeslet and dipole with poles at the center of the sphere that are given by 
\begin{eqnarray}
\label{VelocityEqn_3} 
\mathbf{G} &=& \dfrac{\mathbf{ \delta}}{r} +
\dfrac{\mathbf{r}\mathbf{r}}{r^3} \\ 
\label{VelocityEqn_4} 
{\mathbf D} &=& -\dfrac{\mathbf{ \delta}}{ r^3} + 3\dfrac{\mathbf{r}\mathbf{r}}{r^5}
\end{eqnarray}
where $\mathbf{x}_0$ is the center of the sphere, $\mathbf{r} = \mathbf{x} - \mathbf{x}_0$ and $r = | \mathbf{r}
|$. $\mathbf G$ is called the Oseen tensor. The velocity field  is then 
\begin{eqnarray}\label{VelocityEqn_2} u_i (\mathbf{x}) = G_{ij} (\mathbf{x},
\mathbf{x}_0) \Big( \dfrac{3}{4} R U_j \Big) - D_{ij} (\mathbf{x}, \mathbf{x}_0)
\Big( \dfrac{1}{4} R^3 U_j \Big).
\end{eqnarray}
Here $\mathbf{U}$ is the
translational  velocity of the sphere, i.e., $\mathbf{U} (t) = \dot{\mathbf{x}}_0
(t)$. %
The
 relation between the drag force $\mathbf{F}$ and $\mathbf{U}$ is
\begin{eqnarray}\label{VelocityEqn_5} \mathbf{F} = 6 \pi \mu R \mathbf{U}
\end{eqnarray}
Using the above one obtains the fluid velocity 
\begin{eqnarray*} u_i &=& \dfrac{1}{24 \pi \mu} \Big[ 3 \Big(
\dfrac{\delta_{ij}}{r} + \dfrac{r_i r_j}{r^3} \Big) F_j - \Big( -
\dfrac{\delta_{ij}}{r^3} + 3 \dfrac{r_i r_j}{r^5} \Big) R^2 F_j \Big] \\ &=&
\dfrac{1}{24 \pi \mu} \Big[ 3 \Big( \dfrac{F_i}{r} + \dfrac{(\mathbf{F} \cdot
\mathbf{r}) r_i}{r^3} \Big) + \Big( \dfrac{F_i}{r^3} - 3 \dfrac{(\mathbf{F}
\cdot \mathbf{r}) r_i}{r^5} \Big) R^2\Big] \\ &=& \dfrac{1}{24 \pi \mu r} \Big[
\Big( 3 + \dfrac{R^2}{r^2} \Big) F_i + 3 (\dfrac{1}{r^2} - \dfrac{R^2}{r^4})
(\mathbf{F} \cdot \mathbf{r}) r_i \Big]
\end{eqnarray*}
or alternatively 
\begin{eqnarray}\label{VelocityEqn_6} \mathbf{u} \{ \mathbf{F} \}  =
\dfrac{1}{24 \pi \mu r} \Big[ \Big( 3 + \dfrac{R^2}{r^2} \Big) \mathbf{F} + 3 \Big(1 -
\dfrac{R^2}{r^2} \Big) (\mathbf{F} \cdot \hat{\mathbf{r}}) \hat{\mathbf{r}} \Big]
\end{eqnarray}
where $\hat{\mathbf{r}} = \mathbf{r} / |\mathbf{r}|$.

The velocity field $\mathbf{u} \{ \dot{R} \}$ of the flow
 produced by a radially expanding sphere can  be represented
 by a point source with pole at the center $\mathbf{x}_0$ of the sphere, 
 and thus has the form 
\begin{eqnarray}\label{VelocityEqn_7} \mathbf{u} = \alpha  \dfrac{
\hat{\mathbf{r}}}{r^2}
\end{eqnarray}
where the strength of the source ($\alpha$)  is a constant to be determined.  The no-slip boundary
condition on the sphere surface implies that 
\begin{eqnarray}\label{VelocityEqn_8} \mathbf{u} (\mathbf{x}) =
\dfrac{\textrm{d} R}{\textrm{d} t} \hat{\mathbf{r}}.
\end{eqnarray}
From ~($\ref{VelocityEqn_7}, \ref{VelocityEqn_8}$) we find that  $\alpha
= \dot{R} R^2$, and therefore 
\begin{eqnarray}\label{VelocityEqn_9} \mathbf{u} \{ \dot{R} \} = \dot{R}
\Big( \dfrac{R}{r} \Big)^2 \hat{\mathbf{r}}.
\end{eqnarray}
If we represent it in terms of the sphere volume $v = 4 \pi R^3 /3$ this reads 
\begin{eqnarray}\label{VelocityEqn_10} \mathbf{u} \{ \dot{R} \} =
\dfrac{\dot{v}}{4 \pi r^2} \hat{\mathbf{r}}
\end{eqnarray}

This leads to the combined flow n~($\ref{eq.VelocityField_SingleSphere}$):
\begin{eqnarray*} \mathbf{u} \big( \mathbf{r}; R, \mathbf{F}, \dot{R} \big) =
\dfrac{1}{24 \pi \mu r} \Big[ \big( 3 + \dfrac{R^2}{r^2} \big) \mathbf{F} + 3
\big( 1 - \dfrac{R^2}{r^2} \big) \big( \mathbf{F} \cdot \widehat{ \mathbf{r} }
\big) \widehat{\mathbf{r}} \Big] +\dot{R} \Big( \dfrac{R}{r} \Big)^2
\widehat{\mathbf{r}}.
\end{eqnarray*}
%


\section{Accurary of $U_i^{(e)}$ after incorporating the second
reflection}
\label{Appendix.C} 
\setcounter{equation}{0}
\renewcommand{\theequation}{\ref{Appendix.C}.\arabic{equation}}

Next we determine  how a second reflection l contributes to the rigid motions of two
radially expanding spheres.

\textbf{$0$th Reflection.} We consider two spheres with radii $R_1 (t)$ and $R_2
(t)$ resp., centered at $\mathbf{x}_1 (t)$ and $\mathbf{x}_2 (t)$ so that
$\mathbf{x}_2 - \mathbf{x}_1$ points in the  $\mathbf{e}_x$ direction. The radial
expansion of the $i$th sphere alone generates a flow
\begin{eqnarray}\label{eq.Ref0_VelocityField} \mathbf{u}_i^{(0)} ( \mathbf{x}) = R_i^2
\dot{R}_i \dfrac{\mathbf{x} - \mathbf{x}_i}{ | \mathbf{x} - \mathbf{x}_i |^3}
\end{eqnarray}
and the rigid motion of  the sphere vanishes
\begin{eqnarray}\label{eq.Ref0_RigidMotion} \mathbf{U}_i^{(0)} = \mathbf{0},
\qquad \mathbf{\omega}_i^{(0)} = \mathbf{0}.
\end{eqnarray}

\textbf{$1$st Reflection.} Next we put sphere 1 into the flow $\mathbf{u}_2$
generated by the radially expanding sphere 2.  We calculate the resulting
{\it translational velocity} $\mathbf{U}_1^{(1)}$, {\it angular velocity} 
$\mathbf{\omega}_1^{(1)}$, and the {\it stresslet} $\mathbf{S}_1^{(1)}$, from which we
obtain the velocity field $\mathbf{u}_{21} (\mathbf{x})$.

\textit{Translational velocity} $\mathbf{U}_1^{(1)}$.
\begin{eqnarray*} \mathbf{U}_1^{(1)} = \Big( 1 + \dfrac{R_1^2}{6} \nabla^2
\Big) \mathbf{u}_2^{(0)} \Big|_{\mathbf{x} = \mathbf{x}_1} = R_2^2 \dot{R}_2
\Big[ \dfrac{\mathbf{x}_1 - \mathbf{x}_2}{ | \mathbf{x}_1 - \mathbf{x}_2 |^3} +
\dfrac{R_1^2}{6} \nabla^2 \Big( \dfrac{\mathbf{x} - \mathbf{x}_2}{ | \mathbf{x}
- \mathbf{x}_2 |^3} \Big) \Big|_{\mathbf{x} = \mathbf{x}_1} \Big] = R_2^2
\dot{R}_2 \Big[ - \dfrac{ \mathbf{e}_z }{ l^2} + \dfrac{R_1^2}{6} \nabla^2 \Big(
\dfrac{\mathbf{x} - \mathbf{x}_2}{ | \mathbf{x} - \mathbf{x}_2 |^3} \Big)
\Big|_{\mathbf{x} = \mathbf{x}_1} \Big]
\end{eqnarray*}
Letting $\mathbf{r} = \mathbf{x} - \mathbf{x}_2 $ and $r = |\mathbf{r}|$, we
must calculate
\begin{eqnarray*} \nabla^2 \Big( \dfrac{\mathbf{r}}{r^3} \Big)
= \partial_k \partial_k \dfrac{r_i}{r^3}
\end{eqnarray*}
Since 
\begin{eqnarray*}
\partial_k \Big( \dfrac{1}{r^n} \Big) = - \dfrac{n r_k}{r^{n+2}}
\end{eqnarray*}
we have
\begin{eqnarray*}
\partial_k \Big( \dfrac{r_i}{r^3} \Big) &=& \dfrac{\delta_{ik}}{r^3} - \dfrac{3
r_i r_k}{r^5} \\
\partial_k \partial_k \Big( \dfrac{r_i}{r^3} \Big) &=& - \dfrac{3 \delta_{ik}
r_k}{r^5} - \dfrac{3 \delta_{ik} r_k}{r^5} - \dfrac{3 r_i \partial_k y_k}{r^5} +
\dfrac{15 r_i r_k r_k}{r^7} = 0.
\end{eqnarray*}
Thus
\begin{eqnarray*} \nabla^2 \Big( \dfrac{\mathbf{r}}{r^3} \Big) = \mathbf{0}
\end{eqnarray*}
and
\begin{eqnarray}\label{eq.Ref1_U} \mathbf{U}_1^{(1)} = - \dfrac{R_2^2
\dot{R}_2}{l^2} \mathbf{e}_x
\end{eqnarray}

\textit{Angular velocity} $\mathbf{\omega}_1^{(1)}$.
\begin{eqnarray}\label{eq.Ref1_Omega} \mathbf{\omega}_1^{(1)} = \dfrac{1}{2}
\nabla \times \mathbf{u}_2^{(0)} \Big|_{\mathbf{x} = \mathbf{x}_1} = \mathbf{0}
\end{eqnarray}

\textit{Stresslet} $\mathbf{S}_1^{(1)}$. The rate of deformation that results  from
$\mathbf{u}_2$ is
\begin{eqnarray*} \mathbf{E}_2^{(0)} = \dfrac{1}{2} \Big[ \nabla \mathbf{u}_2^{(0)} +
\big( \nabla \mathbf{u}_2^{(0)} \big)^T \Big] - \dfrac{1}{3} \mathbf{I} \Big( \nabla
\cdot \mathbf{u}_2^{(0)} \Big) = \dot{R}_2 R_2^2 \Big[ \dfrac{1}{2} \Big( \nabla
\dfrac{\hat{\mathbf{r}}}{r^2} + \big( \nabla \dfrac{\hat{\mathbf{r}}}{r^2}
\big)^T \Big) - \dfrac{1}{3} \mathbf{I} \Big( \nabla \cdot
\dfrac{\hat{\mathbf{r}}}{r^2} \Big) \Big]
\end{eqnarray*}
where $\hat{\mathbf{r}} = \mathbf{r}/r$. Using spherical coordinates, we have
\begin{eqnarray*} \nabla \dfrac{\hat{\mathbf{r}}}{r^2} &=& \Big(
\hat{\mathbf{r}} \dfrac{\partial }{\partial r} + \dfrac{1}{r}
\hat{\mathbf{\theta}} \dfrac{\partial }{\partial \theta} + \dfrac{1}{r \sin
\theta} \hat{\mathbf{\phi}} \dfrac{\partial }{\partial \phi} \Big) \Big(
\dfrac{\hat{\mathbf{r}}}{r^2} \Big) = \Big( \dfrac{\partial }{\partial r}
\dfrac{1}{r^2} \Big) \hat{\mathbf{r}} \hat{\mathbf{r}} = - \dfrac{2}{r^3}
\hat{\mathbf{r}} \hat{\mathbf{r}} \\ \nabla \cdot \dfrac{\hat{\mathbf{r}}}{r^2}
&=& \dfrac{\partial }{\partial r} \dfrac{1}{r^2} + \dfrac{2}{r} \dfrac{1}{r^2} =
0.
\end{eqnarray*}
Thus
\begin{eqnarray}\label{eq.Ref1_RateDef} \mathbf{E}_2^{(0)} = - \dfrac{2 R_2^2
\dot{R}_2}{r^3} \hat{\mathbf{r}} \hat{\mathbf{r}} = - \dfrac{2 R_2^2
\dot{R}_2}{r^5} \mathbf{r} \mathbf{r},
\end{eqnarray}
and the stresslet is given by  
\begin{eqnarray*} \mathbf{S}_1^{(1)} = \dfrac{20}{3} \pi \mu R_1^3 \Big( 1 +
\dfrac{R_1^2}{10} \nabla^2 \Big) \mathbf{E}_2^{(0)} \Big|_{\mathbf{x} = \mathbf{x}_1}
= - \dfrac{40}{3} \pi \mu R_1^3 R_2^2 \dot{R}_2 \Big[ \dfrac{\mathbf{e}_x
\mathbf{e}_x}{l^3} + \dfrac{R_1^2}{10} \nabla^2 \Big( \dfrac{\mathbf{r}
\mathbf{r}}{r^5} \Big) \Big|_{\mathbf{x} = \mathbf{x}_1} \Big]
\end{eqnarray*}
We need to calculate
\begin{eqnarray*} \nabla^2 \Big( \dfrac{\mathbf{r} \mathbf{r}}{r^5} \Big)
= \partial_k \partial_k \Big( \dfrac{ r_i r_j }{r^5} \Big)
\end{eqnarray*}
\begin{eqnarray*}
 \partial_k \Big( \dfrac{ r_i r_j }{r^5} \Big) &=& \dfrac{\delta_{ik} r_j}{r^5}
+ \dfrac{\delta_{jk} r_i}{r^5} - \dfrac{5 r_i r_j r_k}{r^7} \\
 \partial_k \partial_k \Big( \dfrac{ r_i r_j }{r^5} \Big) &=& \Big(
\dfrac{\delta_{ik} \delta_{jk}}{r^5} - \dfrac{5 \delta_{ik} r_j r_k}{r^7} \Big)
+ \Big( \dfrac{\delta_{jk} \delta_{ik}}{r^5} - \dfrac{5 \delta_{jk} r_i
r_k}{r^7} \Big)  + \Big( - \dfrac{5 \delta_{ik} r_j r_k}{r^7} - \dfrac{5
\delta_{jk} r_i r_k}{r^7} - \dfrac{5 r_i r_j \partial_k r_k}{r^7} + \dfrac{35
r_i r_j r_k r_k}{r^9} \Big) = \dfrac{2 \delta_{ij}}{r^5}
\end{eqnarray*}
Thus
\begin{eqnarray}\label{eq.Ref1_S} \mathbf{S}_1^{(1)} = - \dfrac{40}{3} \pi \mu
R_1^3 R_2^2 \dot{R}_2 \Big[ \dfrac{\mathbf{e}_x \mathbf{e}_x}{l^3} +
\dfrac{R_1^2}{5 l^5} \mathbf{I} \Big]
\end{eqnarray}

Finally, the velocity field $\mathbf{u}_{21}^{(1)} (\mathbf{x})$ is given by
\begin{eqnarray}\label{eq.Ref1_VelocityField} \mathbf{u}_{21} (\mathbf{x}) =
\Big( \mathbf{S}_1^{(1)} \cdot \nabla \Big) \cdot \dfrac{\mathbf{G}(\mathbf{x}
- \mathbf{x_1})}{8 \pi \mu} + \cdots
\end{eqnarray}
From equations~($\ref{eq.Ref1_S}, \ref{VelocityEqn_3}$) we have that near
$\mathbf{x}_1$,
\begin{eqnarray*} \mathbf{S}_1^{(1)} \sim O \Big( \dfrac{1}{l^3} \Big), \qquad \mathcal{G} \sim
O \Big( \dfrac{1}{l} \Big)
\end{eqnarray*}
and thus the velocity near $\mathbf{x}_1$ scales as $\mathbf{u}_{21}^{(1)} \sim O
(l^{-5})$.

\textbf{2nd Reflection.} Now we reflect once again and consider sphere 1
immersed in the flow $\mathbf{u}_{12}^{(1)}$.  The translational velocity
$\mathbf{U}_1^{(2)}$ that results  is given by
\begin{eqnarray}
\label{eq.Ref2_U} 
\mathbf{U}_1^{(2)} = \Big( 1 +
\dfrac{R_1^2}{6} \nabla^2 \Big) \mathbf{u}_{12}^{(1)} \Big|_{\mathbf{x} =
\mathbf{x}_1}
\end{eqnarray}
From the discussion of the first  reflection we know that near $\mathbf{x}_1$ the
velocity field $\mathbf{u}_{12}^{(1)}$ is $O (l^{-5})$, and thus
$\mathbf{U}_1^{(2)} \sim O (l^{-5})$ as well.


\section{Calculation of $ \bfO_{3,i}^{(1)}$}\label{Appendix.D}
\setcounter{equation}{0}
\renewcommand{\theequation}{\ref{Appendix.D}.\arabic{equation}} 
The angular
velocity of sphere 3 (i.e., the free-floating sphere) $ \bfO_{3,i}^{(1)}$
consists of two parts: $ \bfO_{3,i}^{(1,f)}$ that results from the drag force on
sphere $i$, and $ \bfO_{3,i}^{(1,e)}$ that results from the expansion of sphere
$i$. Here we calculate them separately.

First, $ \bfO_{3,i}^{(1,e)}$, which  results from the flow generated by the
expansion of sphere $i$ is:
\begin{eqnarray*} \mathbf{u}_i^{(0,e)} = R_i^2 \dot{R}_i \dfrac{\mathbf{x} -
\mathbf{x}_i}{|\mathbf{x} - \mathbf{x}_i|^3},
\end{eqnarray*}
and therefore 
\begin{eqnarray}\label{AppD_Omega1e} \bfO_{3,i}^{(1,e)} = \dfrac{1}{2} \nabla
\times \mathbf{u}_i^{(e)} \Big|_{\mathbf{x} = \mathbf{x}_3} = \dfrac{ R_i^2
\dot{R}_i }{2} \nabla \times \Big( \dfrac{\mathbf{x} - \mathbf{x}_i}{|\mathbf{x}
- \mathbf{x}_i|^3} \Big) \Big|_{\mathbf{x} = \mathbf{x}_3}.
\end{eqnarray}
Let $\mathbf{r} = \mathbf{x} - \mathbf{x}_i$ and $r = |\mathbf{x} -
\mathbf{x}_i| $ -- then the antisymmetric part in equation~$\ref{AppD_Omega1e}$ is
\begin{eqnarray*}
\partial_j \Big( \dfrac{r_k}{r^3} \Big) - \partial_k \Big( \dfrac{r_j}{r^3}
\Big) = \Big( \dfrac{\delta_{jk}}{r^3} - \dfrac{3 r_j r_k}{r^5} \Big) - \Big(
\dfrac{\delta_{kj}}{r^3} - \dfrac{3 r_k r_j}{r^5} \Big) = 0
\end{eqnarray*}
and therefore  $ \bfO_{3,i}^{(1,e)} = 0$.

Next, $ \bfO_{3,i}^{(1,f)}$, which  results from the flow generated by the drag
force $\mathbf{F}_i = F_i \mathbf{e}_x$ on sphere $i$, is given by 
\begin{eqnarray*} \bfO_{3,i}^{(1,f)} &=& \dfrac{1}{2} \nabla \times \mathbf{u}
\{ \mathbf{F}_i \} \Big|_{\mathbf{x} = \mathbf{x}_3} \\ &=& \dfrac{1}{16 \pi
\mu} \nabla \times \Big[ \Big( \dfrac{1}{r} + \dfrac{1}{3 r^3} \Big)F_i
\mathbf{e}_x + \Big( \dfrac{1}{r} - \dfrac{1}{ r^3} \Big) (F_i \mathbf{e}_x
\cdot \dfrac{\mathbf{r}}{r}) \dfrac{\mathbf{r}}{r} \Big] \Big|_{\mathbf{x} =
\mathbf{x}_3} \\ &=& \dfrac{F_i}{16 \pi \mu} \nabla \times \Big[ \Big(
\dfrac{1}{r} + \dfrac{1}{3 r^3} \Big) \mathbf{e}_x + \Big( \dfrac{1}{r} -
\dfrac{1}{ r^3} \Big) \dfrac{r_x \mathbf{r}}{r^2} \Big].
\end{eqnarray*}
Since
\begin{eqnarray*} \nabla \times \Big( \dfrac{1}{r^n} \mathbf{e}_x \Big) &=& -
\dfrac{n r_z}{r^{n+2}} \mathbf{e}_y + \dfrac{n r_y}{r^{n+2}} \mathbf{e}_z \\
\nabla \times \Big(\dfrac{r_x}{r^n} \mathbf{r} \Big) &=& - \dfrac{r_z}{r^n}
\mathbf{e}_y + \dfrac{r_y}{r^n} \mathbf{e}_z
\end{eqnarray*}
we have
\begin{eqnarray} \label{AppD_Omega1f} \bfO_{3,i}^{(1,f)} = \dfrac{F_i}{8 \pi \mu
r^3} \Big[ - r_z \mathbf{e}_y + r_y \mathbf{e}_z \Big] \Big|_{\mathbf{r} =
\mathbf{x}_3 - \mathbf{x}_i} = \dfrac{F_i}{8 \pi \mu l_{i3}^3} \big(
\mathbf{x}_3 - \mathbf{x}_i \big)_y \mathbf{e}_z
\end{eqnarray}
and therefore
\begin{eqnarray}\label{AppD_Omega1} \bfO_3^{(1)} = \sum_{i=1,2}
\bfO_{3,i}^{(1,f)} + \bfO_{3,i}^{(1,e)} = \sum_{i=1,2} \dfrac{F_i}{8 \pi \mu
l_{i3}^3} \big( \mathbf{x}_3 - \mathbf{x}_i \big)_y \mathbf{e}_z.
\end{eqnarray}
%


\section{Numerical scheme of two PMPY models}\label{Appendix.E}
\setcounter{equation}{0}
\renewcommand{\theequation}{\ref{Appendix.E}.\arabic{equation}}

The following numerical scheme is written in terms of unscaled variables. 

When the two PMPY models both lie along the $x$-axis, and the spheres are numbered
1, 2, 3, 4 from the negative to the positive $x$-direction, the
equations~($\ref{eq.2PMPY_0Ref}$ - $\ref{eq.2PMPY_RigidMotion_U}$) reduce to
\begin{eqnarray} 
\label{UFsys}
\left(
\begin{array}{ll} A_{11} & A_{12} \\ A_{21} & A_{22}
\end{array}
\right) \left(
\begin{array}{l} \mathbf{U} \\ \mathbf{F}
\end{array}
\right) = \left(
\begin{array}{l} \mathbf{B}_1 \\ \mathbf{B}_2
\end{array}
\right)
\end{eqnarray} 
where
\begin{eqnarray*} \mathbf{U} = \left(
\begin{array}{l} U_1\\ U_2\\ U_3\\ U_4
\end{array}
\right), \qquad \mathbf{F} = \mu^{-1} \left(
\begin{array}{l} F_1\\ F_2\\ F_3\\ F_4
\end{array}
\right). 
\end{eqnarray*} 
Here $A_{11} = - I_4$, where $I_4$ is the $4 \times 4$ identity
matrix,  $A_{12} = (a_{ij})$ is a symmetric $4 \times 4$ matrix, with
\begin{eqnarray*} a_{ii} = \dfrac{1}{6 \pi R_i} , \qquad a_{ij} = \dfrac{1}{4
\pi l_{ij}} \Big( 1 - \dfrac{R_i^2 + R_j^2}{3 l_{ij}^2} \Big)
\end{eqnarray*}
and
\begin{eqnarray*} \big( A_{21} \ A_{22} \big) = \left(
\begin{array}{rrrrrrrr} 1 & -1 & 0 & 0& 0 & 0& 0& 0 \\ 0& 0& 1 & -1 & 0 & 0& 0&
0 \\ 0 & 0& 0& 0 & 1 & 1 & 0 & 0 \\ 0 & 0& 0& 0 & 0 & 0& 1 & 1
\end{array}
\right).
\end{eqnarray*} 
$\mathbf{B}_1 \in \mathbb{R}^4$, is given by 
\begin{eqnarray*} (\mathbf{B}_1)_i = - \sum_{j \neq i} \textrm{sign} (i-j)
\dot{R}_j \Big( \dfrac{R_j}{l_{ij}} \Big)^2
\end{eqnarray*}
and
\begin{eqnarray*} \mathbf{B}_2 = \left(
\begin{array}{l} - \dot{l}_{12} \\ - \dot{l}_{34} \\ 0\\ 0
\end{array}
\right).
\end{eqnarray*}

On the other hand, when the two PMPY models swim in a plane, we approximate the scaled
system to $O (\delta^2)$, hence the rigid motions of the spheres can be
approximated by equations~($\ref{eq.PlanePMPY_U_delta3},
\ref{eq.PlanePMPY_Omega_delta3}$). Returning  to the unscaled equations, to further simplify the system, from
equations~($\ref{eq.2PMPY_ForceFree}$) we know that $\mathbf{F}_{1},
\mathbf{F}_2$ are directed along  $\mathbf{e}_x$  and $\mathbf{F}_{3},
\mathbf{F}_4$ are directed along  $\mathbf{e}_{x'}$.  Moreover,
\begin{eqnarray*} F_2 = - F_1, \qquad F_4 = -F_3
\end{eqnarray*}
Hence the equation system ($\ref{eq.2PMPY_0Ref}$ -
$\ref{eq.2PMPY_RigidMotion_U}$) can be reduced to a system with unknowns
$$U_{1x},  U_{1y}, U_{2x},  U_{2y}, U_{3x'},  U_{3y'}, U_{3x'},  U_{3y'}, F_1, F_3$$
where $U_{1x} = \mathbf{U}_1 \cdot \mathbf{e}_x$, $U_{3x'} = \mathbf{U}_3 \cdot
\mathbf{e}_{x'}$, and so on.  Taking into account the fact that $\mathbf{e}_x =
\mathbf{d}_{12}$ and $\mathbf{e}_{x'} = \mathbf{d}_{34}$, the nondimensional
form of (\ref{UFsys})  can be simplified to
\begin{eqnarray*} &-U_{1x}& + \Big( \dfrac{1}{6 \pi R_1 }- \dfrac{1}{4 \pi
l_{12}} \Big) \dfrac{F_1}{\mu} + \dfrac{1}{8 \pi} \Big[
\dfrac{\mathcal{F}(\mathbf{d}_{31}; \mathbf{e}_{x'}, \mathbf{e}_x) }{l_{13}} -
\dfrac{\mathcal{F}(\mathbf{d}_{41}; \mathbf{e}_{x'}, \mathbf{e}_x) }{l_{14}}
\Big] \dfrac{F_3}{\mu} \\ & & = \dot{R}_2 \Big( \dfrac{R_2}{l_{12}} \Big)^2 -
\dot{R}_3 \Big( \dfrac{R_3}{l_{13}} \Big)^2 \big( \mathbf{e}_x \cdot
\mathbf{d}_{31} \big) - \dot{R}_4 \Big( \dfrac{R_4}{l_{14}} \Big)^2 \big(
\mathbf{e}_x \cdot \mathbf{d}_{41} \big) \\ &-U_{1y}& + \dfrac{1}{8 \pi} \Big[
\dfrac{\mathcal{F}(\mathbf{d}_{31}; \mathbf{e}_{x'}, \mathbf{e}_y) }{l_{13}} -
\dfrac{\mathcal{F}(\mathbf{d}_{41}; \mathbf{e}_{x'}, \mathbf{e}_y) }{l_{14}}
\Big] \dfrac{F_3}{\mu} = - \dot{R}_3 \Big( \dfrac{R_3}{l_{13}} \Big)^2 \big(
\mathbf{e}_y \cdot \mathbf{d}_{31} \big) - \dot{R}_4 \Big( \dfrac{R_4}{l_{14}}
\Big)^2 \big( \mathbf{e}_y \cdot \mathbf{d}_{41} \big) \\ &-U_{2x}& + \Big( -
\dfrac{1}{6 \pi R_2 } + \dfrac{1}{4 \pi l_{21}} \Big) \dfrac{F_1}{\mu} +
\dfrac{1}{8 \pi} \Big[ \dfrac{\mathcal{F}(\mathbf{d}_{32}; \mathbf{e}_{x'},
\mathbf{e}_x) }{l_{23}} - \dfrac{\mathcal{F}(\mathbf{d}_{42}; \mathbf{e}_{x'},
\mathbf{e}_x) }{l_{24}} \Big] \dfrac{F_3}{\mu} \\ & & = - \dot{R}_1 \Big(
\dfrac{R_1}{l_{21}} \Big)^2 - \dot{R}_3 \Big( \dfrac{R_3}{l_{23}} \Big)^2 \big(
\mathbf{e}_x \cdot \mathbf{d}_{32} \big) - \dot{R}_4 \Big( \dfrac{R_4}{l_{24}}
\Big)^2 \big( \mathbf{e}_x \cdot \mathbf{d}_{42} \big) \\ &-U_{2y}& +
\dfrac{1}{8 \pi} \Big[ \dfrac{\mathcal{F}(\mathbf{d}_{32}; \mathbf{e}_{x'},
\mathbf{e}_y) }{l_{23}} - \dfrac{\mathcal{F}(\mathbf{d}_{42}; \mathbf{e}_{x'},
\mathbf{e}_y) }{l_{24}} \Big] \dfrac{F_3}{\mu} = - \dot{R}_3 \Big(
\dfrac{R_3}{l_{23}} \Big)^2 \big( \mathbf{e}_y \cdot \mathbf{d}_{32} \big) -
\dot{R}_4 \Big( \dfrac{R_4}{l_{24}} \Big)^2 \big( \mathbf{e}_y \cdot
\mathbf{d}_{42} \big) \\ &-U_{3x'}& + \Big( \dfrac{1}{6 \pi R_3 } - \dfrac{1}{4
\pi l_{34}} \Big) \dfrac{F_3}{\mu} + \dfrac{1}{8 \pi} \Big[
\dfrac{\mathcal{F}(\mathbf{d}_{13}; \mathbf{e}_{x}, \mathbf{e}_{x'}) }{l_{31}} -
\dfrac{\mathcal{F}(\mathbf{d}_{23}; \mathbf{e}_{x}, \mathbf{e}_{x'}) }{l_{32}}
\Big] \dfrac{F_1}{\mu} \\ & & = - \dot{R}_1 \Big( \dfrac{R_1}{l_{31}} \Big)^2
\big( \mathbf{e}_{x'} \cdot \mathbf{d}_{13} \big) - \dot{R}_2 \Big(
\dfrac{R_2}{l_{32}} \Big)^2 \big( \mathbf{e}_{x'} \cdot \mathbf{d}_{23} \big) +
\dot{R}_4 \Big( \dfrac{R_4}{l_{34}} \Big)^2 \\ &-U_{3y'}& + \dfrac{1}{8 \pi}
\Big[ \dfrac{\mathcal{F}(\mathbf{d}_{13}; \mathbf{e}_{x}, \mathbf{e}_{y'})
}{l_{31}} - \dfrac{\mathcal{F}(\mathbf{d}_{23}; \mathbf{e}_{x}, \mathbf{e}_{y'})
}{l_{32}} \Big] \dfrac{F_1}{\mu} = - \dot{R}_1 \Big( \dfrac{R_1}{l_{31}} \Big)^2
\big( \mathbf{e}_{y'} \cdot \mathbf{d}_{13} \big) - \dot{R}_2 \Big(
\dfrac{R_2}{l_{32}} \Big)^2 \big( \mathbf{e}_{y'} \cdot \mathbf{d}_{23} \big) \\
&-U_{4x'}& + \Big( - \dfrac{1}{6 \pi R_4 } + \dfrac{1}{4 \pi l_{43}} \Big)
\dfrac{F_3}{\mu} + \dfrac{1}{8 \pi} \Big[ \dfrac{\mathcal{F}(\mathbf{d}_{14};
\mathbf{e}_{x}, \mathbf{e}_{x'}) }{l_{41}} - \dfrac{\mathcal{F}(\mathbf{d}_{24};
\mathbf{e}_{x}, \mathbf{e}_{x'}) }{l_{42}} \Big] \dfrac{F_1}{\mu} \\ & & = -
\dot{R}_1 \Big( \dfrac{R_1}{l_{41}} \Big)^2 \big( \mathbf{e}_{x'} \cdot
\mathbf{d}_{14} \big) - \dot{R}_2 \Big( \dfrac{R_2}{l_{42}} \Big)^2 \big(
\mathbf{e}_{x'} \cdot \mathbf{d}_{24} \big) - \dot{R}_3 \Big(
\dfrac{R_3}{l_{43}} \Big)^2 \\ &-U_{4y'}& + \dfrac{1}{8 \pi} \Big[
\dfrac{\mathcal{F}(\mathbf{d}_{14}; \mathbf{e}_{x}, \mathbf{e}_{y'}) }{l_{41}} -
\dfrac{\mathcal{F}(\mathbf{d}_{24}; \mathbf{e}_{x}, \mathbf{e}_{y'}) }{l_{42}}
\Big] \dfrac{F_1}{\mu} = - \dot{R}_1 \Big( \dfrac{R_1}{l_{41}} \Big)^2 \big(
\mathbf{e}_{y'} \cdot \mathbf{d}_{14} \big) - \dot{R}_2 \Big(
\dfrac{R_2}{l_{42}} \Big)^2 \big( \mathbf{e}_{y'} \cdot \mathbf{d}_{24} \big) \\
&U_{2x}& - U_{1x} = \dot{l}_{12} \\ &U_{4x'}& - U_{3x'} = \dot{l}_{34}
\end{eqnarray*}
where
\begin{eqnarray*} \mathcal{F} ( \mathbf{d}; \mathbf{e}_\alpha, \mathbf{e}_\beta)
= \big( \mathbf{e}_\alpha \cdot \mathbf{e}_\beta \big) + \big( \mathbf{e}_\alpha
\cdot \mathbf{d} \big)\big( \mathbf{e}_\beta \cdot \mathbf{d} \big).
\end{eqnarray*}
%


\section{Asymptotic analysis of the three systems consisting of two hobbled  PMPYs}\label{Appendix.F}
\setcounter{equation}{0}
\renewcommand{\theequation}{\ref{Appendix.F}.\arabic{equation}}

For simplicity, we denote the distance between any two spheres $i$ and $j$ by $l_{ij}$. For example, $l_{13} = l_{12} + l_{23}$.


\subsection{System A (two dumb-bells) with controls in $(\dot{l}_I, \dot{l}_{II})$}

The asymptotic behavior of the velocity of the spheres  is
\begin{eqnarray*}
U_1 &\sim & \dfrac{F_1}{6 \pi \mu R_1} + \dfrac{F_2}{4 \pi \mu l_{12}} + \dfrac{F_3}{4 \pi \mu l_{13}} + \dfrac{F_4}{4 \pi \mu l_{14}} \\
U_2 &\sim& \dfrac{F_1}{4 \pi \mu l_{12}} + \dfrac{F_2}{6 \pi \mu R_2} + \dfrac{F_3}{4 \pi \mu l_{23}} + \dfrac{F_4}{4 \pi \mu l_{24}} \\
U_3 &\sim& \dfrac{F_1}{4 \pi \mu l_{13}} + \dfrac{F_2}{4 \pi \mu l_{23}} + \dfrac{F_3}{6 \pi \mu R_3} + \dfrac{F_4}{4 \pi \mu l_{34}} \\
U_4 &\sim& \dfrac{F_1}{4 \pi \mu l_{14}} + \dfrac{F_2}{4 \pi \mu l_{24}} + \dfrac{F_3}{4 \pi \mu l_{34}} + \dfrac{F_4}{6 \pi \mu R_4} 
\end{eqnarray*}
with relations and constraints 
\begin{eqnarray*}
& & U_2 - U_1 = \dot{l}_{12} = \xi_{I}, \quad U_4 - U_3 = \dot{l}_{34} = \xi_{II} \\
& & F_1 + F_2 = 0, \quad F_3 + F_4 = 0.
\end{eqnarray*}
After scaling with hat notation omitted, the system becomes
\begin{eqnarray*}
U_1 &\sim& \dfrac{F_1}{R_1} + \dfrac{3}{2} \Big[ \dfrac{F_2}{l_{12}} + \dfrac{F_3}{l_{13}} + \dfrac{F_4}{l_{14}} \Big] \delta \\
U_2 &\sim& \dfrac{F_2}{R_2} + \dfrac{3}{2} \Big[ \dfrac{F_1}{l_{12}} + \dfrac{F_3}{l_{23}} + \dfrac{F_4}{l_{24}} \Big] \delta  \\
U_3 &\sim& \dfrac{F_3}{R_3} + \dfrac{3}{2} \Big[ \dfrac{F_1}{l_{13}} + \dfrac{F_2}{l_{23}} + \dfrac{F_4}{l_{34}} \Big] \delta  \\
U_4 &\sim& \dfrac{F_4}{R_4} + \dfrac{3}{2} \Big[ \dfrac{F_1}{l_{14}} + \dfrac{F_2}{l_{24}} + \dfrac{F_3}{l_{34}} \Big] \delta \\
 U_2 - U_1 = \dot{l}_{12} &=& \xi_{I}, \quad U_4 - U_3 = \dot{l}_{34} = \xi_{II}  \\
 F_1 + F_2 &=& 0, \quad F_3 + F_4 = 0.
\end{eqnarray*}
With $F_2 = - F_1, \ F_4 = -F_3$, the system is simplified to 
\begin{eqnarray*}
U_1 &\sim& F_1 \Big( \dfrac{1}{R_1} - \dfrac{3 \delta}{2 l_{12}} \Big) + \dfrac{3}{2} F_3 \delta \Big( \dfrac{1}{l_{13}} - \dfrac{1}{l_{14}} \Big) \\
U_2 &\sim& F_1 \Big( \dfrac{3 \delta}{2 l_{12}} - \dfrac{1}{R_2} \Big) + \dfrac{3}{2} F_3 \delta \Big( \dfrac{1}{l_{23}} - \dfrac{1}{l_{24}} \Big)  \\
U_3 &\sim& \dfrac{3}{2} F_1 \delta \Big( \dfrac{1}{l_{13}} - \dfrac{1}{l_{23}} \Big) +  F_3 \Big( \dfrac{1}{R_3} - \dfrac{3 \delta}{2 l_{34}} \Big)   \\
U_4 &\sim& \dfrac{3}{2} F_1 \delta \Big( \dfrac{1}{l_{14}} - \dfrac{1}{l_{24}} \Big) + F_3 \Big( \dfrac{3 \delta}{2 l_{34}}  - \dfrac{1}{R_3}  \Big).
\end{eqnarray*}
When  $R_1 = R_2 = R_3 = R_4 = R$,
\begin{eqnarray*}
\xi_I &=& U_2 - U_1 \sim - F_1 \Big( \dfrac{1}{R_1} + \dfrac{1}{R_2}  \Big)  \quad \Rightarrow \quad F_1 = -\dfrac{R}{2} \xi_I  \\
\xi_{II} &=& U_4 - U_3 \sim - F_3 \Big( \dfrac{1}{R_3} + \dfrac{1}{R_4}  \Big)
\quad \Rightarrow \quad F_3 = -\dfrac{R}{2} \xi_{II} .
\end{eqnarray*}
Velocities of the PMPYs:
\begin{eqnarray*}
\overline{U}_I &=& \dfrac{1}{2} \Big( U_1 + U_2 \Big) 
= \dfrac{1}{2} F_1 \Big( \dfrac{1}{R_1} - \dfrac{1}{R_2} \Big) + \dfrac{3}{4} F_3 \delta \Big( \dfrac{1}{l_{13}} - \dfrac{1}{l_{14}} + \dfrac{1}{l_{23}} - \dfrac{1}{l_{24}} \Big)  
 \sim  - \dfrac{3}{8} R \xi_{II} \delta  \Big( \dfrac{1}{l_{13}} - \dfrac{1}{l_{14}} + \dfrac{1}{l_{23}} - \dfrac{1}{l_{24}} \Big) \\
 \overline{U}_{II} &=& \dfrac{1}{2} \Big( U_3 + U_4 \Big) 
 = \dfrac{3}{4} F_1 \delta \Big( \dfrac{1}{l_{13}} - \dfrac{1}{l_{23}} + \dfrac{1}{l_{14}} - \dfrac{1}{l_{24}}  \Big) + \dfrac{1}{2} F_1 \Big( \dfrac{1}{R_3} - \dfrac{1}{R_4} \Big)
 \sim  - \dfrac{3}{8} R \xi_{I} \delta  \Big( \dfrac{1}{l_{13}} - \dfrac{1}{l_{23}} + \dfrac{1}{l_{14}} - \dfrac{1}{l_{24}}  \Big) 
\end{eqnarray*}
Power of the PMPYs:
\begin{eqnarray*}
P_I &=& F_1 U_1 + F_2 U_2 = F_1 \big( U_1 - U_2 \big) = - F_1 \xi_I = \dfrac{R}{2} \xi_I^2  \\
P_{II} &=& F_3 U_3 + F_4 U_4 = F_3 \big( U_3 - U_4 \big) = - F_3 \xi_{II} = \dfrac{R}{2} \xi_{II}^2
\end{eqnarray*}

However, we observe that although $\overline{U}_{I}, \overline{U}_{II} $ scale
like $O (\delta)$, the net translations $X_{I}, X_{II}$ do not.  Without loss of
generality, we consider the first integral term in $X_{I }$:  
\begin{eqnarray*}
\int \overline{U}_{I} d t =- \dfrac{3}{8} R \delta \int_0^1 \dfrac{\xi_{II} (t)}{ l_{13} (t)} dt
\end{eqnarray*}
$l_{13}$ can be written as $l_{13} = l_{12} + l_{23} = 2 + \Delta l_{12} + \Delta l_{23}$, where $\Delta l_{12}, \Delta l_{23} \sim O(\delta)$, thus
\begin{eqnarray*}
\dfrac{1}{l_{13}} = \dfrac{1}{2 + \Delta l_{12} + \Delta l_{23}} = \dfrac{1}{2} - \dfrac{\Delta l_{12} + \Delta l_{23}}{4} + O (\delta^2)
\end{eqnarray*}
and
\begin{eqnarray*}
- \dfrac{3}{8} R \delta \int_0^1 \dfrac{\xi_{II} (t)}{ l_{13} (t)} dt
&=& - \dfrac{3}{8} R \delta \int_0^1 \xi_{II} \Big(  \dfrac{1}{2} - \dfrac{\Delta l_{12} + \Delta l_{23}}{4} \Big) dt + O (\delta^3) \\
&=& - \dfrac{3}{16} R \delta  \int_0^1 \xi_{II} dt + \dfrac{3}{32} R \delta  \int_0^1 \xi_{II} ( \Delta l_{12} + \Delta l_{23} ) dt + O (\delta^3).
\end{eqnarray*}
The first integral vanishes since it is an exact integral, the second one is an $O (\delta^2)$ term, hence
\begin{eqnarray*}
- \dfrac{3}{8} R \delta \int_0^1 \dfrac{\xi_{II} (t)}{ l_{13} (t)} dt \sim O (\delta^2).
\end{eqnarray*}
A similar argument applies to all other integrals in $X_{I}, X_{II}$, thus at most $X_{I}, X_{II} \sim O (\delta^2)$.

When the radii are not all equal, the leading order term  of
$\overline{U}_{I}, \overline{U}_{II} $ scales as $O (1)$, but  this does
not contribute to the net translations and we still have $X_{I}, X_{II} \sim O
(\delta^2)$. This is because the leading order terms are of the form $\Phi \xi
(t)$, where the coefficient $\Phi$ depends on the radii only, thus $\int \Phi
\xi dt$ is again an exact integral. Also  this case, the $O(\delta)$ terms in
$\overline{U}_{I}, \overline{U}_{II} $ become more complicated, but they are
still of the form $\Phi \xi / l_{ij}$, where $\Phi$ depends on the radii only,
and the same argument above applies, thus in the end we have  $X_{I},
X_{II} \sim O (\delta^2)$.


\subsection{System B with controls in $(\dot{R}_1, \dot{R}_3)$}\label{Appendix.ControlRR}

The asymptotic behavior of the velocity of each sphere is 
\begin{eqnarray*}
U_1 &\sim& \dfrac{F_1}{6 \pi \mu R_1} + \Big[ \dfrac{F_2}{4 \pi \mu l_{12}} - \Big( \dfrac{R_2}{l_{12}} \Big)^2 \dot{R}_2 \Big] + 
\Big[ \dfrac{F_3}{4 \pi \mu l_{13}} - \Big( \dfrac{R_3}{l_{13}} \Big)^2 \dot{R}_3 \Big] + \Big[ \dfrac{F_4}{4 \pi \mu l_{14}} - \Big( \dfrac{R_4}{l_{14}} \Big)^2 \dot{R}_4 \Big]  \\
U_2 &\sim& \dfrac{F_2}{6 \pi \mu R_2} + \Big[ \dfrac{F_1}{4 \pi \mu l_{12}} + \Big( \dfrac{R_1}{l_{12}} \Big)^2 \dot{R}_1 \Big] + 
\Big[ \dfrac{F_3}{4 \pi \mu l_{23}} - \Big( \dfrac{R_3}{l_{23}} \Big)^2 \dot{R}_3 \Big] + \Big[ \dfrac{F_4}{4 \pi \mu l_{24}} - \Big( \dfrac{R_4}{l_{24}} \Big)^2 \dot{R}_4 \Big]  \\
U_3 &\sim& \dfrac{F_3}{6 \pi \mu R_3} + \Big[ \dfrac{F_1}{4 \pi \mu l_{13}} + \Big( \dfrac{R_1}{l_{13}} \Big)^2 \dot{R}_1 \Big] + 
\Big[ \dfrac{F_2}{4 \pi \mu l_{23}} + \Big( \dfrac{R_2}{l_{23}} \Big)^2 \dot{R}_2 \Big] + \Big[ \dfrac{F_4}{4 \pi \mu l_{34}} - \Big( \dfrac{R_4}{l_{34}} \Big)^2 \dot{R}_4 \Big]   \\
U_4 &\sim& \dfrac{F_4}{6 \pi \mu R_4} + \Big[ \dfrac{F_1}{4 \pi \mu l_{14}} + \Big( \dfrac{R_1}{l_{14}} \Big)^2 \dot{R}_1 \Big] + 
\Big[ \dfrac{F_2}{4 \pi \mu l_{24}} + \Big( \dfrac{R_2}{l_{24}} \Big)^2 \dot{R}_2 \Big] + \Big[ \dfrac{F_3}{4 \pi \mu l_{34}} + \Big( \dfrac{R_3}{l_{34}} \Big)^2 \dot{R}_3 \Big] 
\end{eqnarray*}
with relations
\begin{eqnarray*}
& & U_2 - U_1 = 0, \quad U_4 - U_3 = 0 \\
& & F_1 + F_2 = 0, \quad F_3 + F_4 = 0 \\
& & R_1^2 \dot{R}_1 + R_2^2 \dot{R}_2 = 0 \quad \Rightarrow \quad R_2^2 \zeta_2 = - R_1^2 \zeta_1  \\
& & R_3^2 \dot{R}_3 + R_4^2 \dot{R}_4 = 0 \quad \Rightarrow \quad R_4^2 \zeta_4 = - R_3^2 \zeta_3 .
\end{eqnarray*}
After scaling with hat notation omitted
\begin{eqnarray*}
U_1 &\sim& \dfrac{F_1}{R_1} +  \Big[ \dfrac{3 F_2}{2 l_{12}} \delta - \Big(\dfrac{R_2}{l_{12}} \Big)^2 \zeta_2 \delta^2 \Big]
+  \Big[ \dfrac{3 F_3}{2 l_{13}} \delta - \Big(\dfrac{R_3}{l_{13}} \Big)^2 \zeta_3 \delta^2 \Big] +  \Big[ \dfrac{3 F_4}{2 l_{14}} \delta - \Big(\dfrac{R_4}{l_{14}} \Big)^2 \zeta_4 \delta^2 \Big]  \\
U_2 &\sim& \dfrac{F_2}{R_2} +  \Big[ \dfrac{3 F_1}{2 l_{12}} \delta + \Big(\dfrac{R_1}{l_{12}} \Big)^2 \zeta_1 \delta^2 \Big]
+  \Big[ \dfrac{3 F_3}{2 l_{23}} \delta - \Big(\dfrac{R_3}{l_{23}} \Big)^2 \zeta_3 \delta^2 \Big] +  \Big[ \dfrac{3 F_4}{2 l_{24}} \delta - \Big(\dfrac{R_4}{l_{24}} \Big)^2 \zeta_4 \delta^2 \Big]  \\
U_3 &\sim& \dfrac{F_3}{R_3} +  \Big[ \dfrac{3 F_1}{2 l_{13}} \delta + \Big(\dfrac{R_1}{l_{13}} \Big)^2 \zeta_1 \delta^2 \Big]
+  \Big[ \dfrac{3 F_2}{2 l_{23}} \delta + \Big(\dfrac{R_3}{l_{23}} \Big)^2 \zeta_2 \delta^2 \Big] +  \Big[ \dfrac{3 F_4}{2 l_{34}} \delta - \Big(\dfrac{R_4}{l_{34}} \Big)^2 \zeta_4 \delta^2 \Big]  \\
U_4 &\sim& \dfrac{F_4}{R_4} +  \Big[ \dfrac{3 F_1}{2 l_{14}} \delta + \Big(\dfrac{R_1}{l_{14}} \Big)^2 \zeta_1 \delta^2 \Big]
+  \Big[ \dfrac{3 F_2}{2 l_{24}} \delta + \Big(\dfrac{R_3}{l_{24}} \Big)^2 \zeta_2 \delta^2 \Big] +  \Big[ \dfrac{3 F_3}{2 l_{34}} \delta + \Big(\dfrac{R_3}{l_{34}} \Big)^2 \zeta_3 \delta^2 \Big].
\end{eqnarray*}
With $F_2 = - F_1, \ F_4 = - F_3$, the system becomes
\begin{eqnarray*}
U_1 &\sim& F_1 \Big( \dfrac{1}{R_1} - \dfrac{3 \delta}{2 l_{12}} \Big) + \dfrac{3 F_3}{2} \delta \Big( \dfrac{1}{l_{13}} - \dfrac{1}{l_{14}} \Big) 
+ \delta^2 \Big[ - \Big( \dfrac{R_2}{l_{12}} \Big)^2 \zeta_2 - \Big( \dfrac{R_3}{l_{13}} \Big)^2 \zeta_3 - \Big( \dfrac{R_4}{l_{14}} \Big)^2 \zeta_4 \Big]  \\
U_2 &\sim& F_1 \Big( - \dfrac{1}{R_2} + \dfrac{3 \delta}{2 l_{12}} \Big) + \dfrac{3 F_3}{2} \delta \Big( \dfrac{1}{l_{23}} - \dfrac{1}{l_{24}} \Big) 
+ \delta^2 \Big[ \Big( \dfrac{R_1}{l_{12}} \Big)^2 \zeta_1 - \Big( \dfrac{R_3}{l_{23}} \Big)^2 \zeta_3 - \Big( \dfrac{R_4}{l_{24}} \Big)^2 \zeta_4 \Big]   \\
U_3 &\sim&  \dfrac{3 F_1}{2} \delta \Big( \dfrac{1}{l_{13}} - \dfrac{1}{l_{23}} \Big) +  F_3 \Big( \dfrac{1}{R_3} - \dfrac{3 \delta}{2 l_{34}} \Big) + 
+ \delta^2 \Big[ \Big( \dfrac{R_1}{l_{13}} \Big)^2 \zeta_1 + \Big( \dfrac{R_2}{l_{23}} \Big)^2 \zeta_2 - \Big( \dfrac{R_4}{l_{34}} \Big)^2 \zeta_4 \Big]  \\
U_4 &\sim&  \dfrac{3 F_1}{2} \delta \Big( \dfrac{1}{l_{14}} - \dfrac{1}{l_{24}} \Big) +  F_3 \Big( - \dfrac{1}{R_4} + \dfrac{3 \delta}{2 l_{34}} \Big) + 
+ \delta^2 \Big[ \Big( \dfrac{R_1}{l_{14}} \Big)^2 \zeta_1 + \Big( \dfrac{R_2}{l_{24}} \Big)^2 \zeta_2 + \Big( \dfrac{R_3}{l_{34}} \Big)^2 \zeta_3 \Big].
\end{eqnarray*}
Expand $F_1, F_3$ as
\begin{eqnarray*}
F_1 = F_1^0 + \delta F_1^1 + \delta^2 F_1^2 + O (\delta^3), \quad F_3 = F_3^0 +
\delta F_3^1 + \delta^2 F_3^2 + O (\delta^3); 
\end{eqnarray*}
then
\begin{eqnarray*}
U_1 &\sim& \dfrac{F_1^0}{R_1} + \delta \dfrac{F_1^1}{R_1} -  \delta \dfrac{3 F_1^0}{2 l_{12}} + \delta^2 \dfrac{F_1^2}{R_1} - \delta^2 \dfrac{3 F_1^1}{2 l_{12}} 
+ \dfrac{3}{2} \delta F_3^0 \Big( \dfrac{1}{l_{13}} - \dfrac{1}{l_{14}} \Big) + \dfrac{3}{2} \delta^2 F_3^1 \Big( \dfrac{1}{l_{13}} - \dfrac{1}{l_{14}} \Big) \\
& & + \delta^2 \Big[ - \Big( \dfrac{R_2}{l_{12}} \Big)^2 \zeta_2 - \Big( \dfrac{R_3}{l_{13}} \Big)^2 \zeta_3 - \Big( \dfrac{R_4}{l_{14}} \Big)^2 \zeta_4 \Big] + O (\delta^3) \\
U_2 &\sim& - \dfrac{F_1^0}{R_2} - \delta \dfrac{F_1^1}{R_2} +  \delta \dfrac{3 F_1^0}{2 l_{12}} - \delta^2 \dfrac{F_1^2}{R_2} + \delta^2 \dfrac{3 F_1^1}{2 l_{12}} 
+ \dfrac{3}{2} \delta F_3^0 \Big( \dfrac{1}{l_{23}} - \dfrac{1}{l_{24}} \Big) + \dfrac{3}{2} \delta^2 F_3^1 \Big( \dfrac{1}{l_{23}} - \dfrac{1}{l_{24}} \Big) \\
& & + \delta^2 \Big[  \Big( \dfrac{R_1}{l_{12}} \Big)^2 \zeta_1 - \Big( \dfrac{R_3}{l_{23}} \Big)^2 \zeta_3 - \Big( \dfrac{R_4}{l_{24}} \Big)^2 \zeta_4 \Big] + O (\delta^3) \\
U_3 &\sim& \dfrac{F_3^0}{R_3} + \delta \dfrac{F_3^1}{R_3} -  \delta \dfrac{3 F_3^0}{2 l_{34}} + \delta^2 \dfrac{F_3^2}{R_3} - \delta^2 \dfrac{3 F_3^1}{2 l_{34}} 
+ \dfrac{3}{2} \delta F_1^0 \Big( \dfrac{1}{l_{13}} - \dfrac{1}{l_{23}} \Big) + \dfrac{3}{2} \delta^2 F_1^1 \Big( \dfrac{1}{l_{13}} - \dfrac{1}{l_{23}} \Big) \\
& & + \delta^2 \Big[  \Big( \dfrac{R_1}{l_{13}} \Big)^2 \zeta_1 + \Big( \dfrac{R_2}{l_{23}} \Big)^2 \zeta_2 - \Big( \dfrac{R_4}{l_{34}} \Big)^2 \zeta_4 \Big] + O (\delta^3) \\
U_4 &\sim& -  \dfrac{F_3^0}{R_4} - \delta \dfrac{F_3^1}{R_4} +  \delta \dfrac{3 F_3^0}{2 l_{34}} - \delta^2 \dfrac{F_3^2}{R_4} + \delta^2 \dfrac{3 F_3^2}{2 l_{34}} 
+ \dfrac{3}{2} \delta F_1^0 \Big( \dfrac{1}{l_{14}} - \dfrac{1}{l_{24}} \Big) + \dfrac{3}{2} \delta^2 F_1^1 \Big( \dfrac{1}{l_{14}} - \dfrac{1}{l_{24}} \Big) \\
& & + \delta^2 \Big[  \Big( \dfrac{R_1}{l_{14}} \Big)^2 \zeta_1 + \Big( \dfrac{R_2}{l_{24}} \Big)^2 \zeta_2 + \Big( \dfrac{R_3}{l_{34}} \Big)^2 \zeta_3 \Big] + O (\delta^3).
\end{eqnarray*}
Compare the $O(1)$  terms:
\begin{eqnarray*}
U_1^0 = U_2^0 \ \Rightarrow \ \dfrac{F_1^0}{R_1} = - \dfrac{F_1^0}{R_2} \ \Rightarrow \ F_1^0 = 0, \quad \textrm{similarly, } \quad F_3^0 = 0.
\end{eqnarray*}
Compare the $O (\delta)$  terms:
\begin{eqnarray*}
U_1^1  = U_2^1 \ \Rightarrow \  \dfrac{F_1^1}{R_1} =  - \dfrac{F_1^1}{R_2} \ \Rightarrow \  F_1^1 = 0, \quad \textrm{similarly, } \quad F_3^1 = 0.
\end{eqnarray*}
Compare the $O (\delta^2)$  terms:
\begin{eqnarray*}
U_1^2 &=& \dfrac{F_1^2}{R_1} + \Big[ - \Big( \dfrac{R_2}{l_{12}} \Big)^2 \zeta_2 - \Big( \dfrac{R_3}{l_{13}} \Big)^2 \zeta_3 - \Big( \dfrac{R_4}{l_{14}} \Big)^2 \zeta_4 \Big]  \\
U_2^2 &=& - \dfrac{F_1^2}{R_2} + \Big[ \Big( \dfrac{R_1}{l_{12}} \Big)^2 \zeta_1 - \Big( \dfrac{R_3}{l_{23}} \Big)^2 \zeta_3 - \Big( \dfrac{R_4}{l_{24}} \Big)^2 \zeta_4 \Big]  \\
U_1^2 = U_2^2 &\Rightarrow& F_1^2 = \dfrac{R_1 R_2}{R_1 + R_2} \Big( \dfrac{1}{l_{13}^2} - \dfrac{1}{l_{23}^2} - \dfrac{1}{l_{14}^2} + \dfrac{1}{l_{24}^2} \Big) R_3^2 \zeta_3
\end{eqnarray*}
and similarly,
\begin{eqnarray*}
U_3^2 &=& \dfrac{F_3^2}{R_3} + \Big[  \Big( \dfrac{R_1}{l_{13}} \Big)^2 \zeta_1 + \Big( \dfrac{R_2}{l_{23}} \Big)^2 \zeta_2 - \Big( \dfrac{R_4}{l_{34}} \Big)^2 \zeta_4 \Big]  \\
U_4^2&=& -  \dfrac{F_3^2}{R_4} + \Big[  \Big( \dfrac{R_1}{l_{14}} \Big)^2 \zeta_1 + \Big( \dfrac{R_2}{l_{24}} \Big)^2 \zeta_2 + \Big( \dfrac{R_3}{l_{34}} \Big)^2 \zeta_3 \Big]  \\
U_3^2 = U_4^2 &\Rightarrow& F_3^2 = \dfrac{R_3 R_4}{R_3 + R_4} \Big( \dfrac{1}{l_{14}^2} - \dfrac{1}{l_{13}^2} - \dfrac{1}{l_{24}^2} + \dfrac{1}{l_{23}^2} \Big) R_1^2 \zeta_1.
\end{eqnarray*}
Velocities of the PMPYs:
\begin{eqnarray*}
U_I &=& U_1 = U_2 = \dfrac{R_1^2}{l_{12}^2} \zeta_1 \delta^2 + R_3^2 \zeta_3 \delta^2 \Big[ \dfrac{R_1}{R_1 + R_2} \Big( \dfrac{1}{l_{14}^2} - \dfrac{1}{l_{13}^2} \Big) 
+   \dfrac{R_2}{R_1 + R_2} \Big( \dfrac{1}{l_{24}^2} - \dfrac{1}{l_{23}^2} \Big)  \Big]  \\
U_{II} &=& U_3 = U_4 = \dfrac{R_3^2}{l_{34}^2} \zeta_3 \delta^2 + R_1^2 \zeta_1 \delta^2 \Big[ \dfrac{R_3}{R_3 + R_4} \Big( \dfrac{1}{l_{13}^2} - \dfrac{1}{l_{23}^2} \Big) 
+   \dfrac{R_4}{R_3 + R_4} \Big( \dfrac{1}{l_{14}^2} - \dfrac{1}{l_{24}^2} \Big)  \Big] 
\end{eqnarray*}
Power of the PMPYs:
\begin{eqnarray*}
P_I &=& \dfrac{8}{3} \big( R_1 \zeta_1^2 + R_2 \zeta_2^2 \big) = \dfrac{8}{3} R_1 \dfrac{R_1^3 + R_2^3}{R_2^3} \zeta_1^2 \\
P_{II} &=& \dfrac{8}{3} \big( R_3 \zeta_3^2 + R_4 \zeta_4^2 \big) = \dfrac{8}{3} R_3 \dfrac{R_3^3 + R_4^3}{R_4^3} \zeta_3^2 
\end{eqnarray*}


\subsection{System C with controls in $(\dot{l}_{I}, \dot{R}_3)$}

The asymptotic behavior of the velocity of each sphere is
\begin{eqnarray*}
U_1 &\sim& \dfrac{F_1}{6 \pi \mu R_1} + \dfrac{F_2}{4 \pi \mu l_{12}} + \Big[ \dfrac{F_3}{4 \pi \mu l_{13}} - \Big( \dfrac{R_3}{l_{13}} \Big)^2 \dot{R}_3 \Big]
+ \Big[ \dfrac{F_4}{4 \pi \mu l_{14}} - \Big( \dfrac{R_4}{l_{14}} \Big)^2 \dot{R}_4 \Big]  \\
U_2 &\sim& \dfrac{F_2}{6 \pi \mu R_2} + \dfrac{F_1}{4 \pi \mu l_{12}} + \Big[ \dfrac{F_3}{4 \pi \mu l_{23}} - \Big( \dfrac{R_3}{l_{23}} \Big)^2 \dot{R}_3 \Big]
+ \Big[ \dfrac{F_4}{4 \pi \mu l_{24}} - \Big( \dfrac{R_4}{l_{24}} \Big)^2 \dot{R}_4 \Big]  \\
U_3 &\sim& \dfrac{F_3}{6 \pi \mu R_3} + \dfrac{F_1}{4 \pi \mu l_{13}} + \dfrac{F_2}{4 \pi \mu l_{23}} + \Big[ \dfrac{F_4}{4 \pi \mu l_{34}} - \Big( \dfrac{R_4}{l_{34}} \Big)^2 \dot{R}_4 \Big]  \\
U_4 &\sim& \dfrac{F_4}{6 \pi \mu R_4} + \dfrac{F_1}{4 \pi \mu l_{14}} + \dfrac{F_2}{4 \pi \mu l_{24}} + \Big[ \dfrac{F_3}{4 \pi \mu l_{34}} + \Big( \dfrac{R_3}{l_{34}} \Big)^2 \dot{R}_3 \Big] 
\end{eqnarray*}  
with relations
\begin{eqnarray*}
& & U_2 - U_1 = \dot{l}_{12} = \xi_I, \quad U_4 - U_3 = 0 \\
& & F_1 + F_2 = 0 , \quad F_3 + F_4 = 0 \\
& & R_3^2 \zeta_3 + R_4^2 \zeta_4 = 0 \ \Rightarrow \ R_4^2 \zeta_4 = - R_3^2 \zeta_3.
\end{eqnarray*}
After scaling
\begin{eqnarray*}
U_1 &\sim& \dfrac{F_1}{R_1} + \dfrac{3 F_2}{2 l_{12}} \delta + \Big[ \dfrac{2 F_3}{2 l _{13}} \delta - \Big( \dfrac{R_3}{l_{13}} \Big)^2 \zeta_3 \delta^2 \Big]
+ \Big[ \dfrac{3 F_4}{2 l_{14}} \delta - \Big( \dfrac{R_4}{l_{14}} \Big)^2 \zeta_4 \delta^2 \Big]  \\
U_2 &\sim& \dfrac{F_2}{R_2} + \dfrac{3 F_1}{2 l_{12}} \delta + \Big[ \dfrac{2 F_3}{2 l _{23}} \delta - \Big( \dfrac{R_3}{l_{23}} \Big)^2 \zeta_3 \delta^2 \Big]
+ \Big[ \dfrac{3 F_4}{2 l_{24}} \delta - \Big( \dfrac{R_4}{l_{24}} \Big)^2 \zeta_4 \delta^2 \Big] \\
U_3 &\sim& \dfrac{F_3}{R_3} + \dfrac{3 F_1}{2 l_{13}} \delta + \dfrac{3 F_2}{2 l_{23}} \delta
+ \Big[ \dfrac{3 F_4}{2 l_{34}} \delta - \Big( \dfrac{R_4}{l_{34}} \Big)^2 \zeta_4 \delta^2 \Big]  \\
U_4 &\sim& \dfrac{F_4}{R_4} + \dfrac{3 F_1}{2 l_{14}} \delta + \dfrac{3 F_2}{2 l_{24}} \delta
+ \Big[ \dfrac{3 F_3}{2 l_{34}} \delta + \Big( \dfrac{R_3}{l_{34}} \Big)^2 \zeta_3 \delta^2 \Big].
\end{eqnarray*}
With  $F_2 = - F_1, \ F_4 = - F_3$, the system is simplified to
\begin{eqnarray*}
U_1 &\sim& \dfrac{F_1}{R_1} - \dfrac{3 F_1}{2 l_{12}} \delta + \Big[ \dfrac{3 F_3}{2 l_{13}} \delta - \Big( \dfrac{R_3}{l_{13}} \Big)^2 \zeta_3 \delta^2 \Big] 
+ \Big[ - \dfrac{3 F_3}{2 l_{14}} \delta + \Big( \dfrac{R_3}{l_{14}} \Big)^2 \zeta_3 \delta^2 \Big]  \\
U_2 &\sim& - \dfrac{F_1}{R_2} + \dfrac{3 F_1}{2 l_{12}} \delta + \Big[ \dfrac{3 F_3}{2 l_{23}} \delta - \Big( \dfrac{R_3}{l_{13}} \Big)^2 \zeta_3 \delta^2 \Big] 
+ \Big[ - \dfrac{3 F_3}{2 l_{24}} \delta + \Big( \dfrac{R_3}{l_{24}} \Big)^2 \zeta_3 \delta^2 \Big]  \\
U_3 &\sim&  \dfrac{F_3}{R_3} + \dfrac{3 F_1}{2 l_{13}} \delta - \dfrac{3 F_1}{2 l_{23}} \delta +  
+ \Big[ - \dfrac{3 F_3}{2 l_{34}} \delta + \Big( \dfrac{R_3}{l_{34}} \Big)^2 \zeta_3 \delta^2 \Big]  \\
U_4 &\sim& - \dfrac{F_3}{R_4} + \dfrac{3 F_1}{2 l_{14}} \delta - \dfrac{3 F_1}{2 l_{24}} \delta +  
+ \Big[ \dfrac{3 F_3}{2 l_{34}} \delta + \Big( \dfrac{R_3}{l_{34}} \Big)^2 \zeta_3 \delta^2 \Big]. 
\end{eqnarray*}
For PMPY I, 
\begin{eqnarray*}
\xi_I = U_2 - U_1 \sim - F_1 \big( \dfrac{1}{R_1} + \dfrac{1}{R_2} \big) \ \Rightarrow \ F_1 = - \dfrac{R_1 R_2}{R_1 + R_2} \xi_I,
\end{eqnarray*}
thus for PMPY II,
\begin{eqnarray*}
U_3 &\sim& \dfrac{F_3}{R_3} - \dfrac{3}{2} \dfrac{R_1 R_2}{R_1 + R_2} \xi_I \delta \Big( \dfrac{1}{l_{13}} - \dfrac{1}{l_{23}} \Big)
- \dfrac{3 F_3}{2 l_{34}} \delta  + \Big( \dfrac{R_3}{l_{34}} \Big)^2 \zeta_3 \delta^2 \\
U_4 &\sim& - \dfrac{F_3}{R_4} - \dfrac{3}{2} \dfrac{R_1 R_2}{R_1 + R_2} \xi_I \delta \Big( \dfrac{1}{l_{14}} - \dfrac{1}{l_{24}} \Big)
+ \dfrac{3 F_3}{2 l_{34}} \delta  + \Big( \dfrac{R_3}{l_{34}} \Big)^2 \zeta_3 \delta^2 .
\end{eqnarray*}
Expand $F_3$ as $F_3 = F_3^0 + \delta F_3^1 + \delta^2 F_3^2 + O (\delta^3)$,
\begin{eqnarray*}
U_3 &\sim& \dfrac{F_3^0}{R_3} + \delta \dfrac{F_3^1}{R_3} + \delta^2 \dfrac{F_3^2}{R_3} - \dfrac{3}{2} \dfrac{R_1 R_2}{R_1 + R_2} \xi_I \delta \Big( \dfrac{1}{l_{13}} - \dfrac{1}{l_{23}} \Big)
- \dfrac{3 F_3^0}{2 l _{34}} \delta - \dfrac{3 F_3^1}{2 l_{34}} \delta^2 + \Big( \dfrac{R_3}{l_{34}} \Big)^2 \zeta_3 \delta^2  \\
U_4 &\sim& - \dfrac{F_3^0}{R_4} - \delta \dfrac{F_3^1}{R_4} - \delta^2 \dfrac{F_3^2}{R_4} - \dfrac{3}{2} \dfrac{R_1 R_2}{R_1 + R_2} \xi_I \delta \Big( \dfrac{1}{l_{14}} - \dfrac{1}{l_{24}} \Big)
+ \dfrac{3 F_3^0}{2 l _{34}} \delta + \dfrac{3 F_3^1}{2 l_{34}} \delta^2 + \Big( \dfrac{R_3}{l_{34}} \Big)^2 \zeta_3 \delta^2 
\end{eqnarray*}
Compare the $O (1)$ terms
\begin{eqnarray*}
U_3^0 = U_4^0 \ \Rightarrow \ \dfrac{F_3^0}{R_3} = - \dfrac{F_3^0}{R_4} \ \Rightarrow \ F_3^0 = 0.
\end{eqnarray*}
Compare the $O ( \delta )$ terms
\begin{eqnarray*}
U_3^1 = U_4^1 \ \Rightarrow \ 
F_3^1 = \dfrac{3}{2} \dfrac{R_1 R_2}{R_1 + R_2} \dfrac{R_3 R_4}{R_3 + R_4} \xi_I \Big( \dfrac{1}{l_{13}} - \dfrac{1}{l_{23}} - \dfrac{1}{l_{14}} + \dfrac{1}{l_{24}} \Big)
\end{eqnarray*}
Velocities of the PMPYs:
\begin{eqnarray*}
\overline{U}_I &=& \dfrac{1}{2} (U_1 + U_2) \\
&=& \dfrac{9}{8} \dfrac{R_1 R_2 R_3 R_4}{(R_1 + R_2) (R_3 + R_4)} \xi_I \delta^2 \Big( \dfrac{1}{l_{13}} - \dfrac{1}{l_{14}} + \dfrac{1}{l_{23}} - \dfrac{1}{l_{24}} \Big)^2
+ \dfrac{1}{2} R_3^2 \zeta_3 \delta^2\Big( \dfrac{1}{l_{14}^2} - \dfrac{1}{l_{13}^2} + \dfrac{1}{l_{24}^2} - \dfrac{1}{l_{23}^2} \Big) \\
\overline{U}_{II} &=& U_3 = U_4 \\
&\sim& \dfrac{3}{2} \dfrac{R_1 R_2}{R_1 + R_2} \xi_I \delta \Big[ \dfrac{R_3}{R_3 + R_4} \Big( \dfrac{1}{l_{23}} - \dfrac{1}{l_{13}} \Big)
+ \dfrac{R_4}{R_3 + R_4} \Big( \dfrac{1}{l_{24}} - \dfrac{1}{l_{14}} \Big) \Big]
\end{eqnarray*}
Power of the PMPYs:
\begin{eqnarray*}
P_I &=& F_1 U_1 + F_2 U_2 = F_1 (U_1 - U_2) = - F_1 \xi_I = \dfrac{R_1 R_2}{R_1 + R_2} \xi_I^2 \\
P_{II} &=&\dfrac{8}{3} (R_3 \zeta_3^2 + R_4 \zeta_4^2) = \dfrac{8}{3} R_3 \dfrac{R_3^3 + R_4^3}{R_4^3} \zeta_3^2
\end{eqnarray*}


\bibliographystyle{jmb}
\bibliography{Qixuan_10052016}

\end{document}